\documentclass[letterpaper,twocolumn,10pt]{article}
\usepackage{usenix2019_v3}
\usepackage[pdftex]{graphicx}
\usepackage{subfigure}
\usepackage{soul}
\usepackage{tikz}
\usepackage{amsmath}
\usepackage{xspace}
\usepackage{array}
\usepackage{bm}  
\usepackage{enumitem}
\usepackage[english]{babel}
\usepackage{blindtext}
\usepackage{microtype}
\usepackage{graphicx}
\usepackage{subfigure}
\usepackage{booktabs} % for professional tables
\usepackage[T1]{fontenc}
\usepackage{color}
\usepackage{hyperref}
\usepackage{filecontents}

\usepackage[absolute]{textpos}

\newenvironment{myitemize}
{
   \vspace{0mm}
    \begin{list}{$\bullet$ }{}
        \setlength{\topsep}{0em}
        \setlength{\parskip}{0pt}
        \setlength{\partopsep}{0pt}
        \setlength{\parsep}{0pt}
        \setlength{\itemsep}{1mm}
}
{
    \end{list}
}

\newcommand{\dgxp}{DGX-1P\xspace}
\newcommand{\dgxv}{DGX-1V\xspace}

\newcommand{\sys}{Blink\xspace}
\newcommand{\system}{Blink\xspace}
\newcommand{\msr}{{\large$^\dag$}}
\newcommand{\uwm}{{\large$^\ddag$}}
\newcommand{\ucb}{{\large$^\star$}}

\begin{document}
%-------------------------------------------------------------------------------
%don't want date printed
%\date{}

\title{\Large \bf {\sys}: Fast and Generic Collectives for Distributed ML}

%for single author (just remove % characters)
 \author{
 \rm{Guanhua Wang\ucb, 
 Shivaram Venkataraman\uwm, 
 Amar Phanishayee\msr,}\\
 \rm{Jorgen Thelin\msr, 
 Nikhil Devanur\msr, 
 Ion Stoica\ucb}\\
 \rm{\textit{\msr Microsoft Research
\space\space\space\space\space \ucb UC Berkeley
\space\space\space\space\space \uwm University of Wisconsin-Madison}}
}

\maketitle

%\ToAppear
%\input{tex/abstract.tex}
%-------------------------------------------------------------------------------
%-------------------------------------------------------------------------------
\begin{abstract}
%Rich datasets and massive compute clusters enable machine learning to tackle hard problems like image classification, object detection, question answering, and translation. 
%To reduce training time of machine learning models, the most widely adopted method for distributed training is data parallelism. 
Model parameter synchronization across GPUs introduces high overheads for data-parallel training at scale. Existing parameter synchronization protocols cannot effectively leverage available network resources in the face of ever increasing hardware heterogeneity. 
To address this, we propose {\tt \system{}}, a collective communication library that
dynamically generates optimal communication primitives by \emph{packing spanning trees}. We propose techniques to minimize the number of trees generated and extend {\tt \system{}} to leverage heterogeneous communication channels for faster data transfers.
Evaluations show that compared to the state-of-the-art (NCCL), {\tt \system{}} can achieve up to $8\times$ faster model synchronization, and reduce end-to-end training time for image classification tasks by up to 40\%.
%\todo{multi-machine, dgx-2} 
%{\color{red}{\tt \system{}} achieves almost linear scalability in distributed settings.}
% for any given topology. {\tt \system{}} achieves the optimal communication rate .
%To address this issue, we propose {\tt \system{}}, a heterogeneity-aware collective communication library. 
%Given any topology, {\tt \system{}} not only provides an optimal communication schedule for homogeneous links, but also leverages heterogeneous communication channels for hybrid, and faster, data transfers.
\end{abstract}

\section{Introduction}
\label{sec:intro}

Large high-quality datasets and massive compute clusters have enabled machine learning algorithms, such as Deep Neural Networks (DNNs), to tackle hard problems in a number of domains including image classification, object detection, machine translation, and speech processing.
Models developed for such tasks can take a long time to train; for example, models for image classification tasks~\cite{image1k} can often take days or even weeks to train on a single GPU.
Thus, fast training of large deep learning models requires distributed training on many GPUs.
The most widely used method for reducing DNN training time is to perform data-parallel training~\cite{tensorflow,facebook}.
In data-parallel training, each GPU has a full copy of the model parameters and
%and trains independently on a subset of the input data.
GPUs frequently exchange parameters with other GPUs involved in training.

Parameter synchronization across GPUs introduces significant overheads when training at scale with communication overheads that can range from 50\% to 90\% for popular ML models~\cite{pipedream-sosp19}.
This problem is accentuated by the fact that GPU computation is getting faster and model sizes are growing larger, thus making communication overheads stand out.
But two recent trends seem to \textit{suggest} that their arrival might alleviate, or even eliminate, such communication bottlenecks for DNN training.
First, on the hardware front, state-of-the-art multi-GPU servers, like NVIDIA's DGX-1~\cite{dgx1} and DGX-2~\cite{dgx2}, now have fast interconnects between GPUs \--- NVLink offers 20-25GBps pairwise and bi-directional peak throughput~\cite{nvlink,nvswitch}.
Second, modern communication libraries such as NVIDIA's Collective Communications Library (NCCL)~\cite{nccl}, Uber's Horovod~\cite{horovod}, and Baidu's Ring AllReduce~\cite{baidu}, with techniques such as wait-free backpropagation designed to hide communication overheads~\cite{PoseidonATC2017}, are solutions specifically targeted at speeding up parameter synchronization.
%IBM PowerAI DDL~\cite{powerai},
% Facebook's Gloo~\cite{gloo}

% (NVIDIA's DGX-1)

\begin{figure}[t!]
  %\vspace{-0.1in}
  \centerline{\includegraphics[width=0.85\columnwidth]{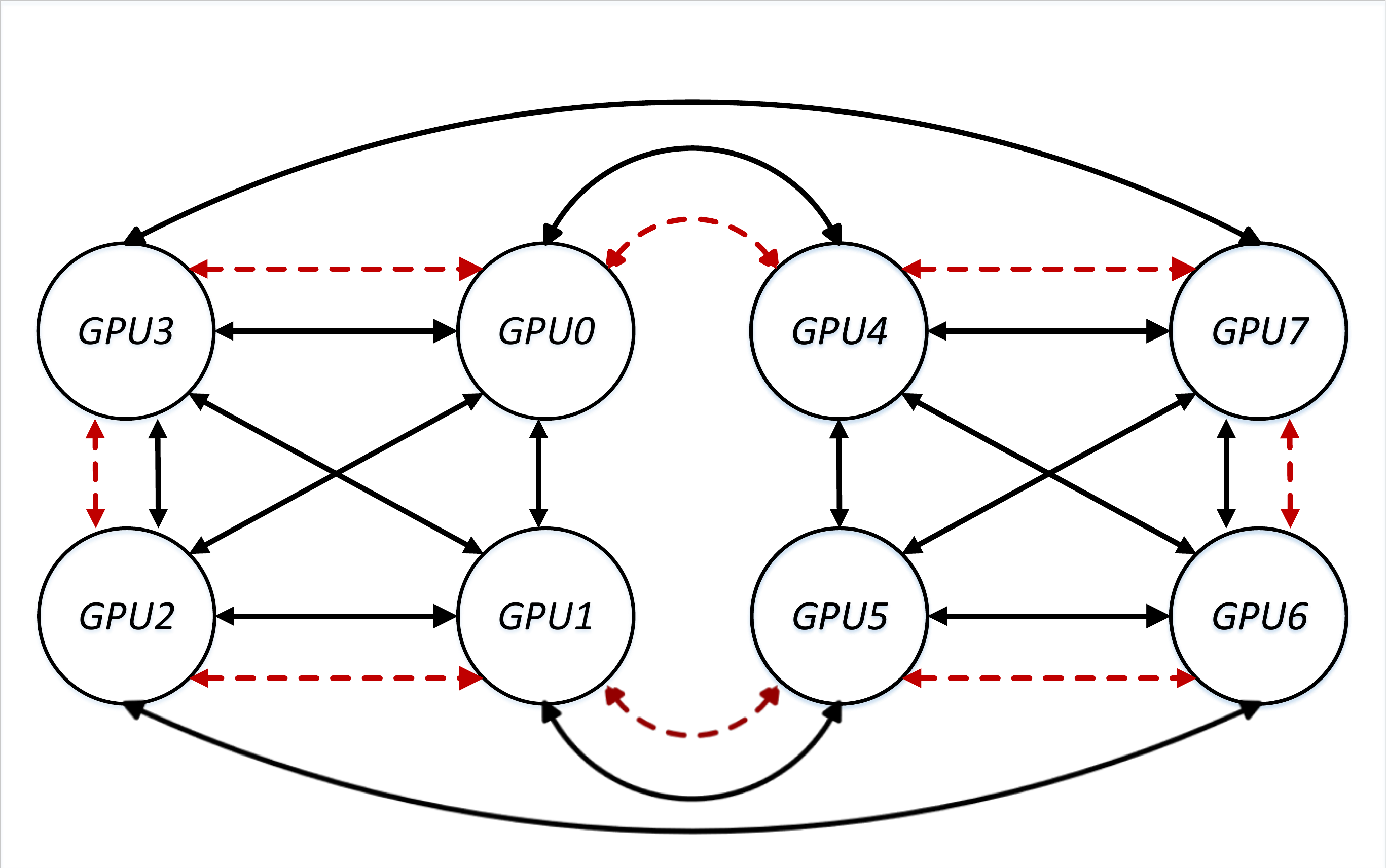}}
%  \vspace{-0.1in}
  \caption{Hybrid mesh-cube topology of NVLink in the DGX-1 8-GPU server.  Solid lines here indicate the bi-directional NVLinks on the DGX-1-P100, red dashed-lines are the additional NVLinks in DGX-1-V100 servers.   NVLink Gen1 has bi-directional pairwise throughput of 18-20GB/s (DGX-1-P100); Gen2 goes up to 22-25GB/s (DGX-1-V100).}
%  \vspace{-0.1in}
  \label{fig:dgx1-topo}
\end{figure}

%\noindent\textbf{Challenges.}  
In this paper, we focus on multi-GPU servers with NVLink~\cite{nvlink} / NVSwitch~\cite{nvswitch} and find that despite recent advances, modern communication libraries for parameter exchange are unable to fully mitigate communication bottlenecks in data-parallel training.
The central hurdle in achieving peak performance for inter-GPU collectives is link under-utilization due to \textit{topology heterogeneity}. We find this occurs due to three main reasons:

First, topology heterogeneity can occur due to \textit{differing server configurations}. Figure~\ref{fig:dgx1-topo} shows an example of two  generations of servers, the DGX-1-P100 (\dgxp) and DGX-1-V100 (\dgxv), and their NVLink topologies.  Protocols have to be topology aware to effectively use hardware.

\begin{figure}[t]
\centering
\subfigure[Fully connected]{\label{fig:3gpu-ring}
\includegraphics[width=0.48\columnwidth]{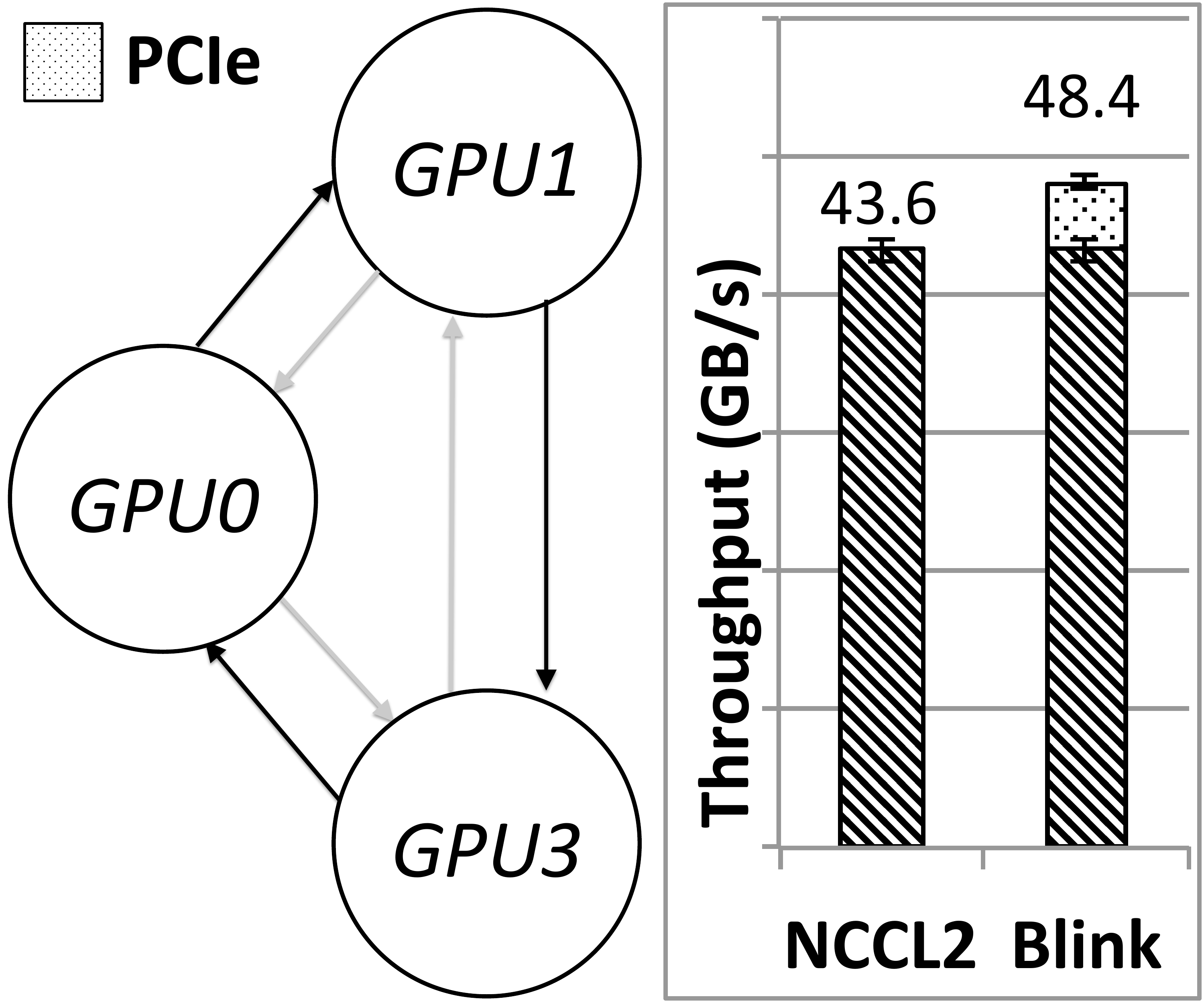}}
\subfigure[Partially connected]{\label{fig:3gpu-chain}
\includegraphics[width=0.48\columnwidth]{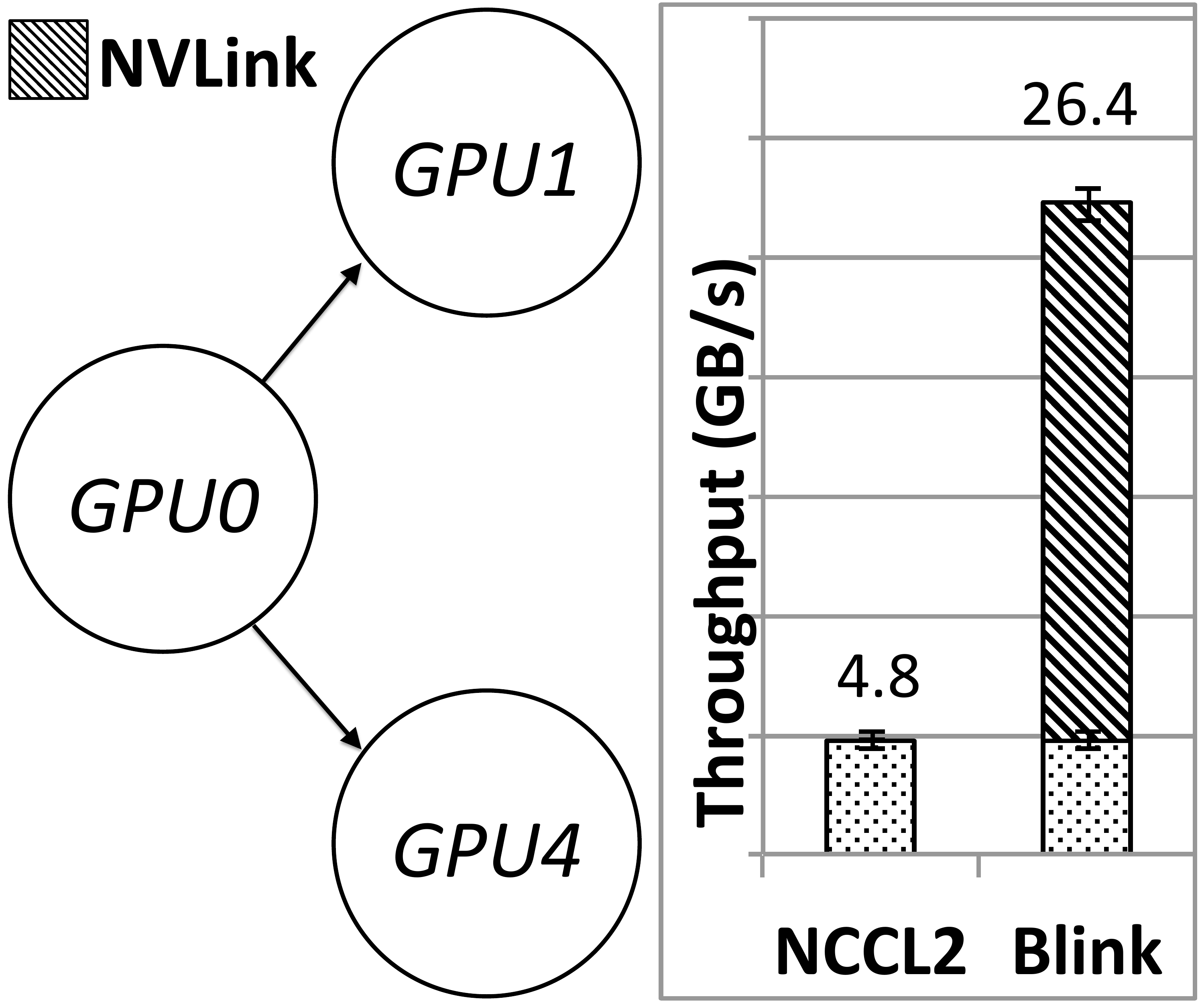}}
\caption{Broadcast throughput, from GPU 0, using both NCCL and {\tt\system{}} on a \dgxp.}
\label{fig:3gpu-broadcast}
\end{figure}

Second, existing schemes do not exploit \textit{link heterogeneity}. For intra-node communication, servers such as the DGX-1 have both inter-GPU point-to-point (P2P) interconnects such as NVLink (20-25GB/s)~\cite{nvlink} and shared interconnects such as PCIe (8-12GB/s)~\cite{pcie}. 
PCIe connects multiple GPUs to each other within a machine, and to the CPU and IO deices, through a PCIe switch hierarchy.
%For inter-node communication, servers are equipped with multiple Ethernet or InfiniBand ports with a throughput of 3GB/s and 7GB/s per-port respectively.
State-of-the-art collectives, such as NCCL and Horovod, all use ring-based protocols which fail to leverage link heterogeneity. The throughput of a ring is limited by the link with lowest bandwidth and hence these protocols either restrict themselves to high bandwidth, homogeneous links, or limit throughput to the link with lowest bandwidth in the ring. For example, for multi-GPU communication within a machine, NCCL prioritizes using only NVLink over PCIe, as PCIe will be the bottleneck if included in a NVLink ring. Figure~\ref{fig:3gpu-ring} shows an example 3 GPU setup for a \texttt{Broadcast} from GPU 0: when fully connected with NVLink, NCCL builds two rings (0->1->3->0 \& 0->3->1->0) using bi-directional NVLinks, and ignores PCIe. 

%{\color{red}For cross-machine communication, NCCL just sets the throughput of a ring to Ethernet/InfiniBand bandwidth which underutilizes high bandwidth intra-node links like NVLink.}\\

\begin{figure}[t]
   \centerline{\includegraphics[width=0.8\columnwidth]{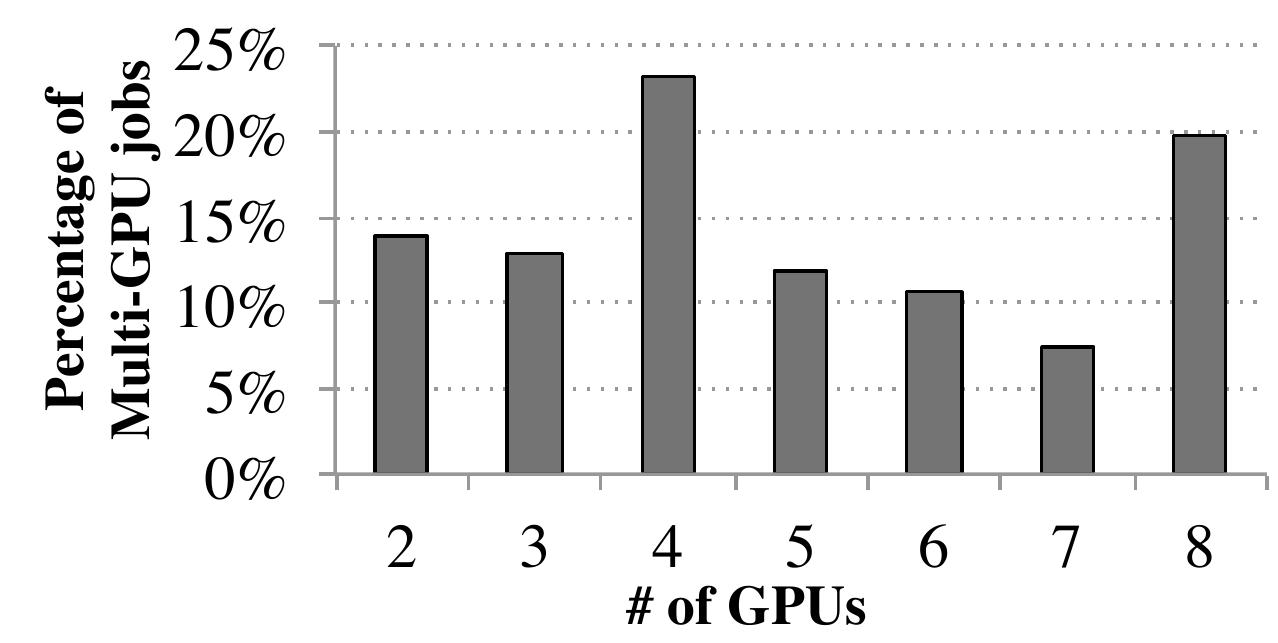}}
%   \vspace{-3mm}
   \caption{Number of GPUs within each 8-GPU server on a cluster allocated to 40,000 multi-GPU jobs on Cloud-X.}
   \label{fig:gpu_dist}
%   \vspace{-4mm}
 \end{figure}

Third, schedulers that allocate GPUs to jobs, especially in \textit{multi-tenant} clusters, are oblivious to interconnect topologies between GPUs.
Many jobs can potentially be co-located on the same machine. Furthermore, even topology aware schedulers must embrace \textit{fragmentation} to avoid queuing delays (e.g., a 8-GPU job might have to contend with 3 GPUs on one machine and 5 GPUs on another)~\cite{philly}.
In an analysis of over 40,000 multi-GPU jobs over a three month period on a multi-tenant cluster at Cloud-X (Figure~\ref{fig:gpu_dist}), we find that it is common for jobs to be allocated 3, 5, 6, or 7 GPUs on individual 8-GPU servers despite multi-GPU jobs overwhelmingly requesting GPUs in powers of 2.
While, fragmentation can be mitigated, not avoided, by making schedulers topology aware \textit{and} capable of migration~\cite{gandiva-osdi18}, such solutions face a higher barrier of entry as there are many independent scheduling frameworks that all need to be changed and not all jobs can be placed appropriately given variable arrival rates.

\begin{figure*}
\centering
\subfigure[6-GPU group topology]{\label{fig:6gpus}
\includegraphics[width=0.3\textwidth]{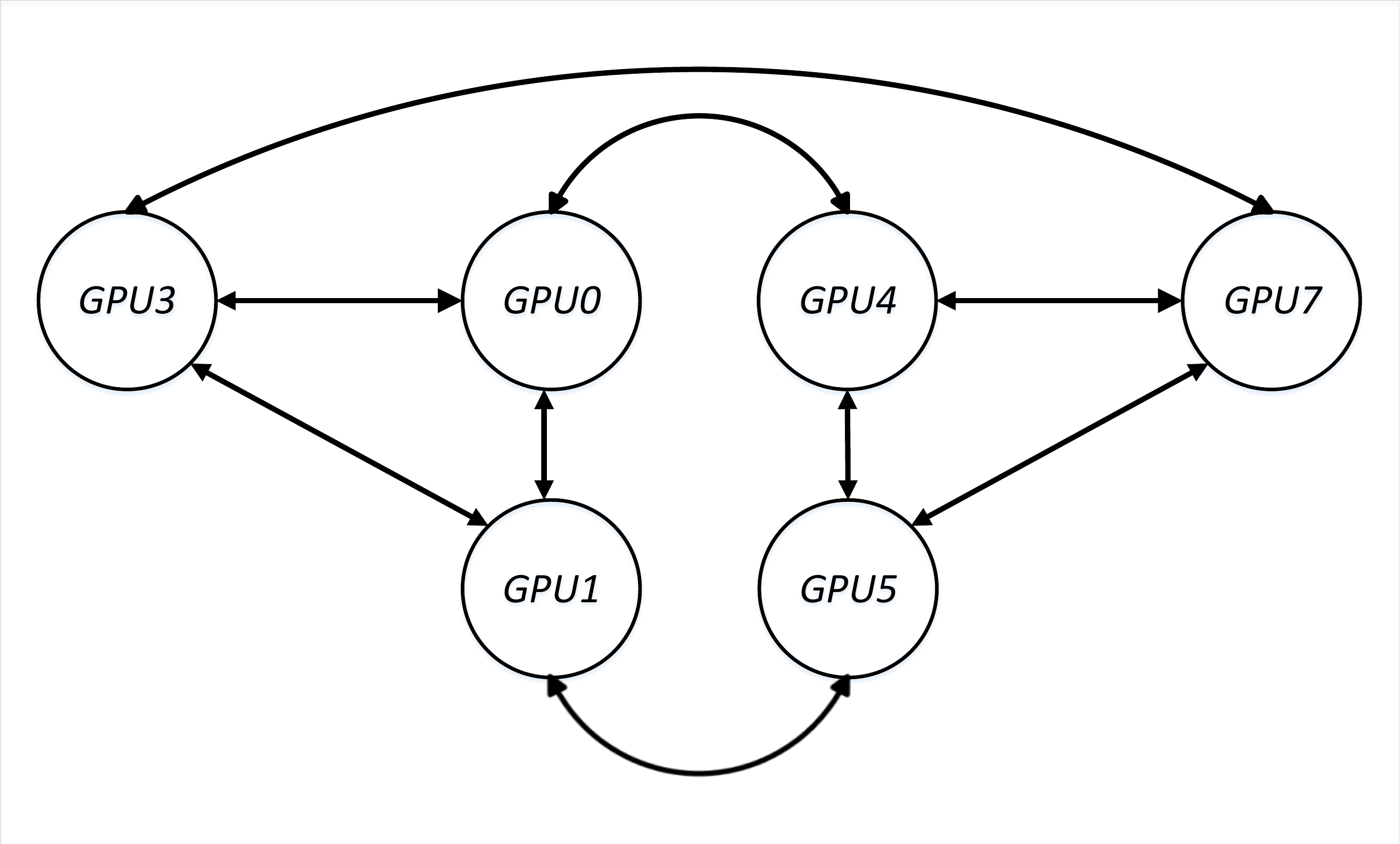}}
\subfigure[NCCL 6-GPU rings.  Links between GPU 1 \& 3, 5 \& 7, and 0 \& 4 are unused.]{\label{fig:6gpus-nccl}
\includegraphics[width=0.6\textwidth]{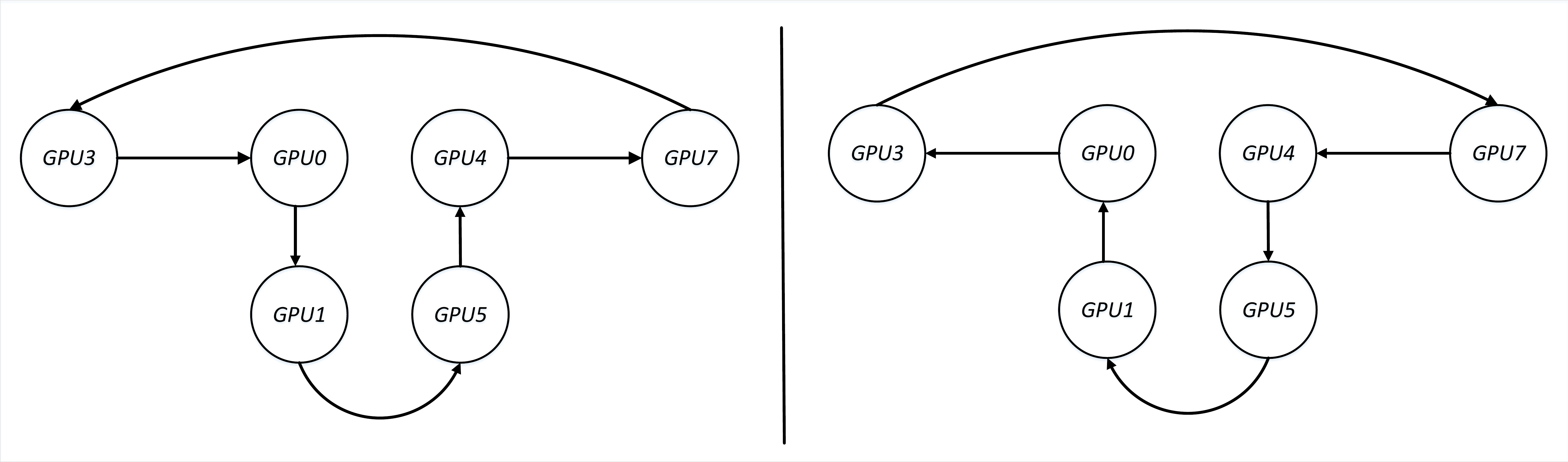}}
\subfigure[{\tt\system{}} 6-GPU spanning trees.]{\label{fig:6gpus-blink} 
\includegraphics[width=0.9\textwidth]{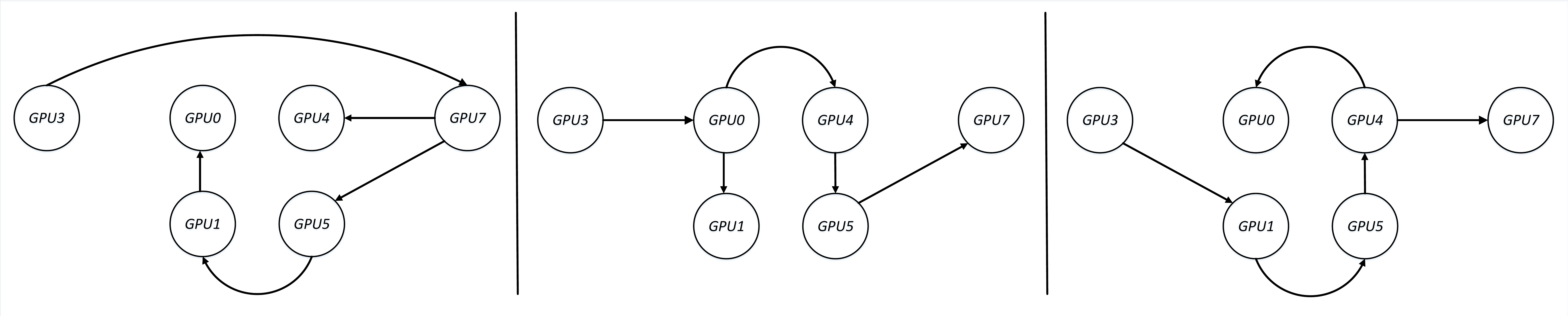}}
%\vspace{-4mm}
\caption{Broadcast comparison between NCCL and {\tt\system{}} over 6-GPUs in \dgxp.}
\label{fig:6gpu-compare}
%\vspace{-2mm}
\end{figure*}

The resulting topology heterogeneity caused by scheduler allocation can result in link under-utilization in current ring-based protocols for parameter exchange.  
For example, in Figure~\ref{fig:3gpu-chain}, 
%if we replace GPU3 with GPU4, 
NCCL is unable to utilize the bi-directional NVLinks between the 3-GPUs;  the lack of NVLink between GPUs 1 and 4 prevents NCCL from constructing NVLink-only rings and it has to fall back on PCIe based communication.   But link under-utilization can also occur even when rings can be constructed using NVLink.  Figure~\ref{fig:6gpu-compare} shows a 6 GPU allocation on a \dgxp, where despite being able to construct two NVLink-based rings, NCCL has to drop some of the links connecting the GPUs as they don't contribute to ring construction.

%To tackle these challenges, we propose {\tt \system{}}, a collectives communication library that achieves near-optimal link utilization regardless of link and topology heterogeneity. 

\noindent\textbf{Contributions.}  In this paper, we propose {\tt \system{}}, a communication library for inter-GPU parameter exchange that achieves near-optimal link utilization.
To handle topology heterogeneity from hardware generations or partial allocations from cluster schedulers, {\tt \system{}} dynamically generates optimal communication primitives for a given topology. 
{\tt \system{}} probes the set of links available for a given job at runtime and builds a topology with appropriate link capacities. Given the topology, {\tt \system{}} achieves the optimal communication rate by \emph{packing spanning trees}, that can utilize more links~\cite{lovasz1976two,edmonds1973edge} when compared to rings. We use a multiplicative-weight update based approximation algorithm to quickly compute the maximal packing and extend the algorithm to further minimize the number of trees generated. We also describe how this scheme can handle one-to-many primitives like Broadcast or Gather and how we can extend this to many-to-many primitives like AllReduce using bi-directional links and hardware capability to compute at line rate. To handle heterogeneous links, {\tt \system{}} simultaneously transfers data on PCIe and NVLink within a machine and balances the amount of data transferred across hybrid
links. {\tt \system{}}'s collectives extend across multiple machines effectively utilizing all available network interfaces.

Based on the spanning trees chosen, {\tt \system{}} dynamically generates code to implement common collective primitives. Our generated code automatically chunks data and uses CUDA streams to 
efficiently pipeline transfer and computation. From the programmer's perspective, {\tt \system{}} provides NCCL-compatible API.
It can be seamlessly plugged into distributed ML frameworks like TensorFlow~\cite{tensorflow}, PyTorch~\cite{pytorch}, etc.
{\tt \system{}} does not requires user program modifications and only relies on \textit{preloading} (\texttt{LD\_PRELOAD}).

%For topology heterogeneity, {\tt \system{}} automatically adapts to new inter-connect topologies (especially for the irregular topology due to partial allocated GPU group) and provide optimal solutions. For link heterogeneity, {\tt \system{}} achieves hybrid data transfer over heterogeneous links simultaneously (PCIe, NVLink). From the programmer perspective, {\tt \system{}} provides NCCL-compatible API, so that there is no need of user-level code changes and {\tt \system{}} can be seamlessly plugged into distributed ML frameworks like TensorFlow~\cite{tensorflow}, PyTorch~\cite{pytorch}, CNTK~\cite{cntk}, MxNet~\cite{mxnet}, etc.

We evaluate {\tt \system{}}'s performance on a number of multi-GPU platforms including \dgxp, and \dgxv and DGX-2. 
Results show that, compared with NCCL, on \dgxv,  {\tt \system{}} achieves up to 6$\times$ speed-up in all-to-one/one-to-all collective communications (e.g. Broadcast, Gather), and is up to 8$\times$ faster in all-to-all collective communications (e.g. AllReduce). On DGX-2, we show that single-hop trees in 
{\tt \system{}} are especially effective for smaller data sizes offering up to 3x lower latency and higher throughput,
compared to NCCL's double-binary trees and rings~\cite{nccl-binary}.
%we show that compared to double-binary trees used by NCCL~\cite{nccl-binary}, multiple one-hop spanning trees used by {\system{}} can lead to up to 3.5x better throughput for AllReduce.
Finally, we also find that {\tt \system{}} can accelerate DNNs training on single and multi-machine setups.
%experiments using both NCCL and {\tt \system{}}. 
For instance, on a single \dgxv machine, compared to NCCL, {\tt \system{}} can reduce communication cost up to 87\% (51\% on average), and speeds up end-to-end training by up to 40\%. %overall. %On multi-machine cases we find that {\tt \system{}} outperforms Horovod with NCCL by 10\%. %\todo{Write about multi-machine result here}.
%where {\tt \system{}} can achieve near linear scalability for distributed DNN training.

\section{Motivation}
\label{sec:motivation}
%\vspace{-3mm}

%the need for tree-based collective communication

In this section, we first discuss the need for more efficient communication primitives and  why ring-based solutions like NCCL cannot handle topology heterogeneity. We highlight the case for spanning tree-based protocols and the need to pack trees to achieve peak performance in the face of topology heterogeneity.  We then present micro-benchmarks characterizing the capabilities of modern GPU hardware that helps guide {\tt \system{}}'s design.

\subsection{The case for packing trees}
The motivation for our work stems from the high communication overheads experienced by deep learning workloads when running data-parallel training even on fast multi-GPU servers like the NVIDIA DGX-1~\cite{pipedream-sosp19}.
These overheads occur despite setting per-GPU minibatch sizes to the largest values that fit in GPU memory, using state of the art libraries like NCCL, and using optimizations common in modern frameworks such as  Wait-free Backpropagation~\cite{PoseidonATC2017}.
Communication overheads arise from a number of factors including increased model sizes and faster computation on newer hardware generations.
%As shown in Figure~\ref{fig:comm-overhead}, 
%Despite using state of the art libraries like NCCL and setting per-GPU minibatch size to the largest values that fit in GPU memory (with \textit{per-GPU batch sizes} of 32 or 64 making them consistent with the source papers for the models), we find significant communication overheads. 
Recent work has made the case for large batch sizes for ResNet~\cite{facebook, IncreaseBS}, which indirectly affects communication overhead by reducing the number of synchronization rounds per-epoch.  
However, these techniques lack generality when it comes to diverse DNN workloads and there continues to be a debate in the machine learning community with regard to their efficacy~\cite{SmallBS, lecunn-smallbs}.

%\amar{Mention that our experiments set the per-GPU minibatch to the largest value that fits in one GPU’s memory - anything larger yields out-of-memory exceptions.  To load batch sizes larger than what the memory on the GPU can support, one can use LARS.  But argument of generality still holds.}

\begin{figure*}[t]
\centering
\begin{minipage}{0.45\textwidth}
  \centering
    \includegraphics[width=\textwidth]{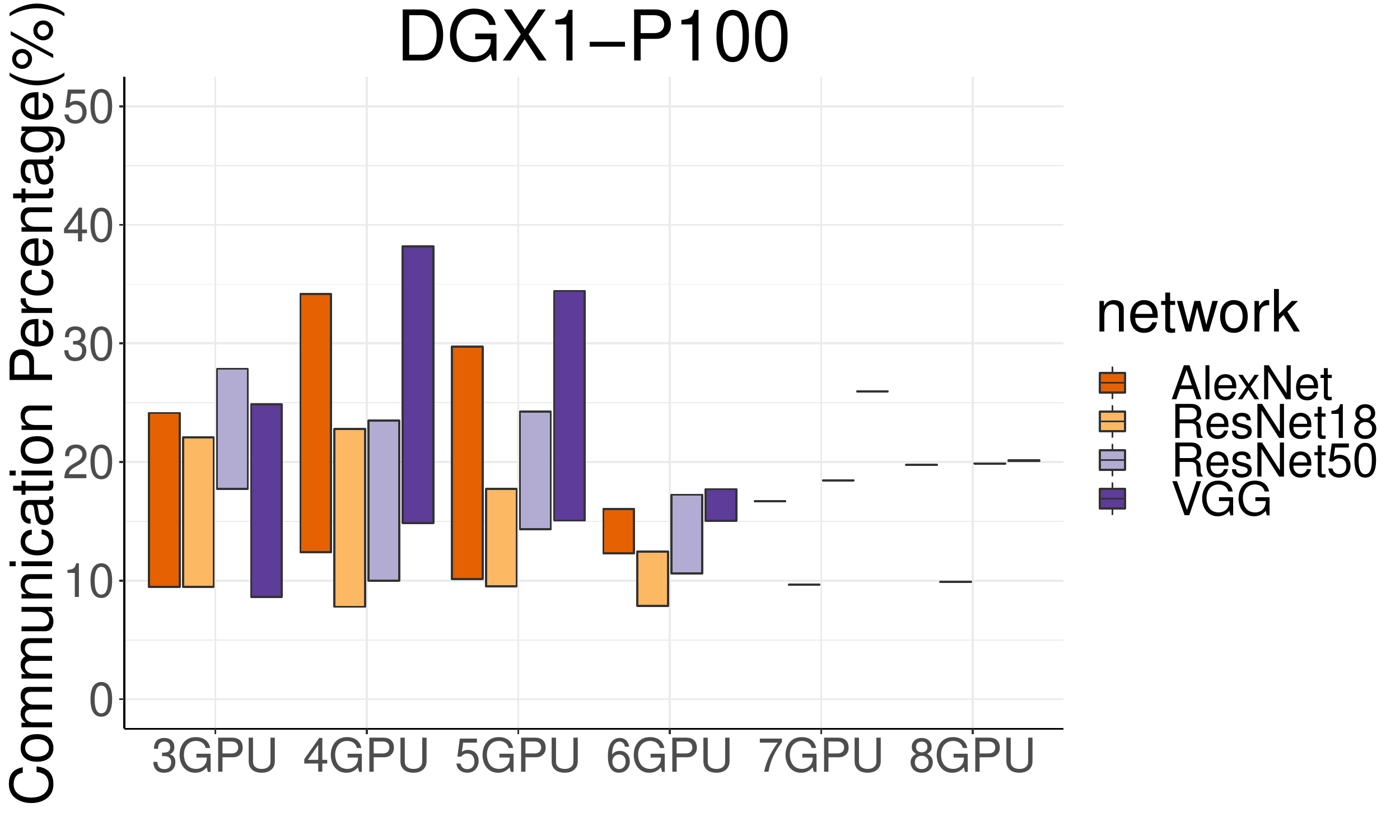}
\end{minipage}
\begin{minipage}{0.45\textwidth}
    \includegraphics[width=\textwidth]{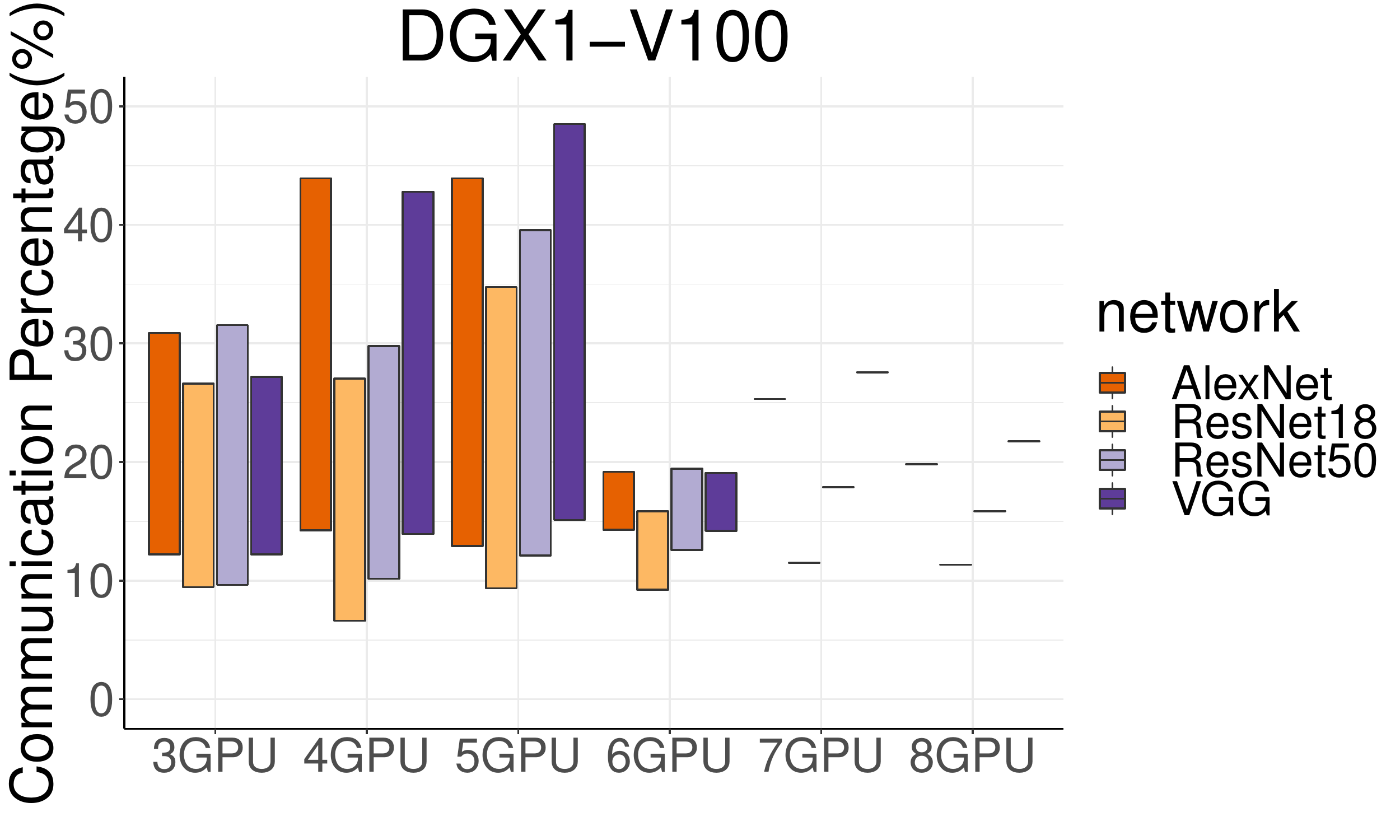}
\end{minipage}
\caption{Best-Worst case communication overhead (percentage of overall runtime) for different DNNs on \dgxp (left) and \dgxv (right) based DGX-1 server when using NCCL.}
\label{fig:comm-overhead}
\end{figure*}

Crucially, we find that even within a single high-performance server like the DGX-1, communication overheads are amplified due to one of the main shortcoming of existing communication libraries like NCCL or Horovod: their inability to handle topology heterogeneity. 
These libraries typically use a fixed ring-based scheme for doing data transfers. 
However, ring-based protocols have structural limitations: for each ring, every node can only have one input and one output. 
This strong restriction makes it impossible for rings to fit into irregular topologies caused due to scheduler allocations (Figure~\ref{fig:gpu_dist}) and this leads to link under-utilization as shown in Figures~\ref{fig:3gpu-chain} and~\ref{fig:6gpu-compare}.
%\todo{Saying something about what we observe (especially with topology heterogeneity) by showing a graph might be very strong.}

Figure~\ref{fig:comm-overhead} shows the communication overhead (best-to-worst-case range), as a percentage of per-iteration time, for four popular image classification DNNs within a \dgxp or a \dgxv when using NCCL\footnote{We use NCCL and NCCL2 interchangeably for v2.4.2}.  
%For a given number of GPUs, say $n$, 
Given $n$ GPUs there could be many $n$ GPU configurations.  We bin these configurations by \textit{topology uniqueness}. For example, a 4 GPU configuration consisting of GPUs [0, 1, 2, 3] is in the same bin as the [4, 5, 6, 7] configuration.  We pick one representative configuration from each bin %\textit{topologically-unique configuration},
and report the best and worst case overheads for each of the $n$ GPU configuration.  Figure~\ref{fig:comm-overhead} highlights that the communication overheads can be as high as 50\% for these DNNs on a \dgxv.

By modeling the links between GPUs as a graph, classic results from Edmonds~\cite{edmonds1973edge} and Lovasz~\cite{lovasz1976two} show that \textit{packing spanning trees} leads to the maximum flow from a chosen root vertex to all the other vertices in a directed graph.
Thus, one-to-many protocols like Broadcast using spanning trees from the root node is a potential option to overcome link under-utilization.
In addition to operations like Broadcast that just forward data, communication libraries also need to implement primitives like AllReduce which can be modeled as a reduce-and-forward in one direction (towards the root) followed by a Broadcast in the other direction.
But this introduces two important questions which we explore next: How close to line rate can GPUs perform computation on data that is being transferred, and can GPUs support multiple transfer trees efficiently?

\subsection{Micro Benchmarks}
\label{sec:ubench}
%Before introducing our {\tt \system{}} collectives, we conduct several groups of micro-benchmark measurements for unique traffic patterns which are involved into our protocol design. Based on topology and link differences, we define mainly 5 basic cases. For NVLink's different spanning tree structures, we test 4 basic toplogies, namely, Depth, Breadth, Multiple-Input Multiple-Output (MIMO), Multiple Chain Aggregation (MCA). For PCIe, we measure throughput for a single ring created over PCIe bus hierarchy.
%\shivaram{It'll be good to say why these tests are important / relevant. The thing to say imho is that this is important because it lets us design trees that are either deep or broad etc.  The other point to make is that MIMO and MCA test out how multiple trees will perform.}

We validate the potential of computing inline with communication over spanning trees on modern GPU hardware. 
We do this using a series of micro-benchmarks mimicking transfer patterns when using spanning trees. 
First we test how deep spanning trees perform as number of the GPUs increases (\emph{depth} tests). 
%Further, to test if trees can have a high branching factor we perform \emph{breadth} tests. 
Next we test how well \emph{multiple trees} passing through a GPU can transfer data at the same time.

%Within these settings, there are mainly three kinds of data transfer patterns we study: data forward, reduce-forward, and reduce-broadcast. Data forward is defined as, on each node, it first receives data from its predecessor/s and then forwards the same data to its successor/s. Data reduce-forward means that, on each node, when receiving data from its predecessor/s, it conducts a reduce function "\textcircled{+}" (e.g. sum, min, max, etc.) between received data and its own local data, and then forwards the reduced result to its successor/s. Data reduce-broadcast is just a combination of forward and reduce+forward. In reduce-broadcast, given a spanning tree, we first do reduce-forward in one direction towards a single node (root), and then forward data from the root in the reverse direction. For consistency, here we use sum as the reduce function for all micro-benchmark tests. Note that it is easy to change reduce function from sum to other functions like min, max, etc.

We present our test results from AWS P3.16xlarge EC2 instance, a \dgxv with 8x NVIDIA V100 GPUs connected over an NVLink topology shown in Figure~\ref{fig:dgx1-topo}. We also ran the same group of experiments on a \dgxp machine. For the sake of brevity, we do not include those results here. 
% To increase communication efficiency, in the following micro benchmark test, we open up 4 streams on each GPU to enable concurrent data transfer across these streams.
%For link heterogeneity in a single machine multi-GPU environment, we measure throughput and overhead for achieving hybrid data transfer using both PCIe and NVLink.

\begin{figure}[!t]
\centering
\includegraphics[width=0.4\textwidth]{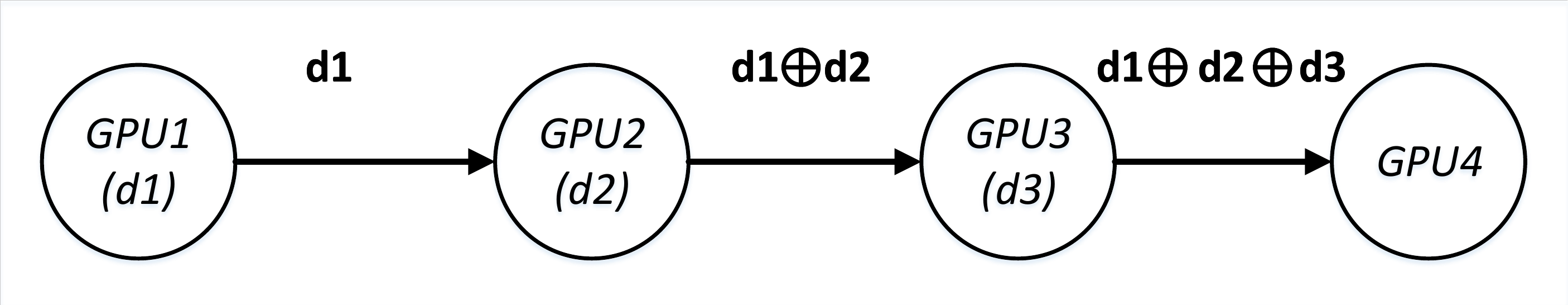}
%\vspace{-4mm}
\caption{Depth test of reduce+forward, over a chain of GPUs.}
\label{fig:depth-add-forward}
%\vspace{-2mm}
\end{figure}

\begin{figure}[!t]
\centering
\includegraphics[width=0.4\textwidth]{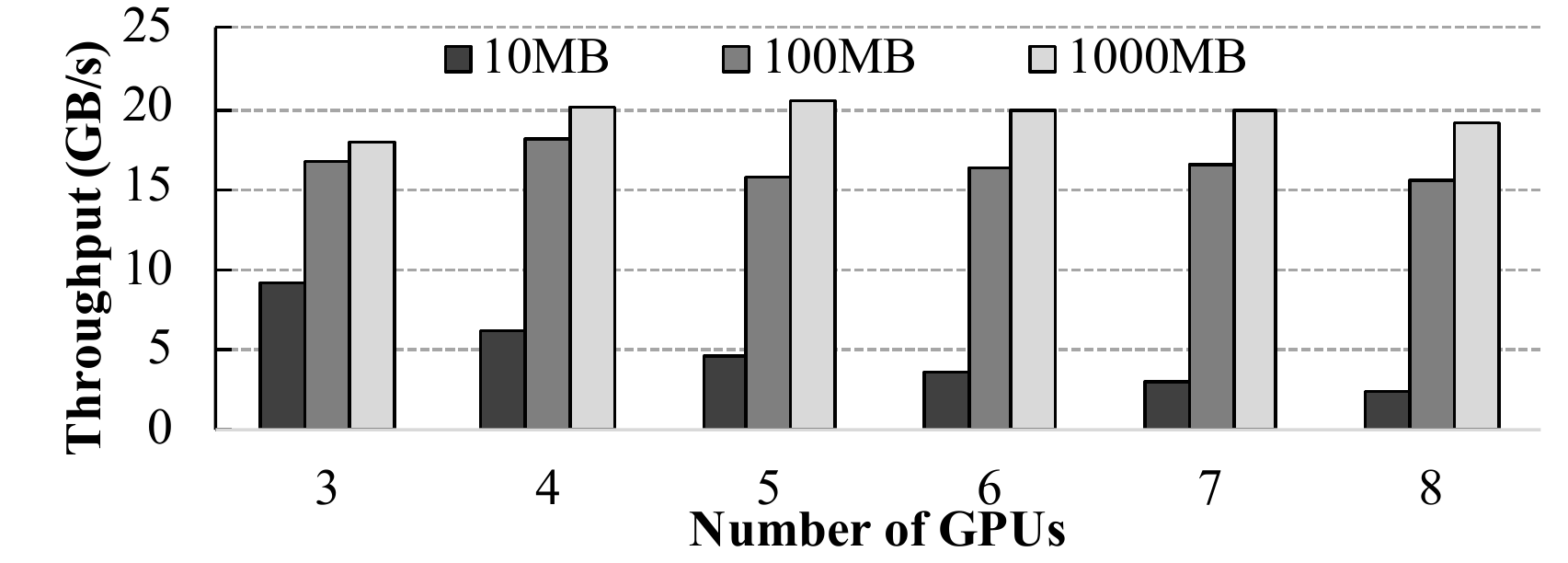}
%\vspace{-4mm}
\caption{Throughput for reduce+forward over a chain of GPUs.}
\label{fig:depth-tput}
%\vspace{-4mm}
\end{figure}

\begin{figure*}[t!]
\centering
\subfigure[Multi-Input-Output (MIMO)]{\label{fig:MIMO}
\includegraphics[width=0.3\textwidth]{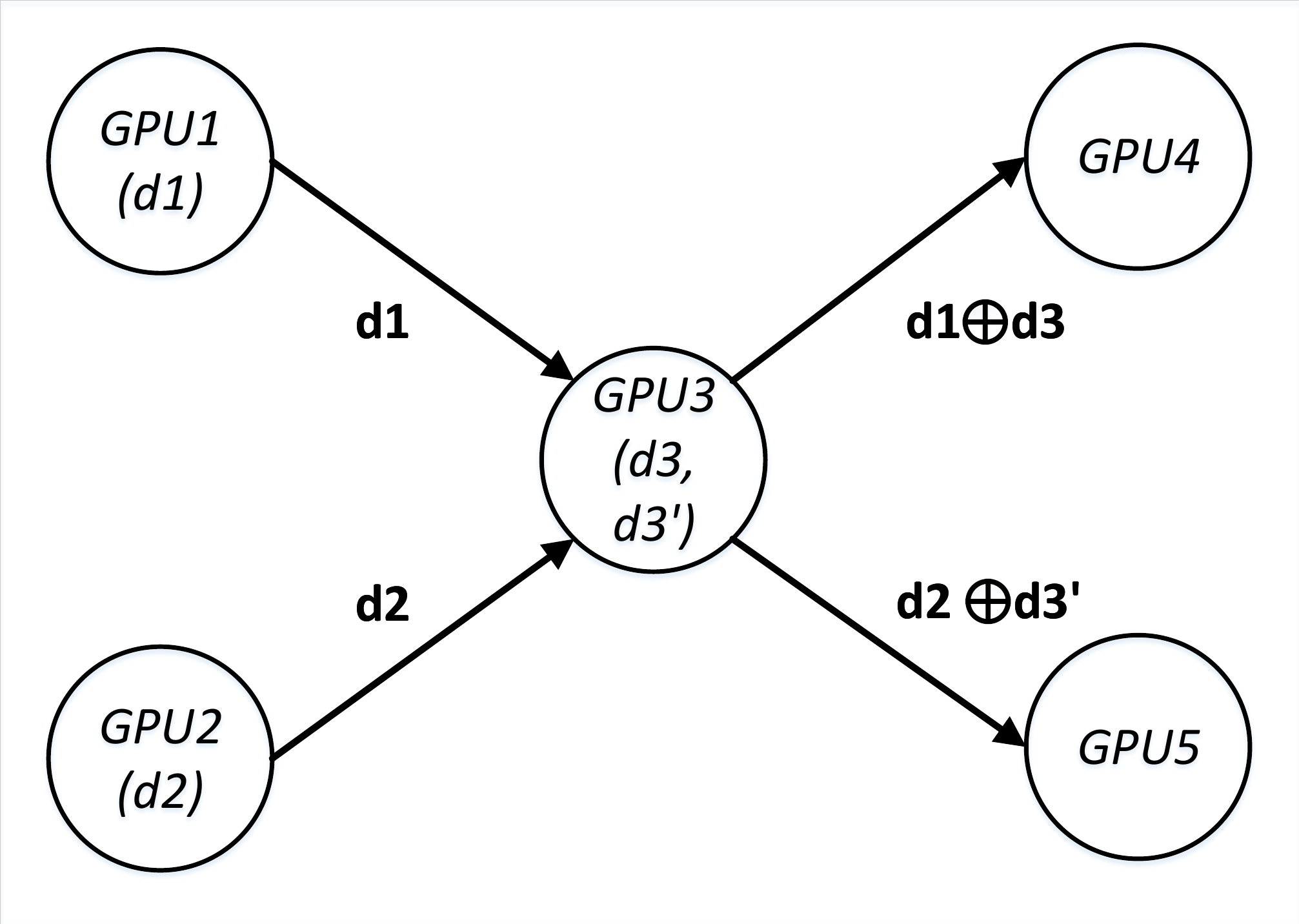}}
\subfigure[Multi-Chain Aggregation (MCA)]{\label{fig:MCA} 
\includegraphics[width=0.3\textwidth]{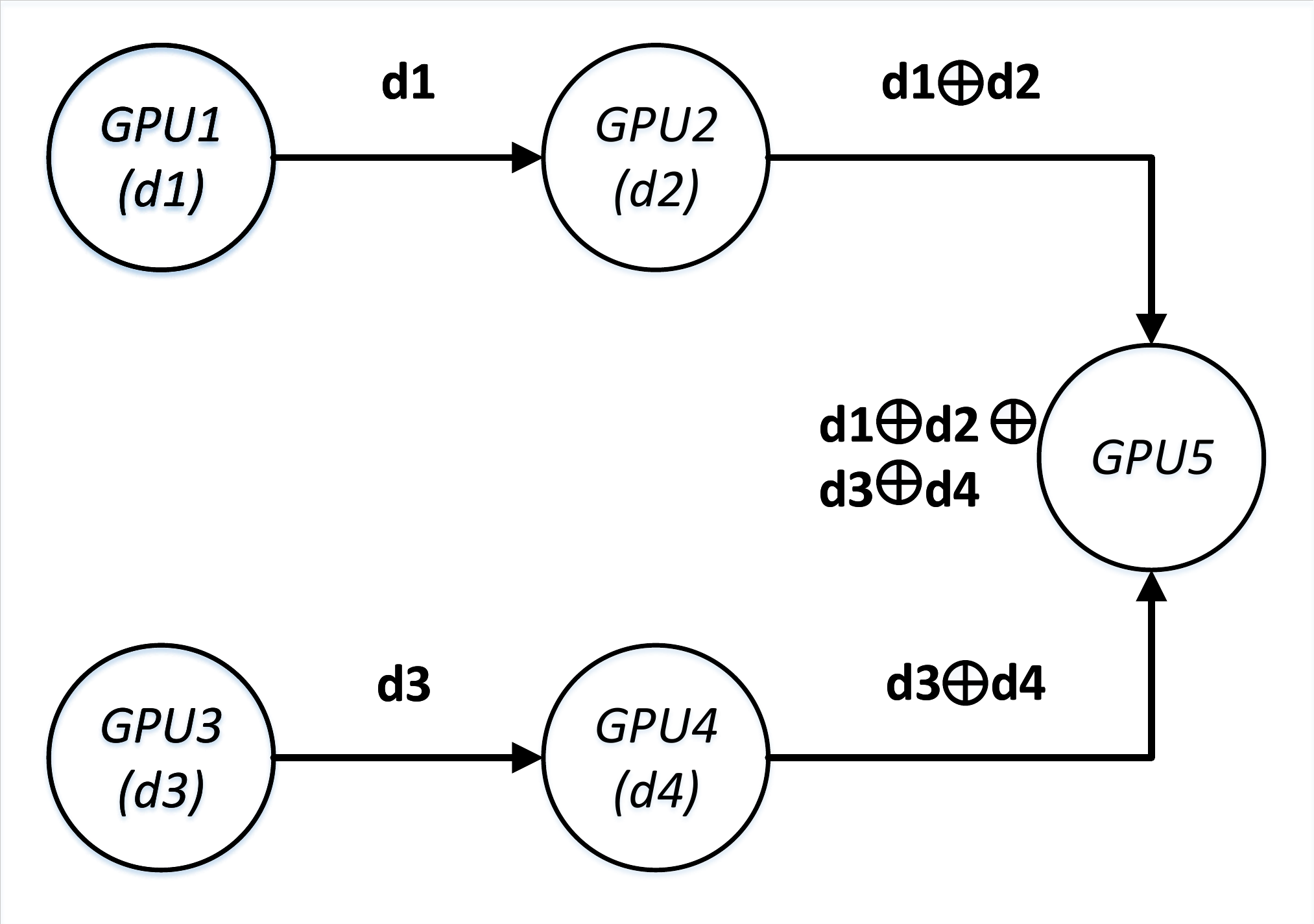}}
\subfigure[MIMO, MCA Throughput]{\label{fig:mimo-mca-pcie}
\includegraphics[width=0.35\textwidth]{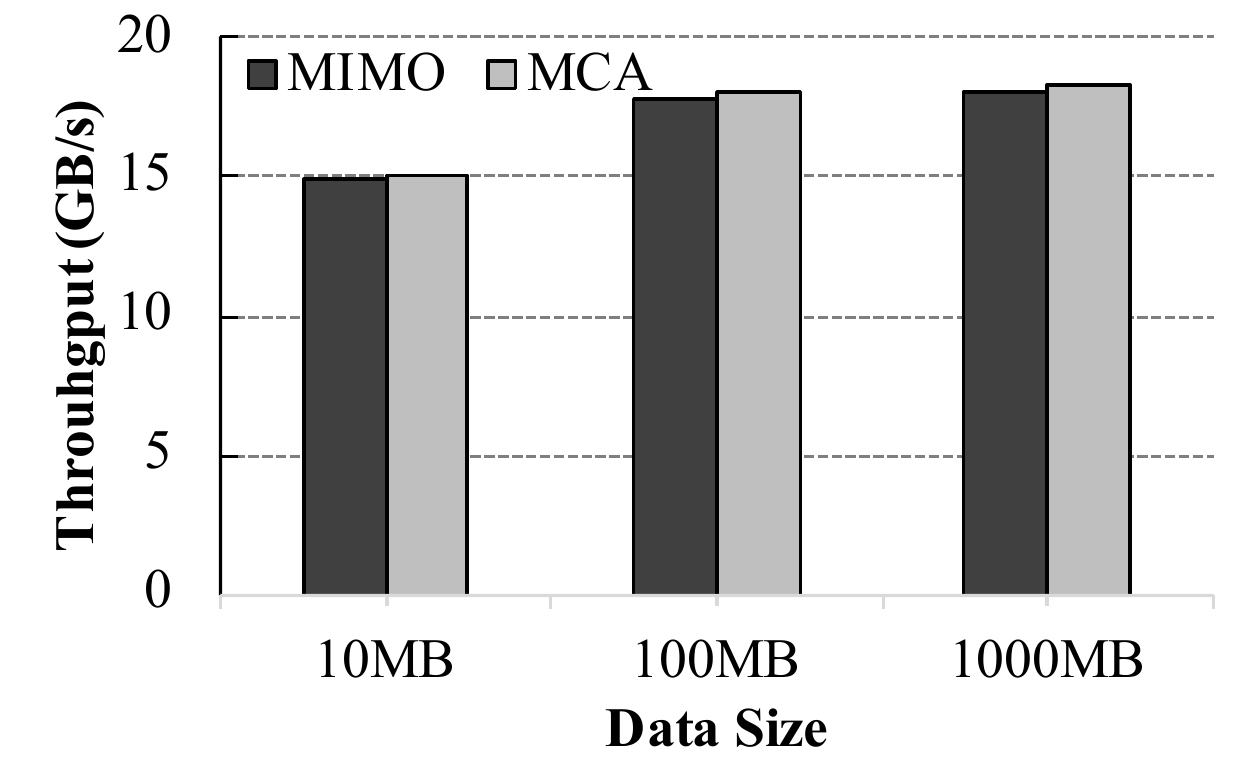}}
%\vspace{-3mm}
\caption{MIMO, MCA topology and test throughput.}
\label{fig:MIMO-MCA}
%\vspace{-3mm}
\end{figure*}

\begin{figure*}[t!]
  \centerline{\includegraphics[width=0.9\textwidth]{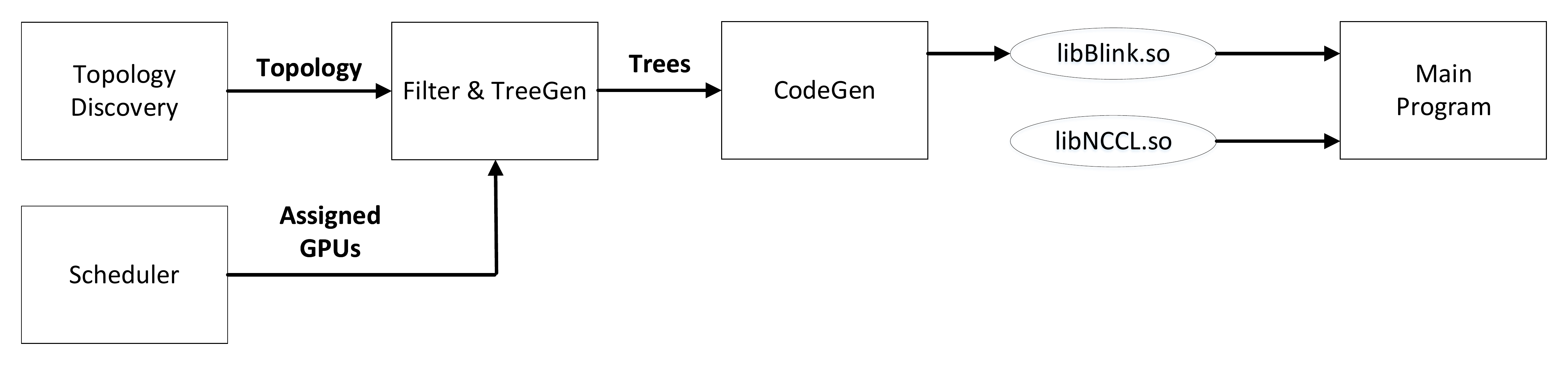}}
%  \vspace{-0.25in}
  \caption{{\tt\system{}} toolchain workflow.}
  \label{fig:workflow}
%  \vspace{-0.2in}
\end{figure*}

%\subsubsection{Depth Test}
\noindent\textbf{Depth Test.}
\label{sec:depthTest}
The first topology class we consider is a depth test where we vary depth of trees that are used. To do this we consider a simple \emph{chain} topology (Figure~\ref{fig:depth-add-forward}).

%Depth test refers to that, given a bunch of GPUs, we use a link chain to connect them all together. Figure~\ref{fig:depth} shows depth test in a 4 gpu case. 

Given a chain topology, we consider a reduce+forward traffic pattern. Results from other traffic patterns (e.g. data forward, reduce-broadcast) are included in Appendix~\ref{appendix:microbench}. 
%As shown in~\ref{fig:chain-forward}, for the forwarding benchmark, GPU1 is the source node with data named d1, and it passes the data d1 to GPU2 and then GPU2 forwards it to GPU3 etc. 
For reduce+forward (Figure~\ref{fig:depth-add-forward}), each GPU has its own data. When a GPU receives data from its predecessor, it invokes a reduction function (denoted as \textcircled{+}) on the received data with its own data, passing the result to its successor. 
%Finally, we implement reduce+broadcast by doing reduce+forward in one direction and forward in the other direction as shown as Figure~\ref{fig:chain-reduce-bcast}, as such a capability can be used for all-to-all reductions.

%We test forward, reduce+forward and reduce-broadcast cases mentioned above. 
We test these operations over different number of GPUs (3-8GPU) and vary data sizes from 10MB to 1000MB (Figure~\ref{fig:depth-tput}). 
%In the case of forward only, as we increase the chain length, the throughput decreases from around 22 GB/s (with 3GPU) to around 20 GB/s (8 GPU case) for 1000MB. 
%The impact is less visible for reduce+forward where 
As we increase the chain length, throughput decreases to around 19 GB/s from around 21 GB/s for 1000MB.
%throughput is around 18GB/s. Finally for reduce-broadcast where the depth of the tree is now doubled, we see the throughput drops from 19GB/s to around 16GB/s for 1000MB
We also see that throughput drops as the dataset size becomes smaller; it is hard to saturate fast links with small data sizes and the constant overheads in invoking CUDA operations are significant at smaller data sizes.

\noindent\textbf{Multi-transfer Test.}
\label{sec:MIMOTest}
\label{sec:MCATest}
Next we consider the effect of having multiple transfers simultaneously take place in a given topology. These tests are important to ascertain if we can have multiple data transfers happen in parallel.
To do this we consider two topologies: a multi-input, multi-output (MIMO) as shown in Figure~\ref{fig:MIMO} and a multi-chain aggregation (MCA) shown in Figure~\ref{fig:MCA}.

In the MIMO topology (Figure~\ref{fig:MIMO}), two nodes on the left concurrently send data to the center node. The center node aggregates its local data (d3, d3') with received data blocks (d1, d2) respectively, and then forwards the aggregated result (d1\textcircled{+}d3, d2\textcircled{+}d3') to two different destinations.
We test performance with multiple dataset sizes as shown in Figure~\ref{fig:mimo-mca-pcie}. We find that for datasets larger than 10MB, we can achieve around 18GB/s throughput, which is around 15\% lower than maximum throughput on NVLink Gen2.

In the MCA topology (Figure~\ref{fig:MCA}), we consider a center node that merges two reduce+forward chains. Figure~\ref{fig:mimo-mca-pcie} shows that MCA has roughly the same throughput as MIMO and achieves around 18 GB/s for data larger than 10 MB.
%\subsubsection{Multi-Chain Aggregation Test}
%\label{sec:MCATest}
%\shivaram{Lets combine this with previous section and call it 'multi-tree' tests. Also at the end of every subsection we should repeat the takeaway from the numbers.}
%For homogeneous links, the last topology we explore is called Multi-Chain Aggregation (MCA). 

%\subsubsection{PCIe ring Test}
%\label{sec:PCIeRingTest}

%We discuss about link heterogeneity in this section. Within a multi-GPU box like DGX-1, we have both PCIe and NVLink communication channels. To measure PCIe bandwidth for collective communications, we conduct a ring-reduce experiment over 4 GPUs. As depicted in Figure~\ref{fig:PCIe-ring}, in a simple PCIe bus hierarchy, from GPU1 (left) to GPU4 (right), we build a reduce chain. Once GPU4 gets the final reduce result of all the four GPUs, it forwards the reduce result in the reverse direction (right to left). Performance result shows in Figure~\ref{fig:mimo-mca-pcie}, it achieves 2-5 GB/s on average with varied data sizes.
%\shivaram{Same point as above. Lets add a takeaway sentence here}

\noindent\textbf{Summary.} From the micro-benchmark results, we see modern GPUs with NVLink interconnects provide good support for deep and broad trees while forwarding data. We also see that GPUs can perform reductions while forwarding data and also support multiple transfers at the same time. While these scenarios do show some drop in performance compared to pairwise NVLink transfers, this drop is only minor, and the resultant throughput is much higher than that achievable when using PCIe. Overall, these results make it promising to explore the use of spanning trees to implement collective communication protocols.

\subsection{{\tt \system{}} Approach}
\label{sec:approach}

%NOTE: Cross machine communication can be completely overlap with in-machine GPU communication. It is because cross machine and in-machine communication are using different channels. Cross machine uses PCIe+IB, whereas in machine using NVLink (NV1 or NV2).

We next outline our approach to building high performance collective communication primitives in {\tt \system{}} and present an end-to-end workflow as shown in Figure~\ref{fig:workflow}.

Our main approach in {\tt \system{}} is to \emph{dynamically} generate the appropriate collective communication primitives to make it best utilize a given topology. We achieve high utilization by packing spanning trees and use algorithms that can maximize the transfer rate achieved while minimizing the number of trees used. Finally, we implement many-to-many algorithms like AllReduce by performing many-to-one and one-to-many operations on each direction of bi-directional links.
The workflow of using {\tt \system{}} consists of:
\begin{myitemize}
    \item At \emph{runtime}, once a deep learning job has been scheduled and assigned a set of GPUs, {\tt \system{}} is able to probe the topology of the machine and infer the interconnect topology across only the GPUs allocated.
    \item Once we have the topology, we model collective communication operations as flows on a directed graph and compute the maximum fractional packing of spanning trees. We denote this step as \texttt{TreeGen} and this step outputs a set of spanning trees and weights corresponding to how much data should be sent over them.
    \item Next, \texttt{CodeGen} parses the spanning trees and \emph{generates} CUDA code. The code generated matches the API offered by NCCL and is packaged into a shared library \texttt{libblink.so}.
    \item Finally we set the \texttt{LD\_PRELOAD} flag to dynamically load the {\tt \system{}} implementations when the main program is invoked. This ensures that existing programs can be run without any modification.
\end{myitemize}
%To handle 

%As shown in Figure~\ref{fig:workflow}, {\tt\system{}} first automatically probes the whole system topology. When receiving meta-data of assigned GPUs from resource scheduler, {\tt\system{}} launches Tree Generator (TreeGen) \textbf{XXX: what is filter in mwu?}. Based on assigned GPU group and  its corresponding topology, treeGen outputs maximum number of spanning trees within the topology of allocated GPU group. 

%After TreeGen passing the network scheduling results (spanning trees) to our Code Generator (CodeGen), CodeGen automatically generate CUDA code for the implementation of {\tt \system{}}'s collective communication primitives based on the treeGen outputs. It is worth mentioning that, the generated data transfer code exposes exactly the same API as NVIDIA NCCL~\cite{nccl}, and use it to generate our blink library binaries (i.e. libBlink.so).

%To seamlessly replace the corresponding NCCL function call, we use \emph{LD\_PRELOAD} trick, which forces our blink library to loaded before NCCL counterpart during dynamic linking process to generate the executable binary for the main program. 

%After the whole process mentioned above, the use code (Main Program) can repeatedly make function call to our blink library.

%\input{tex/approach.tex}
\section{Design}
\label{sec:design}

In this section we outline the design of {\tt \system{}} and describe our techniques for creating protocols that address the dual challenges of high link utilization and heterogeneous topologies. We first study one-to-many protocols like Broadcast or Gather and describe our approach to packing spanning trees and the approximation framework we use to efficiently generate spanning trees. Second, we describe our refining technique that helps  minimize the number of trees generated. Third, we discuss how our techniques can be extended to handle all-to-all protocols like AllReduce. Fourth, we propose techniques to leverage hybrid set of links for example, PCIe~\cite{pcie} and NVLink~\cite{nvlink}. Finally, we extend our design to NVSwitch~\cite{nvswitch} embedded DGX-2 machine~\cite{dgx2} and multi-server settings.

\subsection{Packing Spanning Trees}
\label{sec:packSpanningTree}
We first consider the problem of broadcasting data from one root GPU to all the other GPUs in the system. The topology we infer from the allocated resources can be modeled as a directed graph where every GPU is a vertex $V$ and every link (NVLink or PCIe) is marked as a directed edge $E$. Each directed edge also has a bandwidth proportional capacity.

Given the above model, the optimal rate possible for broadcast is the maximum weight of flows that originate from a given root vertex $r$ and reach all the other vertices in the graph. This problem is well studied in graph theory~\cite{edmonds1973edge} and prior work has shown that the optimal rate can be achieved by finding the maximal packing of a number of \emph{directed spanning trees} or arborescences in the graph~\cite{lovasz1976two}. Each arborescence $T_i$ originates at the root vertex and follows directed links to span every other vertex. Thus the problem of finding the optimal schedule for broadcast can be solved by finding the set of maximum weight arborescences that satisfy the capacity constraints.

\setlength{\abovedisplayskip}{0pt}
\setlength{\belowdisplayskip}{0pt}
\setlength{\abovedisplayshortskip}{0pt}
\setlength{\belowdisplayshortskip}{0pt}

\begin{align}
    &\max \sum_{i} w_i \\
    \text{such that }
        &\forall {e \in E}, \sum_{i} \kappa_{i} * w_i < c_{e}\\
    \text{where  }
    &\kappa_{i} = 
    \begin{cases}
        1, & \text{if } e \in T_i\\
        0,              & \text{otherwise}
    \end{cases}
\end{align}

More formally, our problem statement is given a graph $G$ with vertices $V$, edges $E$ and root vertex $r$ and spanning trees $T_1, T_2, T_3 ... T_i$ we wish to find the weights $w_i$ such that the sum of weights trees passing through any edge does not exceed the capacity of the particular edge.

While the above formulation can be viewed as an optimization problem, the number of arborescences in a graph can be exponentially large ($O(n^{n-2})$ for a complete graph) and hence is not a practical model to use. A number of more efficient exact algorithms have been proposed for this problem but their running time is still $O(n^3mlog(n^2/m))$ for a graph with $n$ vertices and $m$ edges~\cite{gabow1998packing}. In this paper we instead use a recently proposed approximate packing scheme and then discuss how we minimize the number of trees used to achieve the optimal rate.

\subsection{Approximate Packing}
\label{sec:MWU}
The multiplicative weight update (MWU) is an algorithmic technique that is used in a number of domains ranging from optimization to game theory. Our specific use of MWU here follows a recently proposed algorithm to achieve near-linear time approximation for fractional packing problems~\cite{chekuri2017near}. For the case of packing spanning trees, this approach finds a $(1 - \epsilon)$-approximation in $O(m\ln m/\epsilon^2)$, where $m$ is the number of edges.% in the graph. 

%\shivaram{Do we need a pseudo-code block ?}
The MWU procedure for finding the optimal set of packing spanning trees proceeds in the following fashion: We initialize every edge with a capacity and a weight that marks how much of the capacity has been used. Given this, we run an iterative method where at each iteration we find the minimum weight spanning tree given the current assignment. We then increment the weight on this chosen tree by an $\epsilon$ factor and update weights on the graph correspondingly. The algorithm provably converges after $O(\ln m/\epsilon^2)$ iterations and on convergence we get a set of directed spanning trees $T_1...T_i$ and corresponding weights $w_i$ for each of them. The total rate for broadcast will be the sum of weights $\Sigma_i{w_i}$. 

While the MWU procedure has very low execution time and achieves the optimal rate, there is no bound on the number of spanning trees returned. For example we find that with \dgxv topology of 8 GPUs, the MWU procedure returns 181 spanning trees while the minimum number of trees that can be used to achieve the same optimal rate is 6. The weights on the trees generated by MWU vary from 0.002 to 0.899.  Having a larger number of trees will mean that the amount of data transmitted per tree will be much smaller leading to lower throughput (Section~\ref{sec:ubench}) and higher overhead in scheduling transfers in the generated code (Section~\ref{sec:implementation}).
%. Further having a larger number of trees also increases the overhead in generated code. % (Section~\ref{sec:implementation}).  %Thus we next discuss techniques to minimize the number of trees generated. This has a number of implications for our system.

\subsubsection{Minimizing Number of Trees}
\label{sec:ILP}
We design an integer-linear program based solution to minimize the number of spanning trees that are used. From the above described MWU procedure we get the optimal rate $b^*$ and a set of candidate spanning trees $T_1, ... T_k$. 
To minimize the number of spanning trees, we formulate an \emph{integer} linear program (ILP) similar to the one presented before but with each weight is restricted to be 0 or 1. 
This problem can be expressed as

\begin{align}
    &\max \sum_{i=1}^{k} w_i \\
    \text{such that  }
    &\forall {e \in E}, \sum_{i} \kappa_{i} * w_i < c_{e}\\
    &\forall w_i \in \{0, 1\}\\
    \text{where  }
    &\kappa_{i} = 
    \begin{cases}
        1, & \text{if } e \in T_i\\
        0,              & \text{otherwise}
    \end{cases}
\end{align}

$k$ here is controlled by the number of trees returned by the MWU procedure and thus is much smaller than the overall number of spanning trees present in the graph. Solving this ILP will yield $\hat c$, the maximum rate that is feasible by only using integer capacities for each tree. However $\hat c$ might be much lower than $c^*$ and we thus iteratively relax the constraints (i.e. allowing $w_i$ to take fractional values) until $\hat c$ is within a configured threshold (e.g., 5\%) of $c^*$.
% like 0.5

Using this procedure reduces the number of trees from 181 to 6 for the 8-GPU case in \dgxv topology with each tree having a rate of 1.0. In terms of data size, this improves the amount of data transferred through a single tree leading to better link utilization. For a 1000MB transfer, each tree will now transfer 166MB while without the ILP the transfer sizes vary from 0.33MB to 148MB.

\subsection{Handling many-to-many operations}
\label{sec:allReduceDesign}
The above discussion focused on one-to-many operations like Broadcast and Gather where packing directed spanning trees yields the optimal rate. To handle many-to-many operations we exploit the fact that all the links found in these machines are bi-directional in nature and hence we can create an undirected graph to run a many-to-one primitive using one direction of links and correspondingly run a one-to-many primitive in the other direction.
For example, to do an AllReduce operation on the directed graph, we first run a \emph{reduce} operation to a chosen root vertex using the undirected graph and then do a \emph{broadcast} operation from the root vertex using the same tree but with links going in the reverse direction.  

This strategy of using two undirected trees also matches the lower bound of number of messages required for AllReduce operations. As shown in prior work~\cite{ring-allreduce}, the minimum number of messages that need to be sent by a process for AllReduce, is $2\times \lceil\frac{N-1}{N} \rceil$. The spanning tree over $N$ vertices contains $N-1$ edges and accounting for trees in both directions (one for Reduce and one for Broadcast) we similarly have $2 \times (N-1)$ messages. Assuming a continuous forwarding model (similar to our benchmarks in Section~\ref{sec:ubench}),  messages sent by all $N$ processes simultaneously and we can thus achieve a similar bound of $2\times \lceil \frac{N-1}{N} \rceil$ messages per process.

\begin{figure*}[t]
\centering
\centerline{\includegraphics[width=0.99\textwidth]{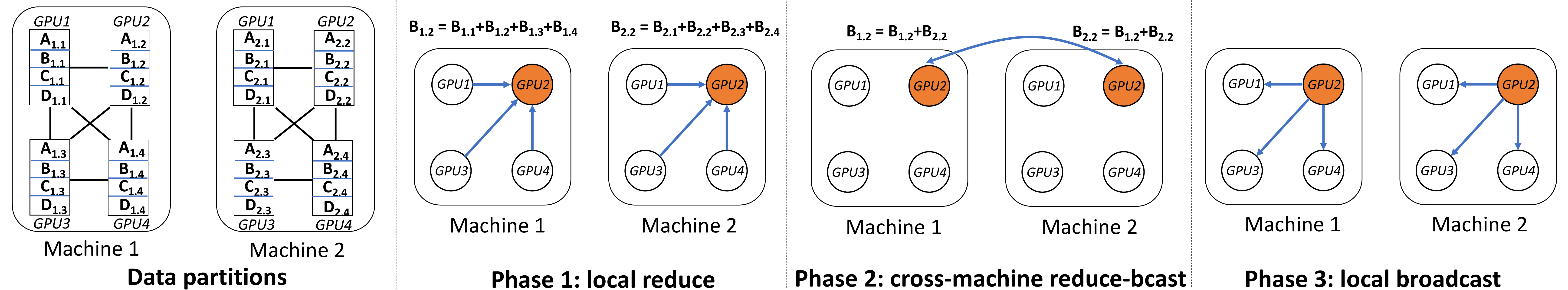}}
%\vspace{-0.15in}
\caption{Three-phase AllReduce protocol for cross-machine settings.  Data item $X_{m.g}$ refers to data partition $X$ on server $m$ and GPU $g$. Each data partition has a distinct server-local root.  The figure above shows the reduction (function is denoted as $+$) for partition $B$ which has a root at $GPU2$. Similar protocol is followed for other data partitions.}
\label{fig:hierarchy-allreduce}
%\vspace{-0.3in}
\end{figure*}

\subsection{Handling hybrid communication}
\label{sec:hybrid}
We next discuss how we handle hybrid PCIe and NVLink topologies in the context of our design presented above. The main challenge in 
using both PCIe and NVLink comes from the fact that NVIDIA driver does not directly allow users to control access to both links and if NVLinks are detected, the system will automatically enable P2P data transfer among GPUs using NVLinks. In our experience we find that 
using \texttt{cudaDeviceDisablePeerAccess} disables NVLinks and forces data transfer through PCIe links. However this still has the limitation that we cannot construct a unified topology with both sets of links. We address this problem by constructing two separate sets of trees, one over PCIe links and another over NVLinks.

One of the challenges with this approach is to balance the amount of data that is transferred over each link type. Our approach here is to minimize the maximum time taken by each of the transfers i.e. minimize $max(T_{PCIe}, T_{NVL})$. 

We denote $D_{total}$ as the total data needs to be transferred, and $D_{PCIe}$, $D_{NVL}$ as the data size assigned on either PCIe or NVLink respectively. $T_{dpa}$ is the latency for calling the 
\texttt{disable\_peer\_access()} and we denote $BW_{PCIe}$ and $BW_{NVL}$ as the bandwidth of PCIe and NVLink trees. Given this notation and objective, we can see that the optimal data split can be achieved by making $T_{PCIe} = T_{NVL}$.

\begin{equation}
\label{eq:splitResult}
\begin{aligned}
\text{Objective } & T_{PCIe} + T_{dpa} = T_{NVL}\\
\implies
D_{PCIe} =& \frac{D_{total} \times BW_{PCIe}}{BW_{PCIe} + BW_{NVL}}\;-\\
&\frac{T_{dpa} \times BW_{PCIe} \times BW_{NVL}} {BW_{PCIe} + BW_{NVL}}\\
D_{NVL} =& D_{total} - D_{PCIe}
\end{aligned}
\end{equation}

%Given this notation, 
The optimal data splits are shown in Equation~\ref{eq:splitResult}. Note that in Equation~\ref{eq:splitResult}, $T_{dpa}$ is empirically measured and may vary depending on number of GPUs. We measure this during the initial few calls into our library.

\subsection{DGX-2 and Multi-server settings}
\label{sec:multi-server-and-dgx2}
We next extend our design to switch-based settings like DGX-2 and multi-machine training. 
The DGX-2 consists of 16 V100 GPUs connected over NVSwitch; 
each GPU is connected to the switch over 6x NVLinks (150GBps bidirectional throughput).
On the DGX-2, NCCL constructs double binary trees~\cite{nccl-binary} for small dataset sizes ($<16KB$) and rings for larger datasets.
In contrast, on the DGX-2, {\tt \system{}}'s generated spanning trees for AllReduce (reduce-broadcast) are deceptively simple: 
with $m$ GPUs, each GPU acts as a root for $1/m$ of the data chunks and each root is directly connected to $(m - 1)$ leaf nodes,
resulting in $m$ one-hop trees.
{\tt\system{}}'s one-hop trees have a significant latency and throughput advantage over NCCL's double-binary trees for smaller dataset sizes; 
we show this quantitatively in Section~\ref{sec:dgx2Allreduce}.

\begin{figure}[t]
  \centerline{\includegraphics[width=0.9\columnwidth]{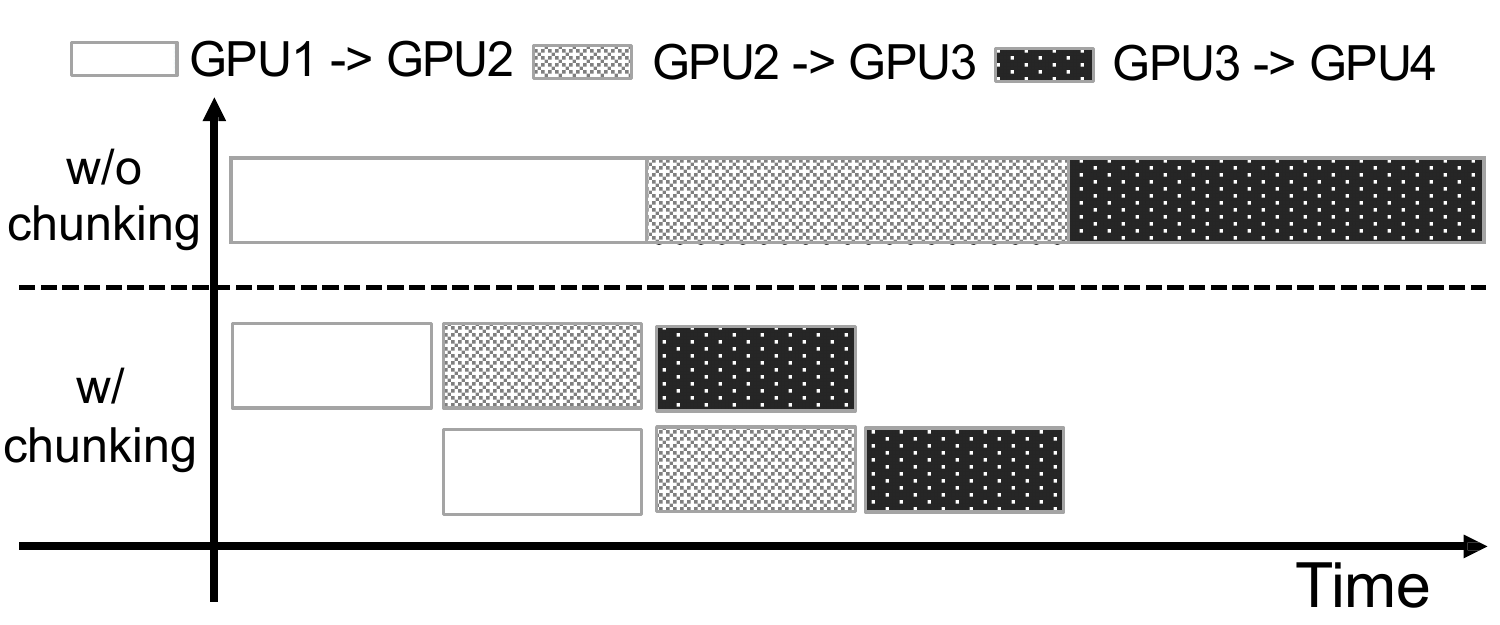}}
%  \vspace{-4mm}
  \caption{Data chunking to reduce multi-hop latency.}
  \label{fig:chunk-data}
%  \vspace{-0.2in}
\end{figure}

When the GPUs of a training task span multiple servers, connected over a switch or a hierarchy of switches, {\tt\system{}} uses a three phase protocol. 
As shown in Figure~\ref{fig:hierarchy-allreduce}, we first partition data based on the number of spanning trees we have (i.e. 4 in this case). The first phase consists of a per-server reduction over local spanning trees \--- the root of each tree within each server aggregates data from its children as before.
The second, new, phase consists of cross-server reduce-broadcast (similar to within the DGX-2) \--- across $n$ servers, 
there are $n$ one-hop cross-server trees, with each server-local root connected to $(n - 1)$ roots on other servers.
%each of them acting as a global root for a distinct $1/n^{th}$ of data.
%each server-local root broadcasts the data from the first phase to its $(n-1)$ counterparts on other servers, 
%while at the same time receiving data from ($n-1$) counterparts and performing a local reduction.
The third phase consists of each server-local root broadcasting the result of the second phase to all nodes in their server. We evaluate our multi-server protocol in Section~\ref{sec:eval-multi-server}.

\section{Implementation}
\label{sec:implementation}

%\iffalse
%\begin{figure}[t]
%\centering
%\subfigure[chain forward]{\label{fig:8gpu-chain-f-chunk}
%\includegraphics[width=0.45\columnwidth]{figs/8gpu-chain-f-chunk.pdf}}
%\subfigure[chain reduce+forward]{\label{fig:8gpu-chain-af-chunk} 
%\includegraphics[width=0.45\columnwidth]{figs/8gpu-chain-af-chunk.pdf}}
%\vspace{-0.1in}
%\caption{Chunk Size selection in 8 GPU chain}
%\label{fig:8gpu-chain-chunk}
%\vspace{-0.25in}
%\end{figure}
%\fi

In this section, we first discuss our code generation implementation and discuss how choosing the appropriate chunk size is important to achieve good performance. % and finally present some of our optimizations to overcome current hardware limitations.

% and specifically detail some of the challenges in memory management when using spanning trees. We also

%Then we discuss about the relationship between throughput and number of CUDA streams within each spanning tree. Within each CUDA stream, we empirically analyze the chunk size selection.

\subsection{CodeGen Implementation}
\label{sec:codeGen}
For ease of illustration, we discuss two types of collective communications: Broadcast and AllReduce. We note that these are the most frequently used primitives by deep learning workloads and other collective primitives follow similar patterns. For example, Gather is the inverse of Broadcast, and AllGather is AllReduce without using a reduction function.
\begin{figure}[t]
\centering
%\subfigure[Automatic selected chunk size]{\label{fig:auto-chunk}
\includegraphics[width=0.8\columnwidth]{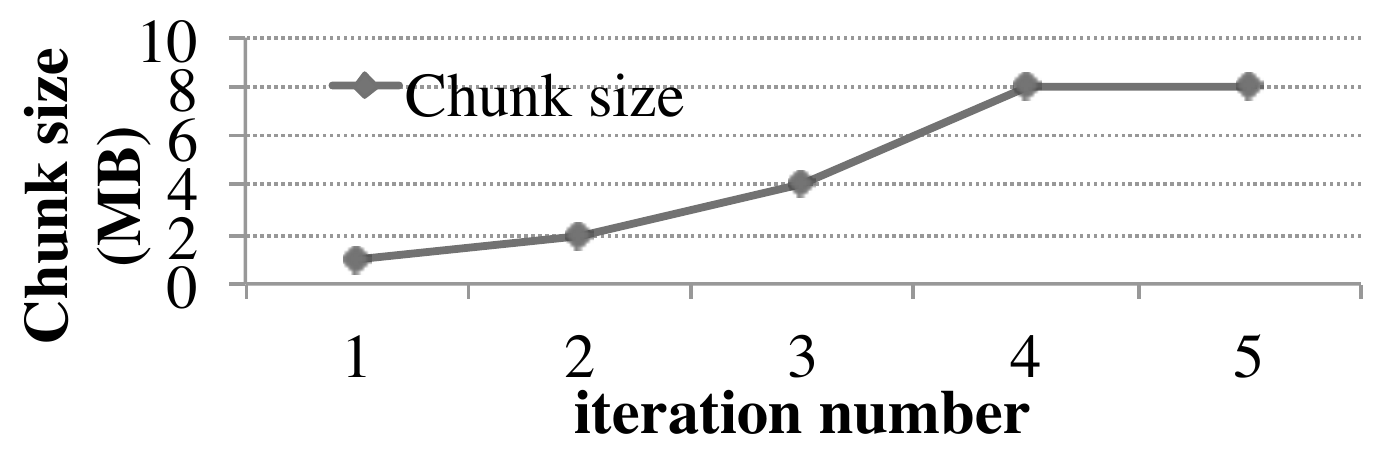}
%\subfigure[Corresponding throughput achieved]{\label{fig:auto-tput} 
\includegraphics[width=0.8\columnwidth]{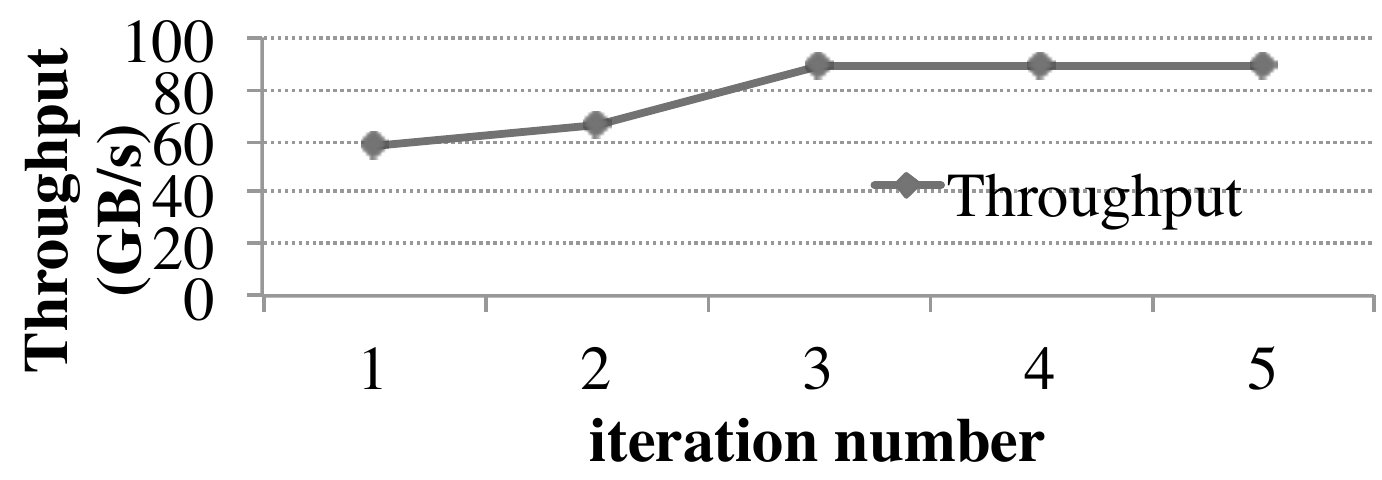}
%\vspace{-3mm}
\caption{Automatic chunk size selection with MIAD (multiple-increase, additive-decrease).}
\label{fig:auto-chunksize}
%\vspace{-0.25in}
\end{figure}
%\subsubsection{NVLink CodeGen}
%\label{sec:nvlinkCodeGen}

\noindent\textbf{Broadcast:} We first parse the spanning trees generated by the procedure described in Section~\ref{sec:design}, with each spanning tree having a different  weight associated with it. Once we receive the input buffer to be broadcast from the root note, we split the buffer among all the spanning trees based on their weights.% and correspondingly calculate the memory offsets at the destination GPUs. 
To perform data transfer on a link in the tree, we issue a \texttt{cudaMemCpy} command from the source to the destination GPU.  To reduce latency, instead of transmitting all the data assigned to this tree at once, we further divide data in each tree into multiple small chunks. Once a chunk has been transferred, we issue a CUDA event to notify the destination. To enable parallel transfers across trees, we use CUDA streams and by using a stream per link, per tree we can achieve high utilization. %Every CUDA stream represents an ordered sequence of command executions and by using a stream per link, per tree we can achieve high utilization.

% within each CUDA stream

%Based on weight, on each GPU, we calculate the corresponding buffer's memory addresses (as the data transfer starting point) for each tree.

%To enhance the throughput performance, for each spanning tree, we launch multiple CUDA streams between each send-recv GPU pair to achieve concurrent data transfer over these streams. And we will discuss more about how to choose the best number of CUDA streams in Section~\ref{sec:streamTree}.

%This is extremely important in the cases like the depth test in Section~\ref{sec:depthTest}. In this chain forwarding case, if we transmit data all at once, each GPU will not start forwarding data until it receives all the data from its predecessor, and this data receiving time introduce huge latency. By chunking the whole data into small pieces, each GPU will start forwarding data when it receives a new small data chunk, which significantly reduce the multi-hop latency. We will further discuss about how to pick the right chunk size in Section~\ref{sec:chunkSizeSelection}.

%\gw{should we also mention that we need to add some garbage collection in ncclGroupEnd?}

\noindent\textbf{AllReduce:} As described in Section~\ref{sec:allReduceDesign}, we execute AllReduce by leveraging bi-directional links. We perform reductions in one direction to a root node. Once the root node computes the final reduce result, it is broadcast in the reverse direction. We implement all the reduction functions supported by NCCL (e.g. min, max, etc.) as CUDA kernels.

\subsection{CodeGen Optimizations}
We next discuss two issues we faced during {\tt\system{}} implementation that stem from limitations of existing hardware. 
%We first discuss our approach to automatically choose an appropriate chunk size and next discuss challenges in sharing links among multiple spanning trees.

\subsubsection{Automatic chunk size selection}
\label{sec:chunkSizeSelection}

Within each CUDA stream, a \emph{chunk} is our atomic unit for data copy / synchronization between sender and receiver. 
%Pair-wise, chunk size determines how long should the sender wait before transmitting the next chunk of data. 
For spanning trees, chunk size is an important factor in determining overall latency, because each node cannot start forwarding until it receives a complete chunk from its predecessor. 
Figure~\ref{fig:chunk-data} shows a simple example in a four GPU scenario. 
Splitting data into two chunks reduces transfer time by a third when compared to a setting with no chunking.
Our goal is to parallelize (pipeline) data transfers while minimizing multi-hop latency.
Thus intuitively, making the chunk size small should improve performance and link utilization.
However for each chunk we need to issue at least three CUDA commands for copying/synchronization
%: a \texttt{cudaStreamWaitEvent} to wait for the incoming transfer, \texttt{cudaMemCpy} to do the forwarding and \texttt{cudaEventRecord} to record outgoing transfer. 
and having a large number of small chunks leads to increased overhead in scheduling these commands.
%and this is particularly severe when there are a large number of GPUs thus leading to lower throughput.

% using a very small chunk size imposes significant scheduling overhead for CUDA commands. 
%as much as possible 
%that is forwarded

% While making the chunk size small lowers latency, there are downsides to making the chunk size too small. A smaller chunk size increases the number of CUDA commands issued and thus has increased scheduling overhead. 

%\noindent{\textbf{High scheduling overhead in 8 GPU case: }}

%Open up too many streams or dividing chunk to be too small will put a lot of pressure on scheduler. This scheduling overhead grows proportionally with the number of GPU in use. And the scheduling overhead becomes more severe in 8 GPU cases, which significantly decreases the performance. 

%should open many CUDA stream to achieve high-degree of data parallel transfer, and reduce the chunk size on each stream to be very small. 
% , we did not consider about the scheduling and launching cost for these streams and events

% Second, a smaller data chunk may not fully saturate the link bandwidth available. 

Thus we use an adaptive scheme to automatically select the chunk size. As machine learning models are typically run for a large number of iterations, we observe that we can use the first few iterations to explore how changing the chunk-size affects overall performance. This is necessary as in our experience the optimal chunk size varies based on the data size, number of spanning trees in the topology and maximum depth of each tree.

\begin{figure}[t]
  \centerline{\includegraphics[width=0.9\columnwidth]{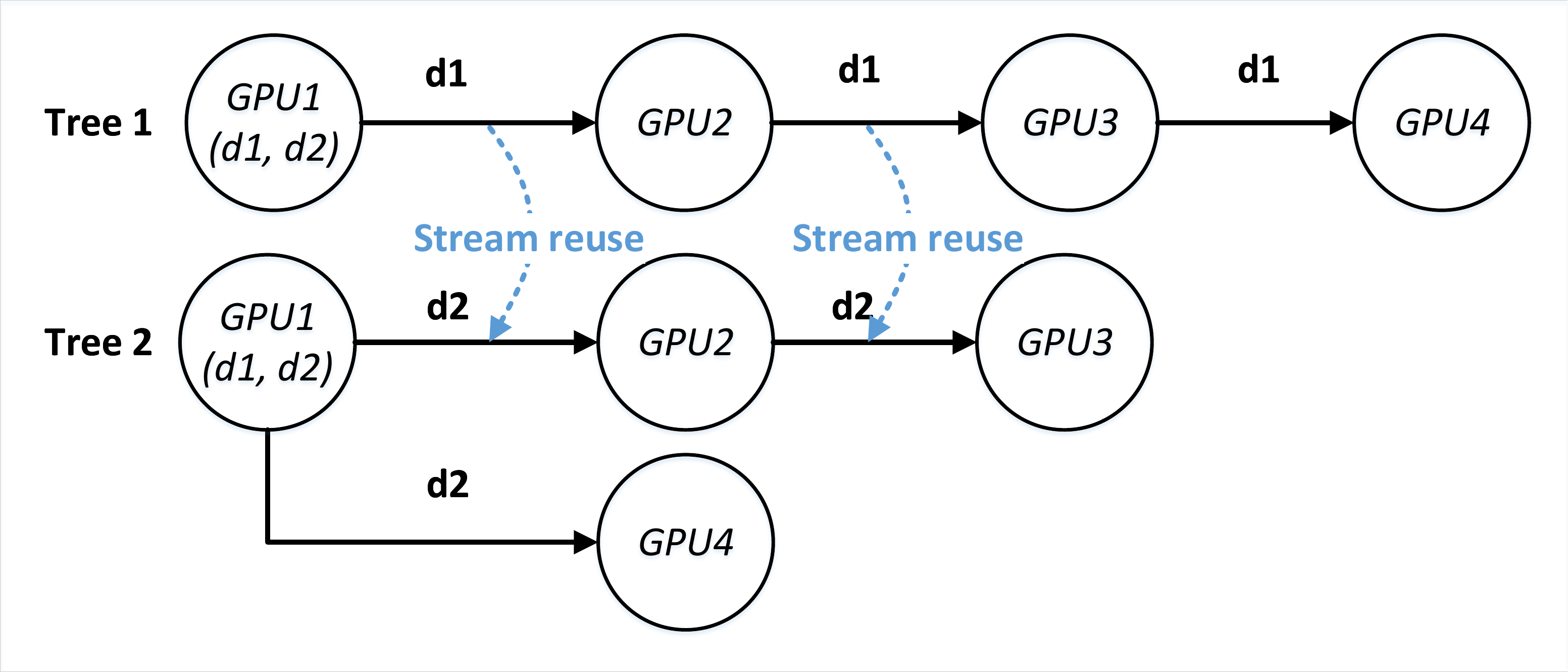}}
%  \vspace{-0.2in}
  \caption{Stream reuse for fair sharing of links.}
  \label{fig:stream-reuse}
%  \vspace{-0.15in}
\end{figure}

Our algorithm follows a \textit{multiplicative increase, additive decrease}  (\textbf{MIAD}) scheme across iterations. We initialize the chunk size with a small value and increase the chunk size by a multiplicative factor as long as the measured throughput is increasing. If the throughput decreases we additively decrease the chunk size until we reach a steady state.
Figure~\ref{fig:auto-chunksize} shows an example execution of our chunk size selection algorithm when running broadcast over 4 GPUs. Here, we start with a chunk size of 1MB and multiplicatively increase it by 2$\times$ on every iteration. We find that after four iterations the throughput stabilizes to the optimal value.

\subsubsection{Link Sharing}
\label{sec:codeGenOpt}
One of the other challenges with using multiple trees on existing hardware is that the CUDA functions do not provide any direct control on how links are shared. For example if say there are two trees with weight 0.5 that are passing through the same link, then a fair sharing scheme would transmit one chunk from the first tree followed by one chunk from second tree. However in our experiments we find that the CUDA implementation does not always result in fair sharing and that chunks from one of the trees could be arbitrarily delayed. This introduces gaps in the forwarding pipeline and harms the effective throughput achieved.

%We mainly discuss about two issues we faced during {\tt\system{}} implementation, namely link split issue, and high scheduling overhead issue.
%\noindent{\textbf{Splitting a link:}} 
%Basically, if multiple trees use the same link, we cannot evenly split the bandwidth to each tree. At each time slot, the full link bandwidth can only be assigned to a stream in a single tree. 

Since ordering guarantees are only provided by CUDA streams, we address this problem by reusing CUDA streams when the same link is used in multiple trees at roughly the same position.
%we propose stream re-use approach. The basic idea is if multiple trees using same send-recv GPU pairs, at roughly the same position, instead of generating new streams, we reuse the stream and relevant data structure initialized in the previous tree. 
For example, as shown in Figure~\ref{fig:stream-reuse}, we have two spanning trees both starting from GPU1, which contain two data pieces (d1 for tree1, d2 for tree2). Once we have created streams for first tree, we compare pairwise link positions between the two trees. Note that link GPU1 <-> GPU2 (first hop from the source) is in the same position on both trees. Thus when creating streams for tree 2, instead of initializing a new stream, we re-use the stream from tree 1 and schedule transfers to ensure fair sharing.
%Having assigned the same stream, we then schedule transfers to ensure fair sharing of link. 
%In our implementation, we also relax the stream reuse constraint to cover streams which are within a hope.
% and GPU2 <-> GPU3 (second hop from the source node) are

% which reduces the number of stream and other relevant (e.g. events, data synchronization) scheduling instructions. 
\begin{figure}[t]
\centering
\centerline{\includegraphics[width=0.95\columnwidth]{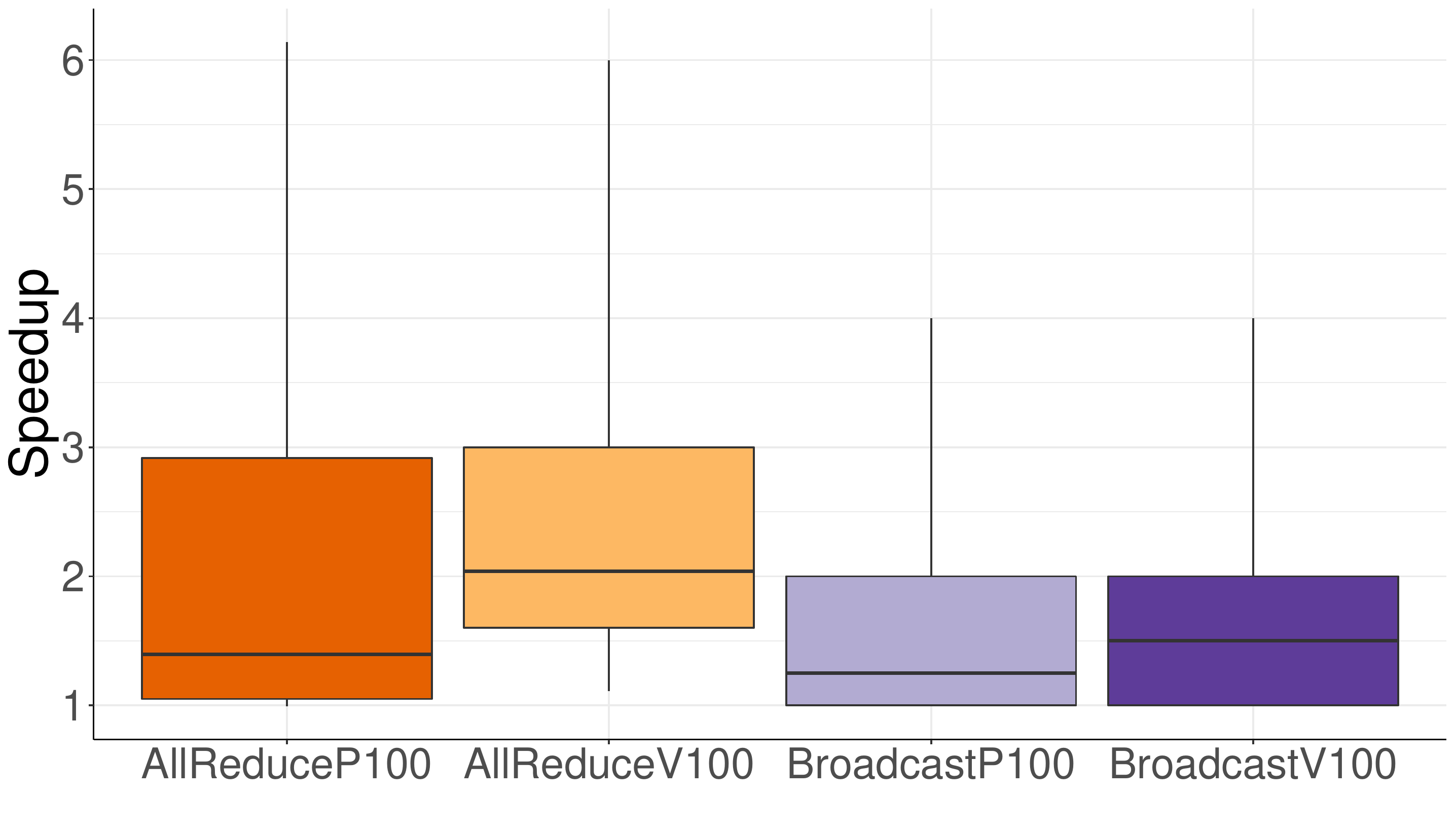}}
%\vspace{-0.2in}
\caption{Theoretical speedups from packing spanning trees compared rings on \dgxp (P100) and \dgxv (V100). Boxplot shows a distribution for possible configurations and whiskers show 5th and 95th percentile.}
\label{fig:th-speedups}
%\vspace{-0.1in}
\end{figure}

\begin{figure*}[t!]
\centering
\includegraphics[width=1\textwidth]{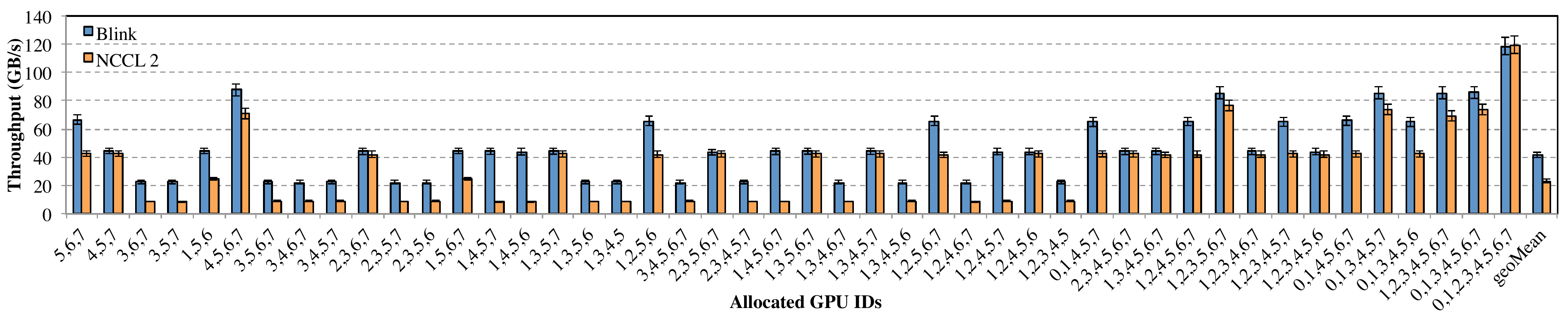}
%\vspace{-0.35in}
\caption{Broadcast throughput comparison between NCCL2 and {\tt Blink} for all unique topologies on \dgxv.}
\label{fig:bcast-nccl-blink}
%\vspace{-0.25in}
\end{figure*}

\begin{figure}[t]
\centering
\centerline{\includegraphics[width=0.9\columnwidth]{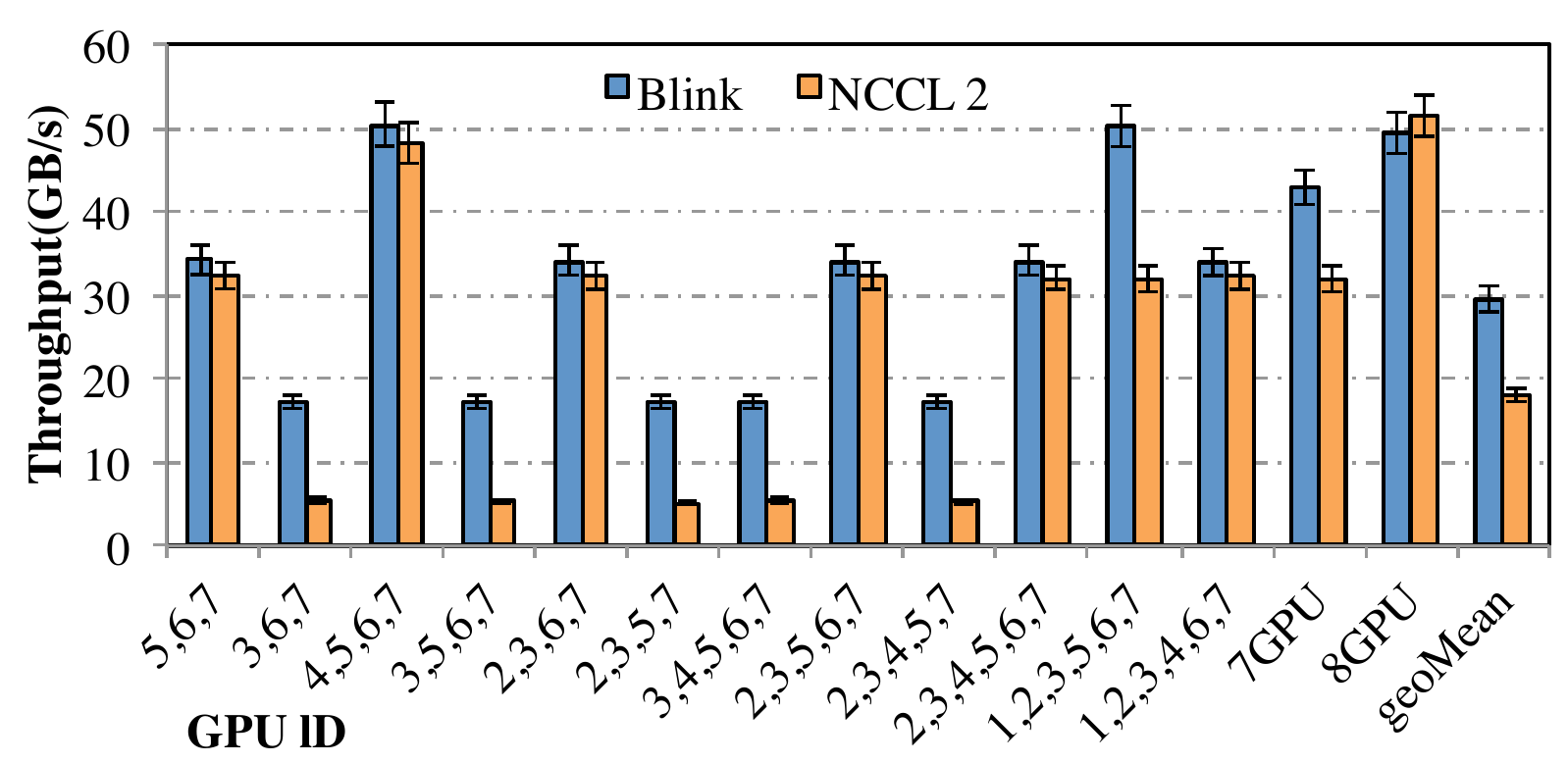}}
%\vspace{-0.15in}
\caption{Broadcast comparison between NCCL2 and {\tt Blink} in all possible topologies on \dgxp.}
\label{fig:msr-bcast}
%\vspace{-0.3in}
\end{figure}

\section{Evaluation}
\label{sec:eval}

\begin{figure*}[!t]
\centering
\includegraphics[width=1\textwidth]{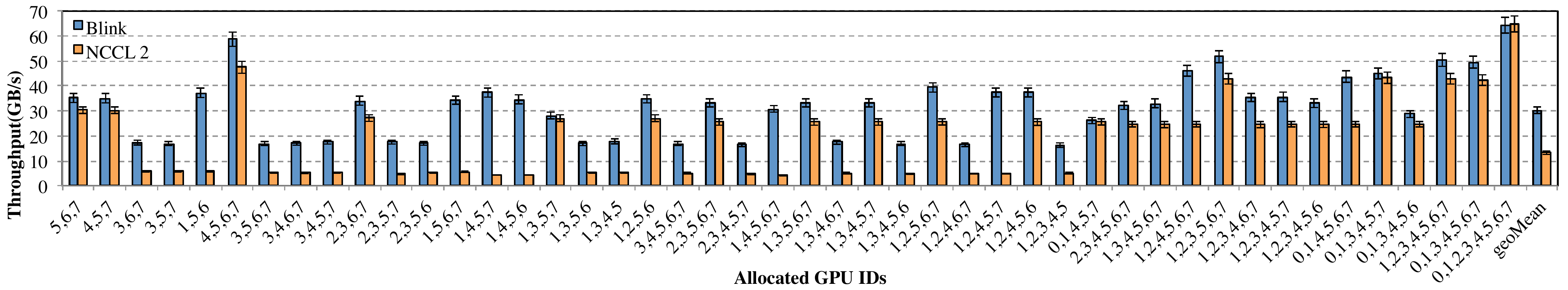}
%\vspace{-0.35in}
\caption{AllReduce throughput comparison between NCCL2 and {\tt Blink} for all unique topologies on \dgxv.}
%\vspace{-0.1in}
\label{fig:Allreduce-nccl-blink}
\end{figure*}

In this section, we evaluate {\tt \system{}}'s performance along three fronts. First, we discuss the benefits of packing trees and present theoretical comparisons between {\tt \system{}} with NVIDIA NCCL, the start-of-the-art ring-based collectives library. Second, we show experimental results highlighting throughput comparison between NCCL and {\tt \system{}} for Broadcast and AllReduce on three different hardware settings (\dgxp, \dgxv, DGX-2). Third, we discuss the trade offs in performing hybrid data transfers using both PCIe and NVLink.  Fourth, we provide end-to-end speed-up results of using {\tt \system{}} with four popular DNNs on both single DGX-1 and multi-DGX-1 settings. %We also present results from combining transfers over PCIe and NVLink in Appendix~\ref{appendix:pcie}.
%\todo{Talk about PCIe in appendix}

\subsection{Tree Packing Benefits}
\label{sec:treePackingBenefits}
\label{sec:theoryNCCLBlink}
We first evaluate the theoretical benefits of packing spanning trees vs. a ring-based approach used by libraries like NCCL. We compare the number of rings that are created in a given topology by NCCL and the total weight of spanning trees packed by {\tt \system{}} for all possible allocations from 3 GPUs to 8 GPUs on both the V100 and P100 machine. We translate this to a broadcast rate using the lower bounds on messages required for Broadcast $\lceil \frac{N-1}{N} \rceil$ and AllReduce ($2\times \lceil \frac{N-1}{N} \rceil$). That is given 4 rings for the 8 GPU case, each ring will operate at $\frac{8}{14}$ of link bandwidth and with 4 such rings our effective rate is $\frac{32}{14}$.  We approximate the bandwidth for PCIe rings to have half as much bandwidth as NVLink.

% We run this analysis

% We verify that the actual rate achieved by NCCL matches our prediction when using NVLinks.
% (we get this information from \texttt{NCCL\_DEBUG=INFO}) 
% across both V100 and P100 architectures

Figure~\ref{fig:th-speedups} shows the distribution of speedups we can achieve by packing spanning trees. We see in all cases packing spanning trees should be at least as fast as using rings and that in some cases (i.e. where rings have to go through PCIe), we can achieve up to 6x speedup. We note that our speedups could be higher in practice due to PCIe performing worse than our model or lower due to chunking overheads.

\subsection{Broadcast, AllReduce Micro-benchmarks}
\label{sec:46ubench}
\label{sec:nvlinkBcast}
\label{sec:nvlinkAllreduce}
\label{sec:dgx2Allreduce}

We next compare the performance of {\tt \system{}} with state-of-the-art NCCL2 on the two most frequently used collective primitives, namely Broadcast and AllReduce. 
Considering the topology (Figure~\ref{fig:dgx1-topo}),
and accounting for the different number of GPUs in use and their positions, we have 46 different topology settings for \dgxv, 
and 14 different topology settings for the \dgxp machine. 
For both Broadcast and AllReduce (Figure~\ref{fig:bcast-nccl-blink}, Figure~\ref{fig:msr-bcast}, Figure~\ref{fig:Allreduce-nccl-blink}), 
the number list on x-axis indicates the allocated GPUs in each configuration and  can be directly mapped to Figure~\ref{fig:dgx1-topo}.

%\begin{figure*}[!t]
%\centering
%\begin{minipage}{0.55\textwidth}
%  \centering
%    \includegraphics[width=\textwidth]{r-plots/iteration-time-cifar-geoMean.pdf}
%    \label{fig:e2e-iter-cifar}
%\end{minipage}
%\begin{minipage}{0.43\textwidth}
%    \includegraphics[width=\textwidth]{r-plots/comm-time-cifar-geoMean.pdf}
%    \label{fig:e2e-comm-cifar}
 %   \vspace{-3mm}
%\end{minipage}
%\vspace{-0.2in}
%\caption{Blink end-to-end training time reduction (CIFAR10) within a DGX-1 (V100) machine.}
%\label{fig:e2e-cifar}
%\vspace{-0.2in}
%\end{figure*}

\begin{figure*}[!t]
\centering
\begin{minipage}{0.55\textwidth}
\centering
%\subfigure[Time reduction for each iteration]{\label{fig:e2e-iter-image1k}
\includegraphics[width=\textwidth]{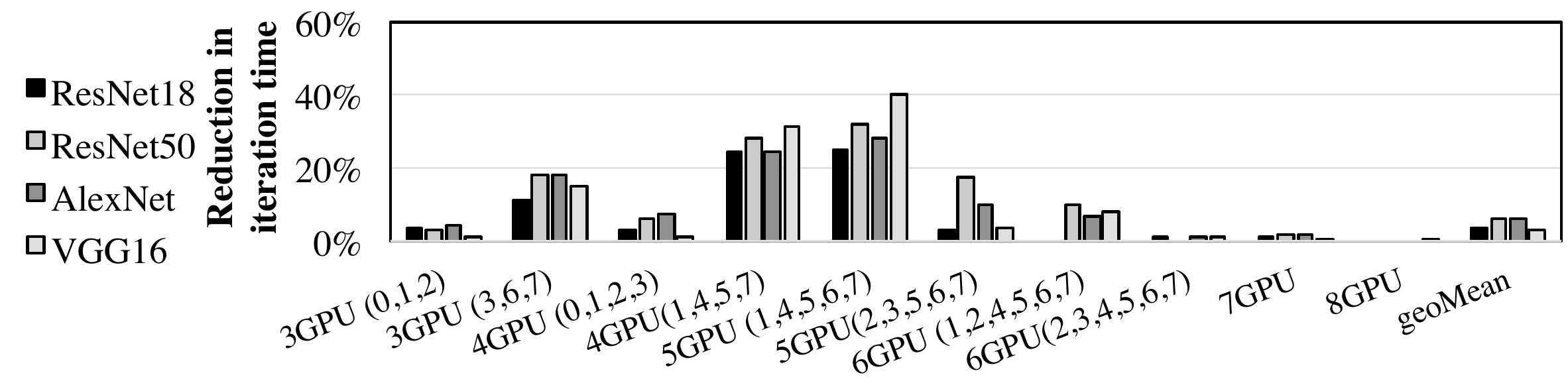}
\end{minipage}
\begin{minipage}{0.43\textwidth}
%\subfigure[Communication time reduction]{\label{fig:e2e-comm-image1k}
\includegraphics[width=\textwidth]{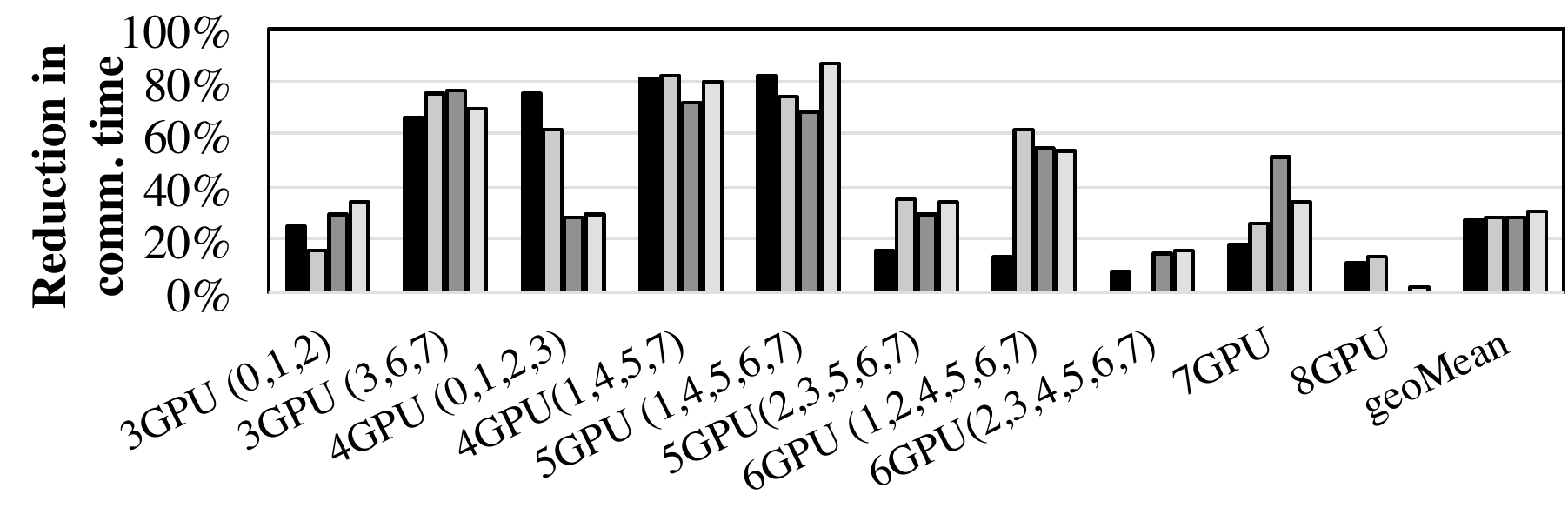}
\end{minipage}
%\vspace{-0.15in}
\caption{{\tt Blink} end-to-end training time reduction (ImageNet1K) within a \dgxv machine.}
%\vspace{-0.15in}
\label{fig:e2e-image1k}
\end{figure*}

\subsubsection{NVLink Broadcast} We provide Broadcast throughput comparison between NCCL and {\tt \system{}} for all possible topologies induced by GPU allocations on a \dgxv on AWS (p3.16xlarge). The number of GPUs we use range from 3 to 8. To fully saturate our interconnects, we test with a total data size of 500MB (50MB to 1000 MB error-bars).

In Figure~\ref{fig:bcast-nccl-blink}, {\tt \system{}} can achieve up to 6$\times$ (2x geometric mean) speed up  in performance compared to NCCL.  In the cases where GPUs are not fully connected over NVLink (e.g. GPU 1,4,5,6, as shown in Figure~\ref{fig:dgx1-topo}), NCCL cannot form NVLink-only rings across these GPUs, thus forcing it to fall back on using PCIe for data transfers.  This results in many NVLink channels going unused, leading to dramatically lower throughput.
NCCL matches {\tt \system{}} when it can form a fully connected NVLink ring and when {\tt \system{}} can only create one spanning tree (e.g., when using GPU 2,3,6,7, as depicted in Figure~\ref{fig:dgx1-topo}, NCCL2 can form one bi-directional ring: GPU2<->GPU6<->GPU7<->GPU3<->GPU2). However, even in these cases, {\tt \system{}} still achieves 3-5 GB/s higher performance due to optimized chunked transfers. 

\begin{figure}[t!]
\centering
%\subfigure[latency($\mu$s)]{\label{fig:dgx2-latency}
%\includegraphics[width=0.99\columnwidth]{figs/dgx2-latency.pdf}}
%\subfigure[throughput]{\label{fig:dgx2-tput} 
\includegraphics[width=0.99\columnwidth]{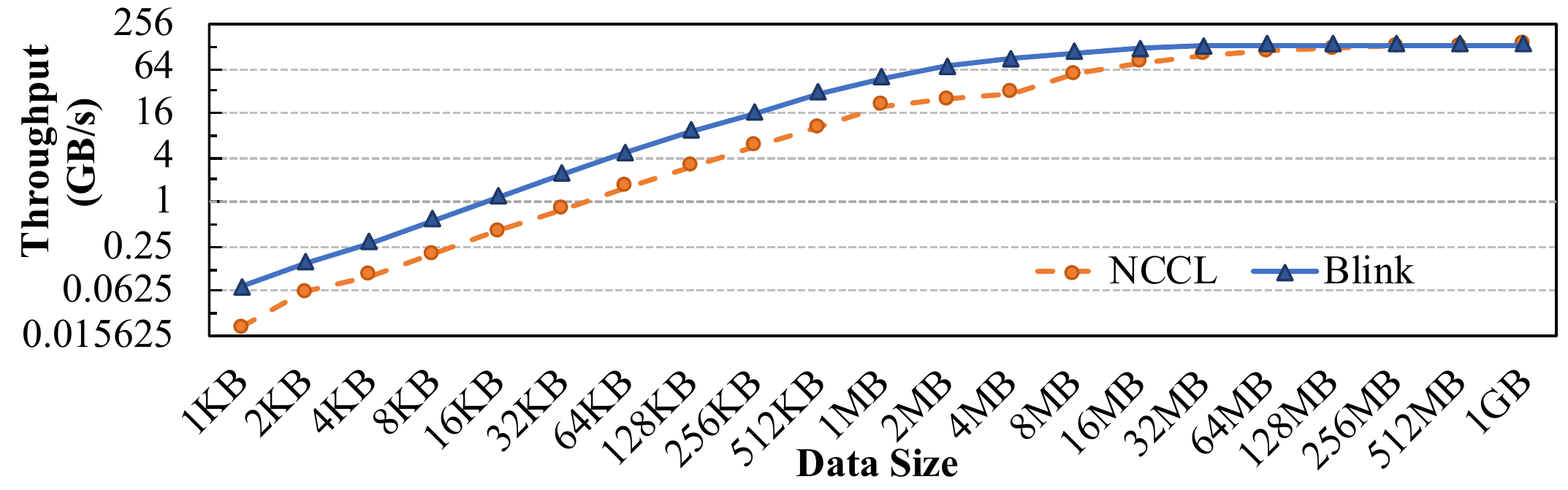}
%}
%\vspace{-0.2in}
\caption{AllReduce throughput ({\tt \system{}} and NCCL2) on a 16-GPU DGX-2.}
\label{fig:dgx2}
%\vspace{-0.2in}
\end{figure}

%\noindent\textbf{DGX-1 (P100) results on Azure:} 
Given the topology difference of \dgxp and \dgxv, we also show the throughput comparison between NCCL and {\tt \system{}} on \dgxp.
As shown in Figure~\ref{fig:msr-bcast}, we only have 14 unique topology configurations, all of which show similar throughput gains as \dgxv. Overall, {\tt \system{}} achieves up to 3x speed up (1.6x geometric mean) over NCCL.

\subsubsection{NVLink AllReduce}  Compared to Broadcast throughput in Figure~\ref{fig:bcast-nccl-blink}, AllReduce achieves lower performance for all 46 configurations for both NCCL and {\tt \system{}} (Figure~\ref{fig:Allreduce-nccl-blink}). 
This is consistent with the micro-benchmark results from Section~\ref{sec:depthTest}.  For example, in the 3 and 4 GPU settings on the \dgxv, AllReduce achieves an average 20-30GB/s less than corresponding Broadcast settings. For the 8 GPU configuration, AllReduce only achieves half of the corresponding Broadcast throughput for both NCCL and {\tt \system{}}. For NCCL's AllReduce, each data chunk needs to go through the ring twice, once for Reduce then for Broadcast, which leads to roughly half the performance. Similarly for {\tt \system{}}, reduction takes place in one direction of the spanning tree, and broadcast in the other direction. 

For AllReduce, {\tt \system{}} outperforms NCCL with up to 8$\times$ (2$\times$ geometric mean) speed up in throughput. Similar to broadcast, {\tt \system{}} has higher throughput gains in the cases where NCCL cannot form NVLink rings over the allocated GPUs or has to drop some links due to the constraint of forming rings.  Results from \dgxp also closely match these findings. %and we omit them here due to space constraints.

\begin{figure}[t!]
\centering
\includegraphics[width=0.99\columnwidth]{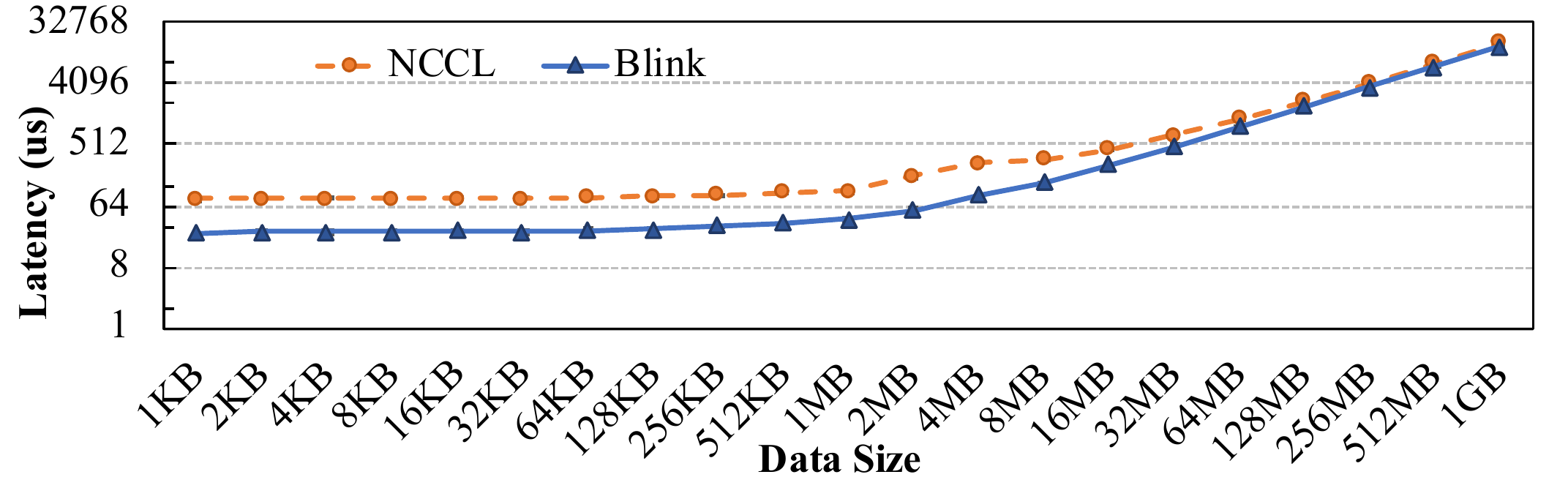}
%\vspace{-0.2in}
\caption{AllReduce Latency in $\mu$s ({\tt \system{}} and NCCL2) on a 16-GPU DGX-2.}
\label{fig:dgx2-latency}
%\vspace{-0.25in}
\end{figure}

%We present above results comparing latency for AllReduce operations when using 16 GPUs on a DGX-2 machine.
%As described in Section~\ref{sec:multi-server-and-dgx2}, {\tt\system{}} uses a number of single-hop trees to perform AllReduce when GPUs are connected using NVSwitch. One of the main advantages of a single-hop tree is that this reduces latency compared to using a ring across the GPUs. To validate this we measure the latency of AllReduce and vary the dataset size from 1KB to 1GB as shown in Figure~\ref{fig:dgx2-latency}. We find that {\tt\system{}} is especially effective for smaller data sizes offering up to 3.32$\times$ lower latency compared to NCCL's double-binary trees and rings.

\subsubsection{DGX-2 AllReduce} We next compare {\tt\system{}} to NCCL when using 16 GPUs on a DGX-2 machine.
As described in Section~\ref{sec:multi-server-and-dgx2}, {\tt\system{}} uses a number of single-hop trees to perform AllReduce when GPUs are connected using NVSwitch on the DGX-2, 
{\tt\system{}} is especially effective for smaller data sizes offering lower latency and higher throughput,
compared to NCCL's double-binary trees and rings.
{\tt\system{}} can get up to 3.32$\times$ lower latency %(Appendix~\ref{appendix:dgx2}, 
(Figure~\ref{fig:dgx2-latency}) and up to 3.5$\times$ better AllReduce throughput (Figure~\ref{fig:dgx2}) than the NCCL's double-binary trees~\cite{nccl-binary} and rings.

\begin{figure}[t]
  \centerline{\includegraphics[width=0.9\columnwidth]{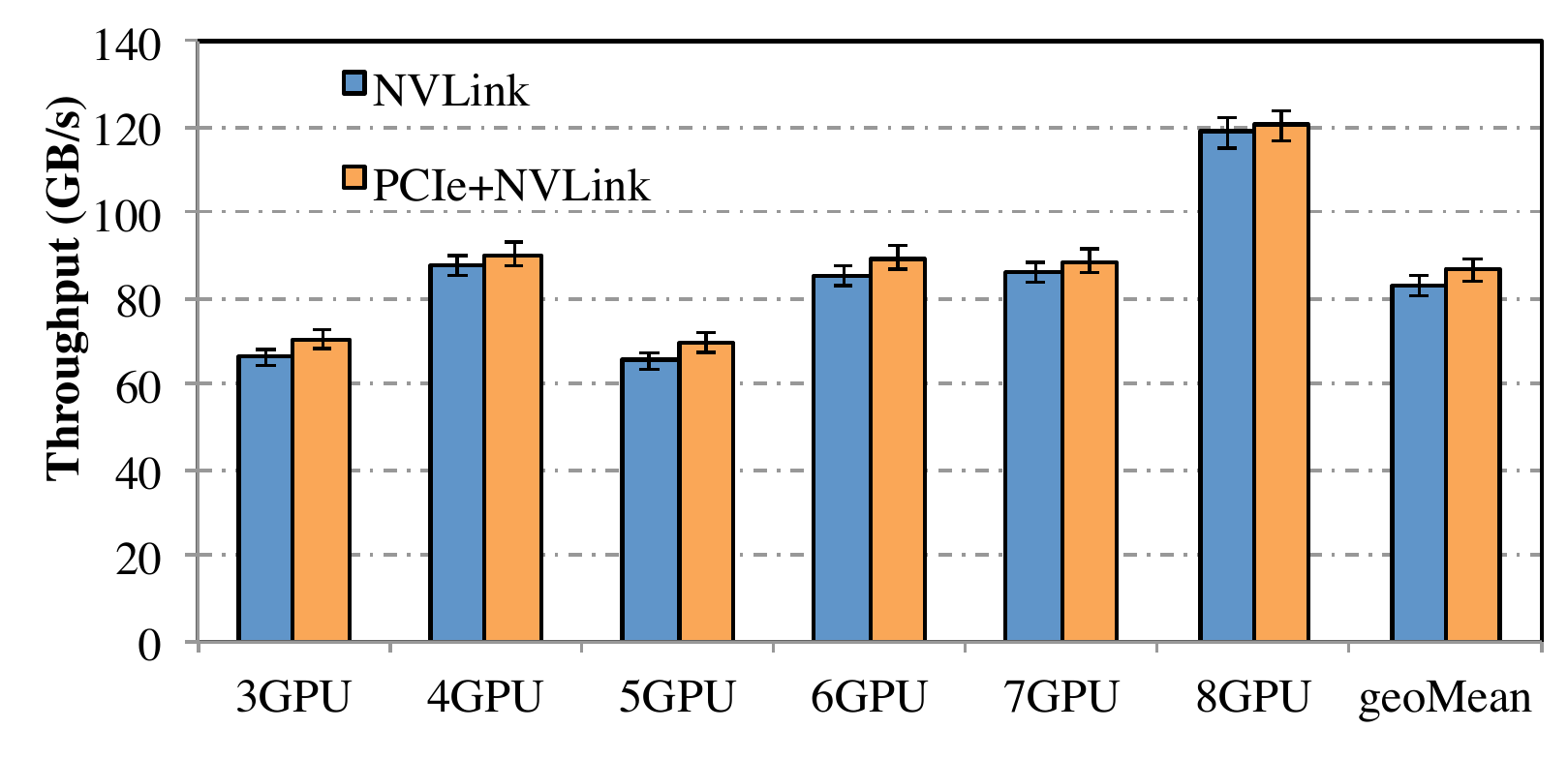}}
%  \vspace{-0.2in}
  \caption{Hybrid and NVLink-only broadcast throughput comparison with varied number of GPUs.}
  \label{fig:hybrid}
  %\vspace{-0.2in}
\end{figure}

\subsection{Hybrid Transfers}
Next, we evaluate hybrid (or combined) data transfers over both PCIe \textit{and} NVLink. For brevity, we only show broadcast results for 3 \--8 GPUs on the AWS \dgxv server. Figure~\ref{fig:hybrid}, highlights the additional 2-5 GB/s performance gain over NVLink only transfers when {\tt \system{}} combines transfers over both NVLink and PCIe.  The time to switch communication channels from NVLink to PCIe increases as the number of GPUs grow. For 3 and 4 GPU settings, compared with NVLink-only broadcast, hybrid transfers can achieve around 5GB/s boost; with 7 and 8 GPUs this bost is only around 2GB/s.  This is because the total time spent on enabling and  disabling peer-access, i.e. switching between PCIe and NVLink, is proportional to the number of GPU in use.
%However, the total time spending on disabl

\subsection{End-to-end Training}
\label{sec:e2eTest}
\label{sec:eval-multi-server}

%\begin{figure}[t!]
%  \centerline{\includegraphics[width=1\columnwidth]{figs/e2e-test.pdf}}
%  \caption{Blink end-to-end DNNs training time reduction over NCCL2 (CIFAR10).}
%  \label{fig:e2e-test}
%\end{figure}
% Large Scale Visual Recognition Challenge 2012 \-- 

We incorporate {\tt \system{}} with PyTorch~\cite{pytorch}, and evaluate the end-to-end performance gains for training.
We use four popular CNNs: AlexNet, ResNet18, ResNet50 and VGG16 and train these models %for two different tasks (with appropriate datasets): CIFAR10~\cite{cifar10} and 
on ImageNet-1K (ILSVRC12) dataset~\cite{image1k}. 
For all models, we use the same \textit{per-GPU} mini-batch size and hyper-parameters used in the original papers. 
%to train them to advertised accuracy
%For example, for Resnet50 we set mini-batchsize to 32 per worker (GPU)~\cite{facebook}; the maximum mini-batch size we use here is 256 %(8 GPU case).  

\noindent\textbf{Single server training.}
We evaluate these models by training them over 3 to 8 GPUs on the \dgxv.  For a fixed number of GPUs, we pick multiple configurations where appropriate, but to save space, we limit ourselves only to a subset of the unique configurations from before.  Specifically, from Figure~\ref{fig:Allreduce-nccl-blink}, for configurations with $n$ GPUs, if we have more than one configuration, we pick ones where the speed-up of {\tt \system{}} over NCCL is unique.

As shown in Figure~\ref{fig:e2e-image1k}, switching collective communication backend from NCCL2 to {\tt \system{}}, can reduce up to 40\% time spent in end-to-end DNN training iterations (6.3\% geometric mean), and achieve up to 87\% communication time reduction (31\% in geometric mean).
%Similar to CIFAR10 we again see comparable behavior across all four deep learning models.

\begin{figure}[t!]
\centering
\subfigure[Using 2 {\dgxv}s]{\label{fig:distE2EHetero} 
\includegraphics[width=0.48\columnwidth]{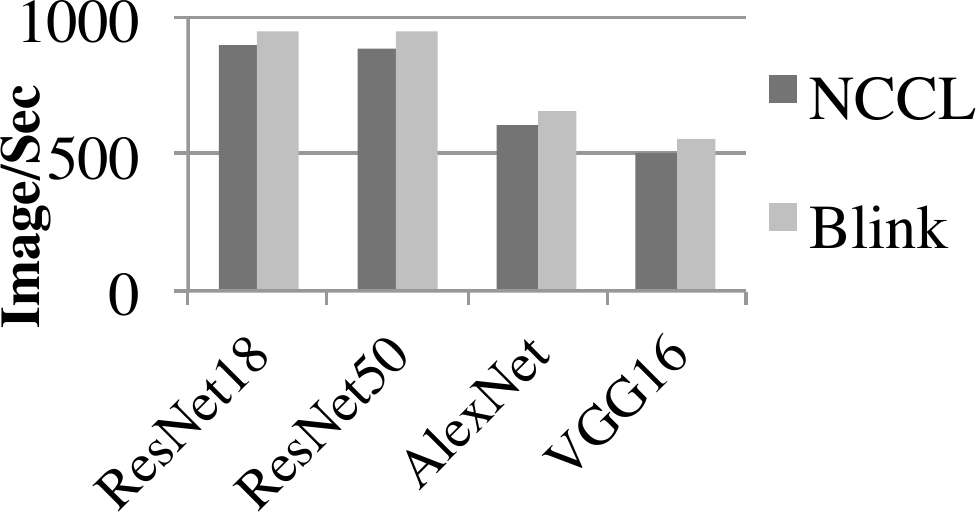}}
\subfigure[AllReduce Projections]{\label{fig:100MB-cross-machine}
\includegraphics[width=0.48\columnwidth]{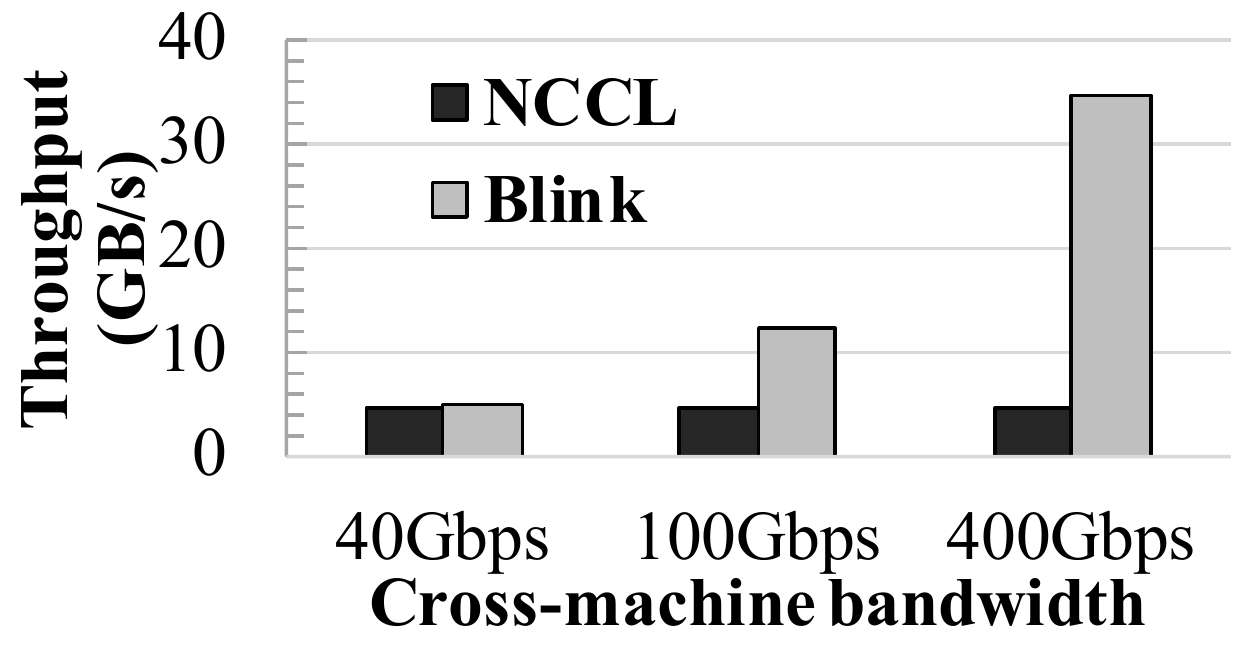}}
%\vspace{-0.2in}
\caption{Multi-DGX-1 DNN training with {\tt \system{}}.}
\label{fig:distE2E}
%\vspace{-0.25in}
\end{figure}

\noindent\textbf{Multi-server training.}
{\tt\system{}}'s multi-server AllReduce consists of a per-server reduction over spanning trees ($t_1$), cross-server broadcast and reduce ($t_2$), followed by a broadcast within each server as before ($t_3$).
%\todo{For multi-server training with no fragmentation, when all the GPUs in a server are allocated to a job, Figure~\ref{fig:distE2E}(a) shows that {\tt\system{}} achieves near linear scaling over 2 to 4 DGX-1 (V100) machines; this is comparable to state-of-the-art solutions~\cite{horovod}.}   
We consider scenarios where the GPU allocation is fragmented across machines, prevalent in multi-tenant clusters as shown in Figure~\ref{fig:gpu_dist}. 
For example we consider a 8GPU job spread across two \dgxv servers with 3 and 5 GPUs allocated respectively.
Figure~\ref{fig:distE2EHetero} shows that {\tt\system{}} outperforms Horovod with NCCL/MPI by up to 11\%.
{\tt\system{}}'s reduction in improvement over NCCL, compared to the gains in single-server training, stem from commodity cloud interconnects.
In commodity networks, inter-server AllReduce throughput (40Gbps) is much lower than intra-server throughput (40GBps).  Thus while {\tt\system{}} can reduce $t_1$ and $t_3$, there isn't much that can be done for $t_2$.

To understand how faster interconnects will change performance,  
we present results from a simulation varying the cross-machine bandwidth (Figure~\ref{fig:100MB-cross-machine}). 
We compare AllReduce throughput for 100MB of data and see that as cross-machine bandwidth increases~\cite{catchcloud,400g-2,400g-1}, {\tt\system{}}'s design will lead to more pronounced end-to-end benefits.  
NCCL is bound by intra-server PCIe throughput where as {\tt Blink} can keep up with inter-server throughput until the intra-\dgxv NVLinks become a bottleneck (for the 3-5 GPU case this is $\sim$300Gbps).

%$\sim$35GB/s or
% throughput
% , such as 400Gbps~\cite{catchcloud,400g-1,400g-2} or multiple IB ports (n$\times$56Gbps), $t_2$ can be shrunk; this will equally benefit NCCL and {\tt\system{}}, but {\tt\system{}}'s reductions in $t_1$ and $t_3$ will result in more pronounced end-to-end benefits (Figure~\ref{fig:e2e-image1k}).

%\begin{figure}[t!]
%\centering
%\subfigure[Using 2 {\dgxv}s]{\label{fig:distE2EHetero} 
%\includegraphics[width=0.45\columnwidth]{figs/gpu_distributed_5_3-crop.pdf}}
%\subfigure[AllReduce Projections]{\label{fig:100MB-cross-machine}
%\includegraphics[width=0.45\columnwidth]{figs/100MB-cross-machine-bw.pdf}}
%\vspace{-0.2in}
%\caption{\small{Multi-DGX-1 DNN training with {\tt \system{}}.}}
%\label{fig:distE2E}
%\vspace{-0.25in}
%\end{figure}

\section{Related Work}
\label{sec:related}
%\vspace{-2mm}
Work on collectives fall in one of two buckets (below):

\noindent\textbf{Topology-fixed Schemes.}
Basic collective operations (e.g. Broadcast, AllReduce) are fully supported in the MPI (Message Passing Interface) standard~\cite{mpitutorial}. 
Earlier work has mainly focused on designing optimal collectives over regular, well-defined network structures like hypercube~\cite{mpi-hypercube,mpi-hypercube2}, full mesh~\cite{mpi-mesh}, etc. 
Recent work has looked at more general networks, with optimizations for scenarios when number of communication nodes are not power of two~\cite{non-power2},
and for auto-tuning of buffer size and algorithm selection for a specific system architecture~\cite{auto-tune-mpi}. 

Under specific network settings, there are many algorithms that achieve better performance than MPI. 
For example, the latency-optimal AllReduce solution, "butterfly algorithm"~\cite{butterfly1,butterfly2,butterfly-mix}, divides AllReduce into two steps: first is a recursive ReduceScatter and then followed by a recursive AllGather. 
But, the communication pattern of butterfly algorithms often cause network contention, which makes it less practical. 
Within a tree or ring topology, ring-based collectives were shown to be bandwidth optimal in homogeneous network settings~\cite{ring-bcast,ring-allreduce}. 
Several companies have developed their own implementations of this algorithm, such as Horovod~\cite{horovod} from Uber, Baidu Ring AllReduce~\cite{baidu}, NVIDIA NCCL~\cite{nccl}, Facebook's Gloo~\cite{gloo}, IBM Power AI DDL~\cite{ibm}.
However, they all operate under the assumption of a fixed topology, which is not a good fit for cloud computing where topology may change dynamically. {\tt \system{}} is designed to handle irregular topologies and yield optimal solutions.

%\todo{cite blueconnect}

\noindent\textbf{Topology-aware Protocols.}
%Traditional MPI-based schemes are mainly designed for communications within a regular and symmetric virtual topology and does not take physical topology heterogeneity into account. 
%There is another branch of work which are topology-aware collective protocols.
Techniques that exploit hierarchy in wide area networks for collective communication center around the idea of minimizing data transfer over slow (wide-area) links~\cite{mpi-topo1,mpi-topo2}. 
The same idea has been extended to cloud environments where node locality is determined by pair-wise network bandwidth measurements~\cite{mpi-net-aware}. Smelt adopts similar idea in NUMA multi-core environment~\cite{kaestle}. 
%Adapt ~\cite{mpi-adapt} increase collective throughput by relaxing synchronization barrier and concurrent execute collective communications without data dependency. However, in distributed ML we cannot loosen synchronization barrier since it may lead to DNN models not converge.
Blueconnect decouples AllReduce into ReduceScatter and AllGather, pipelining these two sub-operations~\cite{blueconnect}.
However it only works on symmetric topologies, making it less flexible than {\tt \system{}} spanning trees.
%To the best of our knowledge, all of them are collectives implemented for CPUs, and none of them address challenges in handling high throughput NVLink interconnects. 
{\tt \system{}} is general and is optimized for multi-GPU collective communication, over symmetric or asymmetric topologies, and can combine heterogeneous links (such as PCIe and NVLink) for data transfer. % (such as combining transfers over PCIe and NVLink). 
%Furthermore, this two level hierarchy (local-area, wide-area) is too coarse-grained to achieve optimal network scheduling, whereas {\tt \system{}} provides topology-aware optimal solutions.

\section{Conclusion}
\label{sec:concl}
%\vspace{-3mm}
{\tt \system{}} is a fast and generic collective communication library to accelerate distributed machine learning. To handle topology heterogeneity prevalent in modern GPU hardware, {\tt \system{}} dynamically packs spanning trees to maximize link utilization. Compared with state-of-the-art, ring-based collective communication protocols like NCCL2, {\tt \system{}} can achieve up to 8$\times$ faster model synchronization and reduce end-to-end DNN model training time by up to 40\%.
%{\color{red}within a single multi-GPU machine.}

% {\tt \system{}} achieves near linear scalability in distributed, multi-machine settings.}

%\noindent\textbf{This work does not raise any ethical issues.}
\section*{Acknowledgements}
We thank the MSR Lab LT, especially Ricardo Bianchini and Donald Kossmann, for
their enthusiastic and unwavering support of Project Fiddle, and for
their generous support in procuring the many resources required
to develop and evaluate Project Blink.
We also thank the MSR GCR staff, especially Jim Jernigan and Steven Dahl, for supporting our DGX-1, DGX-2 needs.

Additionally, Guanhua Wang and Ion Stoica are supported by 
a NSF CISE Expeditions Award CCF-1730628, and their research was also supported by gifts from Alibaba, Amazon Web Services, Ant Financial, CapitalOne, Ericsson, Facebook, Futurewei, Google, Intel, Microsoft, Nvidia, Scotiabank, Splunk, and VMware. 
Shivaram Venkataraman is also supported
by a Facebook faculty research award and support for this research was also provided by the Office of the Vice Chancellor
for Research and Graduate Education at the University of Wisconsin, Madison with funding from the Wisconsin Alumni
Research Foundation.

\bibliographystyle{plain}
\bibliography{blink}

\begin{thebibliography}{10}

\bibitem{tensorflow}
Martin Abadi, Paul Barham, Jianmin Chen, Zhifeng Chen, Andy Davis, Jeffrey
  Dean, Matthieu Devin, Sanjay Ghemawat, Geoffrey Irving, Michael Isard,
  Manjunath Kudlur, Josh Levenberg, Rajat Monga, Sherry Moore, Derek~G. Murray,
  Benoit Steiner, Paul Tucker, Vijay Vasudevan, Pete Warden, Martin Wicke, Yuan
  Yu, and Xiaoqiang Zheng.
\newblock Tensorflow: A system for large-scale machine learning.
\newblock In {\em USENIX OSDI}, 2016.

\bibitem{mpi-mesh}
M.~Barnett, R.~Littlefield, D.~Payne, and R.~van~de Geijn.
\newblock Global combine on mesh architectures with wormhole routing.
\newblock In {\em Proceedings of the 7th International Parallel Processing
  Symposium}, 1993.

\bibitem{mpi-hypercube2}
Laxmi~N. Bhuyan and Dharma~P. Agrawal.
\newblock Generalized hypercube and hyperbus structures for a computer network.
\newblock {\em IEEE Transactions on Computers}, 1984.

\bibitem{mpitutorial}
{Blaise Barney}.
\newblock {Message Passing Interface}.
\newblock \url{https://computing.llnl.gov/tutorials/mpi/}, 2018.

\bibitem{chekuri2017near}
Chandra Chekuri and Kent Quanrud.
\newblock Near-linear time approximation schemes for some implicit fractional
  packing problems.
\newblock In {\em Proceedings of the Twenty-Eighth Annual ACM-SIAM Symposium on
  Discrete Algorithms}, pages 801--820. SIAM, 2017.

\bibitem{blueconnect}
Minsik Cho, Ulrich Finkler, David Kung, and Hillery Hunter.
\newblock Blueconnect: Decomposing all-reduce for deep learning on
  heterogeneous network hierarchy.
\newblock In {\em sysML}, 2019.

\bibitem{dgx1}
{NVIDIA DGX-1}.
\newblock \url{https://www.nvidia.com/en-us/data-center/dgx-1/}, 2017.

\bibitem{dgx2}
{NVIDIA DGX-2}.
\newblock \url{https://www.nvidia.com/en-us/data-center/dgx-2/}, 2018.

\bibitem{edmonds1973edge}
Jack Edmonds.
\newblock Edge-disjoint branchings.
\newblock {\em Combinatorial algorithms}, 1973.

\bibitem{ring-bcast}
A.~Faraj, Pitch Patarasuk, and Xin Yuan.
\newblock Bandwidth efficient all-to-all broadcast on switched clusters.
\newblock {\em International Journal of Parallel Programming}, 2008.

\bibitem{gabow1998packing}
Harold~N Gabow and KS~Manu.
\newblock Packing algorithms for arborescences (and spanning trees) in
  capacitated graphs.
\newblock {\em Mathematical Programming}, 82(1-2):83--109, 1998.

\bibitem{mpi-net-aware}
Yifan Gong, Bingsheng He, and Jianlong Zhong.
\newblock Network performance aware mpi collective communication operations in
  the cloud.
\newblock {\em IEEE Transactions on Parallel and Distributed Systems (TPDS)},
  2015.

\bibitem{facebook}
Priya Goyal, Piotr Dollar, Ross Girshick, Pieter Noordhuis, Lukasz Wesolowski,
  Aapo Kyrola, Andrew Tulloch, Yangqing Jia, and Kaiming He.
\newblock {Accurate, Large Minibatch SGD: Training ImageNet in 1 Hour}.
\newblock {\em arXiv preprint arXiv:1706.02677}, 2017.

\bibitem{ibm}
Hillery Hunter.
\newblock {IBM Research achieves record deep learning performance with new
  software technology}.
\newblock
  \url{https://www.ibm.com/blogs/research/2017/08/distributed-deep-learning/},
  2017.

\bibitem{nccl}
Sylvain Jeaugey.
\newblock {Optimized inter-GPU collective operations with NCCL 2}.
\newblock \url{https://developer.nvidia.com/nccl}, 2017.

\bibitem{philly}
Myeongjae Jeon, Shivaram Venkataraman, Amar Phanishayee, Junjie Qian, Wencong
  Xiao, and Fan Yang.
\newblock {Multi-tenant GPU Clusters for Deep Learning Workloads: Analysis and
  Implications}.
\newblock {\em Microsoft Research Technical Report (MSR-TR-2018-13)}, 2018.

\bibitem{kaestle}
Stefan Kaestle, Reto Achermann, Roni Haecki, Moritz Hoffmann, Sabela Ramos, and
  Timothy Roscoe.
\newblock Machine-aware atomic broadcast trees for multicoress.
\newblock In {\em USENIX OSDI}, 2016.

\bibitem{mpi-topo1}
N.~Karonis, B.~de~Supinski, I.~Foster, W.~Gropp, E.~Lusk, and J.~Bresnahan.
\newblock Exploiting hierarchy in parallel computer networks to optimize
  collective operation performance.
\newblock In {\em Proceedings of the Fourteenth International Parallel and
  Distributed Processing Symposium}, IEEE IPDPS'00, 2000.

\bibitem{mpi-topo2}
T.~Kielmann, R.~F.~H. Hofman, H.~E. Bal, A.~Plaat, and R.~A.~F. Bhoedjang.
\newblock {MagPIe: MPI's collective communication operations for clustered wide
  area systems}.
\newblock In {\em ACM SIGPLAN Symposium on Principles and Practice of Parallel
  Programming}, ACM PPoPP'99, 1999.

\bibitem{lecunn-smallbs}
Yann LeCun.
\newblock {Training with large minibatches is bad for your health.}
\newblock \url{https://twitter.com/ylecun/status/989610208497360896?lang=en},
  2018.

\bibitem{lovasz1976two}
Laszlo Lovasz.
\newblock On two minimax theorems in graph.
\newblock {\em Journal of Combinatorial Theory, Series B}, 21(2):96--103, 1976.

\bibitem{SmallBS}
Dominic Masters and Carlo Luschi.
\newblock Revisiting small batch training for deep neural networks.
\newblock {\em CoRR}, abs/1804.07612, 2018.

\bibitem{pipedream-sosp19}
Deepak Narayanan, Aaron Harlap, Amar Phanishayee, Vivek Seshadri, Nikhil
  Devanur, Greg Granger, Phil Gibbons, and Matei Zaharia.
\newblock Pipedream: Generalized pipeline parallelism for dnn training.
\newblock In {\em ACM Symposium on Operating Systems Principles (SOSP 2019)},
  October 2019.

\bibitem{nccl-binary}
{Massively Scale Your Deep Learning Training with NCCL 2.4 }.
\newblock \url{https://bit.ly/2lFwFQ4}, 2019.

\bibitem{baidu}
Andrew Ng.
\newblock {Bringing HPC Techniques to Deep Learning}.
\newblock
  \url{http://research.baidu.com/bringing-hpc-techniques-deep-learning/}, 2017.

\bibitem{gloo}
Pieter Noordhuis.
\newblock {Accelerating machine learning for computer vision}.
\newblock \url{https://github.com/facebookincubator/gloo}, 2017.

\bibitem{nvlink}
{NVIDIA NVLINK}.
\newblock \url{http://www.nvidia.com/object/nvlink.html}, 2017.

\bibitem{nvswitch}
{NVIDIA NVSWITCH}.
\newblock
  \url{http://images.nvidia.com/content/pdf/nvswitch-technical-overview.pdf},
  2018.

\bibitem{400g-1}
{Removing roadblocks on the path to 400G and beyond }.
\newblock \url{https://bit.ly/2k4PXh9}, 2018.

\bibitem{pytorch}
Adam Paszke, Sam Gross, Soumith Chintala, Gregory Chanan, Edward Yang, Zachary
  DeVito, Zeming Lin, Alban Desmaison, Luca Antiga, and Adam Lerer.
\newblock {Automatic differentiation in PyTorch}.
\newblock In {\em Proceedings of the 31st Conference on Neural Information
  Processing Systems}, NIPS'17, 2017.

\bibitem{ring-allreduce}
Pitch Patarasuk and Xin Yuan.
\newblock Bandwidth optimal all-reduce algorithms for clusters of workstations.
\newblock {\em J. Parallel Distrib. Comput.}, pages 117--124, 2009.

\bibitem{pcie}
{PCI Express: An Overview of the PCI Express Standard}.
\newblock \url{http://www.ni.com/white-paper/3767/en/}, 2014.

\bibitem{butterfly1}
Rolf Rabenseifner.
\newblock Optimization of collective reduction operations.
\newblock In {\em International Conference on Computational Science}, 2004.

\bibitem{image1k}
Olga Russakovsky, Jia Deng, Hao Su, Jonathan Krause, Sanjeev Satheesh, Sean Ma,
  Zhiheng Huang, Andrej Karpathy, Aditya Khosla, Michael Bernstein,
  Alexander~C. Berg, and Li~Fei-Fei.
\newblock Imagenet large scale visual recognition challenge.
\newblock {\em International Journal of Computer Vision}, 2015.

\bibitem{mpi-hypercube}
D.~Scott.
\newblock Efficient all-to-all communication patterns in hypercube and mesh
  topologies.
\newblock In {\em Proceedings of the 6th Distributed Memory Computing
  Conference}, 1991.

\bibitem{horovod}
Alex Sergeev and Mike~Del Balso.
\newblock {Horovod: fast and easy distributed deep learning in TensorFlow}.
\newblock {\em arXiv preprint arXiv:1802.05799}, 2018.

\bibitem{IncreaseBS}
Samuel~L. Smith, Pieter{-}Jan Kindermans, and Quoc~V. Le.
\newblock Don't decay the learning rate, increase the batch size.
\newblock {\em CoRR}, abs/1711.00489, 2017.

\bibitem{non-power2}
Rajeev Thakur, Rolf Rabenseifner, and William Gropp.
\newblock Optimization of collective communication operations in mpich.
\newblock {\em Int. J. High Perform. Comput. Appl.}, 2005.

\bibitem{catchcloud}
Shelby Thomas, Geoffrey~M. Voelker, and George Porter.
\newblock Cachecloud: Towards speed-of-light datacenter communication.
\newblock In {\em USENIX hotcloud 2018}, 2018.

\bibitem{auto-tune-mpi}
Sathish~S. Vadhiyar, Graham~E. Fagg, and Jack Dongarra.
\newblock Automatically tuned collective communications.
\newblock In {\em Proceedings of the 2000 ACM/IEEE Conference on
  Supercomputing}, SC '00, 2000.

\bibitem{butterfly2}
Robert van~de Geijn.
\newblock On global combine operations.
\newblock In {\em Journal of Parallel and Distributed Computing}, 1994.

\bibitem{400g-2}
{Verizon marks milestone with successful 400G technology trial}.
\newblock \url{https://bit.ly/2lKgAs7}, 2018.

\bibitem{gandiva-osdi18}
Wencong Xiao, Romil Bhardwaj, Ramachandran Ramjee, Muthian Sivathanu, Nipun
  Kwatra, Zhenhua Han, Pratyush Patel, Xuan Peng, Hanyu Zhao, Quanlu Zhang, Fan
  Yang, and Lidong Zhou.
\newblock Gandiva: Introspective cluster scheduling for deep learning.
\newblock In {\em 13th {USENIX} Symposium on Operating Systems Design and
  Implementation ({OSDI} 18)}, pages 595--610, Carlsbad, CA, 2018. {USENIX}
  Association.

\bibitem{PoseidonATC2017}
Hao Zhang, Zeyu Zheng, Shizhen Xu, Wei Dai, Qirong Ho, Xiaodan Liang, Zhiting
  Hu, Jinliang Wei, Pengtao Xie, and Eric~P. Xing.
\newblock Poseidon: An efficient communication architecture for distributed
  deep learning on {GPU} clusters.
\newblock In {\em 2017 {USENIX} Annual Technical Conference ({USENIX} {ATC}
  17)}, pages 181--193, Santa Clara, CA, 2017. {USENIX} Association.

\bibitem{butterfly-mix}
Huasha Zhao and John Canny.
\newblock Butterfly mixing: Accelerating incremental-update algorithms on
  clusters.
\newblock In {\em Proceedings of the 2013 SIAM International Conference on Data
  Mining}, 2013.

\end{thebibliography}
%\clearpage

%\input{tex/appendix}
\appendix
\clearpage

%\newpage
\section{Appendix}
\label{appendix}

\subsection{Micro Benchmarks (\dgxv)}
\label{appendix:microbench}

We continue our discussion of micro benchmarks from Section~\ref{sec:ubench}, highlighting results for forwarding on a chain and fan in/out tests.

\subsubsection{Depth Test}
\label{sec:depthTestFull}

\begin{figure}[h]
\centering
\subfigure[chain forward]{\label{fig:chain-forward}
\includegraphics[width=0.4\textwidth]{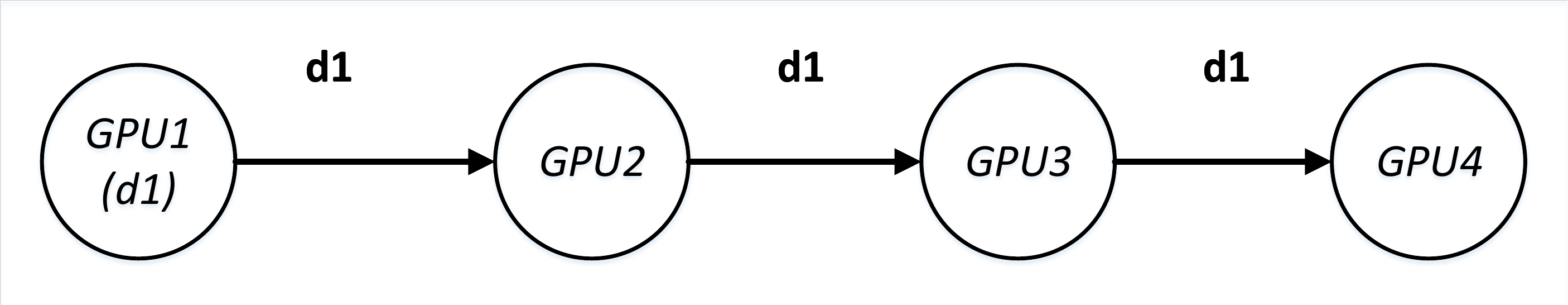}}
\subfigure[chain reduce+forward]{\label{fig:chain-add-forward} 
\includegraphics[width=0.4\textwidth]{figs/chain-add-forward.pdf}}
\subfigure[chain reduce-broadcast]{\label{fig:chain-reduce-bcast}
\includegraphics[width=0.4\textwidth]{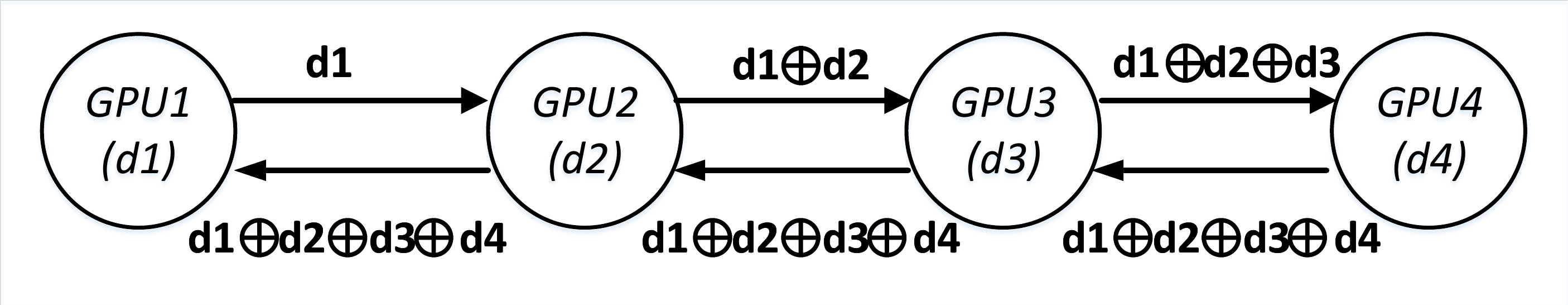}}
%\vspace{-3mm}
\caption{Depth test over a chain of GPUs.}
\label{fig:depth}
%\vspace{-4mm}
\end{figure}

%For the forwarding benchmark (Figure~\ref{fig:chain-forward}), GPU1 is the source node with data named d1, and it passes the data d1 to GPU2 and then GPU2 forwards it to GPU3 etc. 
%For ``reduce+broadcast'' (Figure~\ref{fig:chain-reduce-bcast}), we perform ``reduce+forward'' in one direction and ``forward'' in the other direction, as such a capability can be used for all-to-all reductions.
The first topology class we consider is a depth test where we vary depth of trees that are used. To do this we consider a simple \emph{chain} topology as shown in Figure~\ref{fig:depth}.

%Depth test refers to that, given a bunch of GPUs, we use a link chain to connect them all together. Figure~\ref{fig:depth} shows depth test in a 4 gpu case. 

Given a chain topology, we measure all three kinds of traffic patterns: data forward, reduce+forward, and reduce-broadcast. 
As shown in~\ref{fig:chain-forward}, for the forwarding benchmark, GPU1 is the source node with data named d1, and it passes the data d1 to GPU2 and then GPU2 forwards it to GPU3 etc. For reduce+forward (Figure~\ref{fig:chain-add-forward}), each GPU (except the last one) has its own data. When a GPU receives data from its predecessor, it invokes a reduction function (denoted as \textcircled{+}) on the received data with its own data, passing the result to its successor. Finally, we implement reduce+broadcast by doing reduce+forward in one direction and forward in the other direction as shown as Figure~\ref{fig:chain-reduce-bcast}, as such a capability can be used for all-to-all reductions.

\begin{figure}[h]
\centering
\subfigure[chain forward throughput]{\label{fig:chain-forward-tput}
\includegraphics[width=0.45\textwidth]{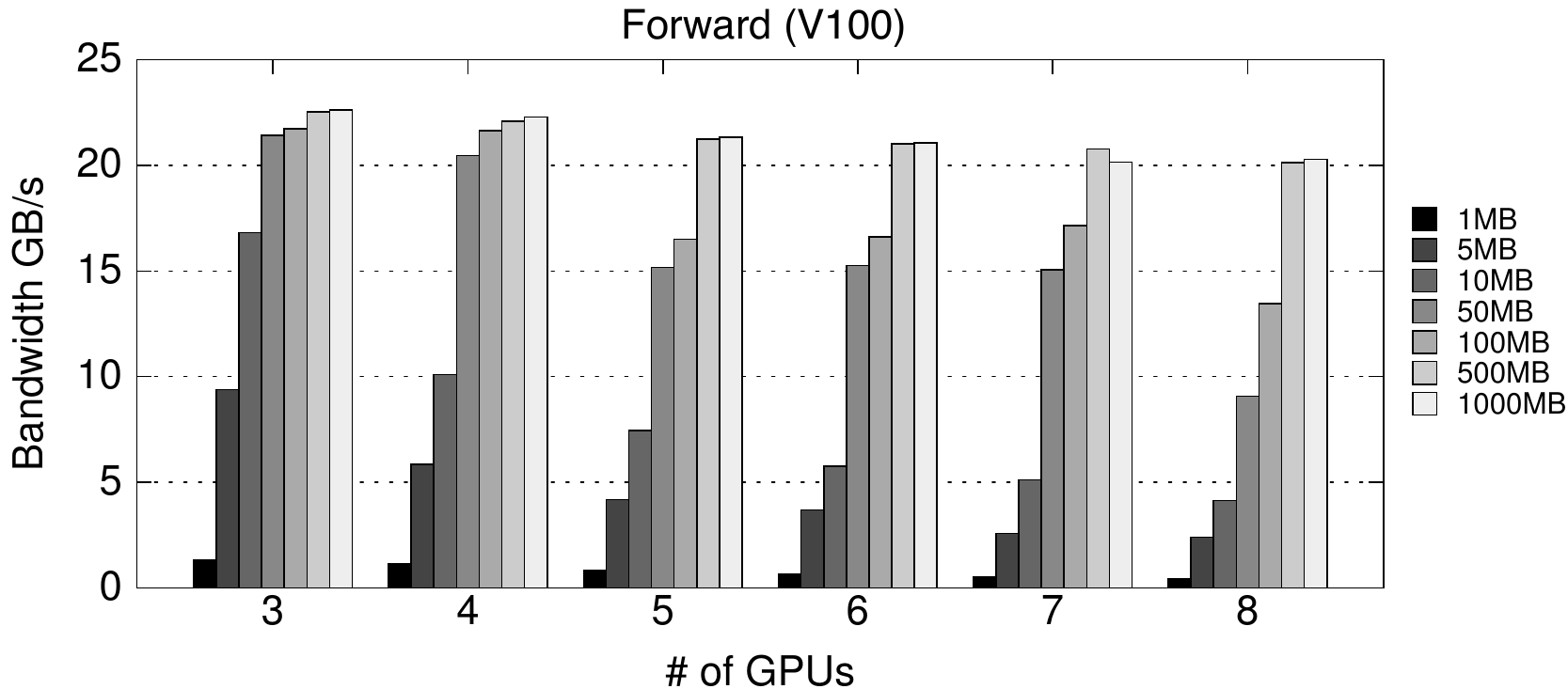}}
\subfigure[chain reduce+forward throughput]{\label{fig:chain-add-forward-tput} 
\includegraphics[width=0.45\textwidth]{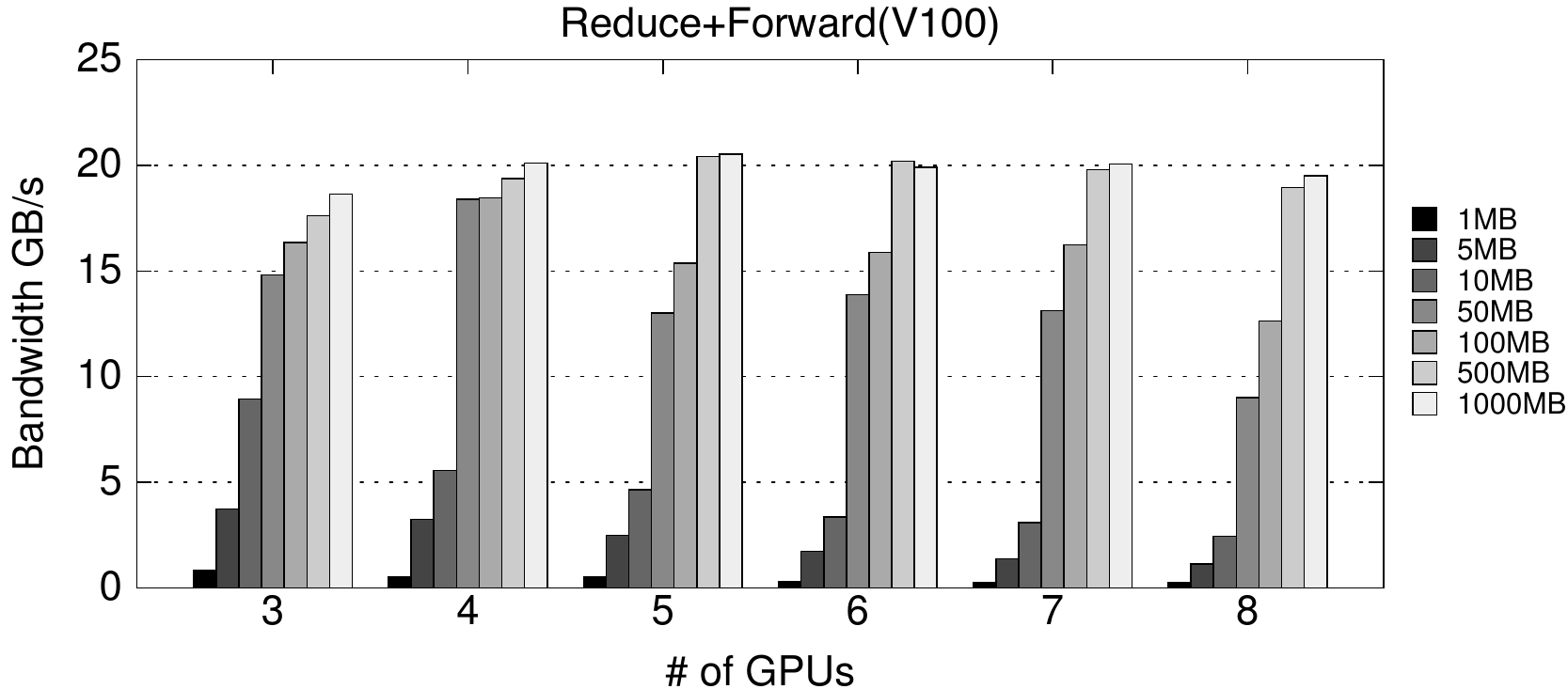}}
\subfigure[chain reduce-broadcast throughput]{\label{fig:chain-reduce-bcast-tput}
\includegraphics[width=0.45\textwidth]{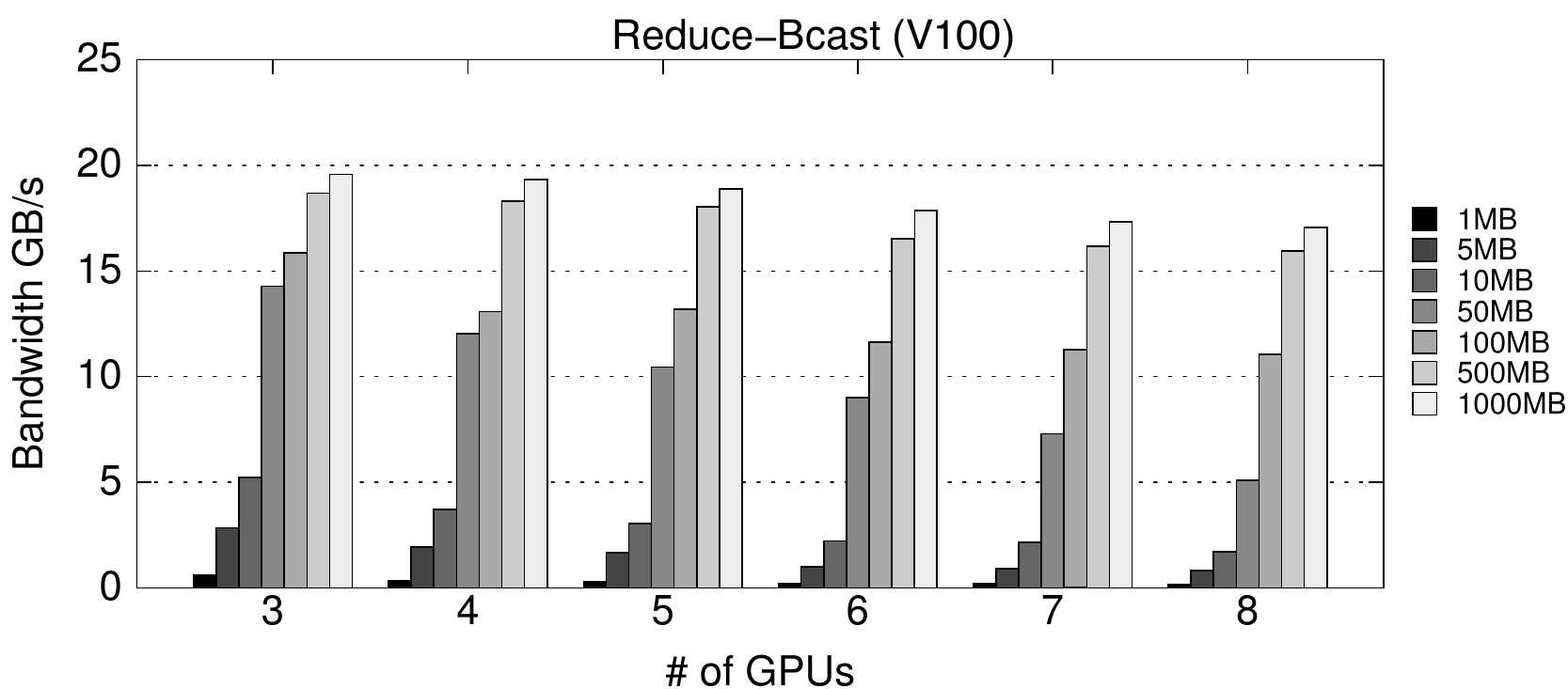}}
%\vspace{-6mm}
\caption{Depth test throughput over a chain of GPUs.}
\label{fig:depth-tput-1}
%\vspace{-4mm}
\end{figure}

%We test forward, reduce+forward and reduce-broadcast cases mentioned above. 
We test these operations over different number of GPUs (3-8GPU) and vary data sizes from 1MB to 1000MB. Results from the experiment are shown in  Figure~\ref{fig:depth-tput-1}. In the case of forward only, as we increase the chain length, the throughput decreases from around 22 GB/s (with 3GPU) to around 20 GB/s (8 GPU case) for 1000MB. The impact is less visible for reduce+forward where throughput is around 18GB/s. Finally for reduce-broadcast where the depth of the tree is now doubled, we see the throughput drops from 19GB/s to around 16GB/s for 1000MB.

We also see that in all three cases the throughput drops as the dataset size becomes smaller. This is related to the fact that it is hard to saturate very fast links with small data sizes and constant overheads in invoking CUDA operations become more significant at smaller data sizes.
 \subsubsection{Breadth Test}
 \label{appendix:breadthTest}
% Orthogonal to the depth test, we next consider breadth tests which measure the throughput for fan-in (many-to-one) and fan-out (one-to-many) communication patterns. We mainly include three types of tests, namely, fan-in forward, fan-in reduce+forward, fan-out forward.

\begin{figure*}[h]
\begin{minipage}{1\textwidth}
   \centering
     \subfigure[Fan-in forward]{\label{fig:fan-in-forward}
     \includegraphics[width=0.32\textwidth]{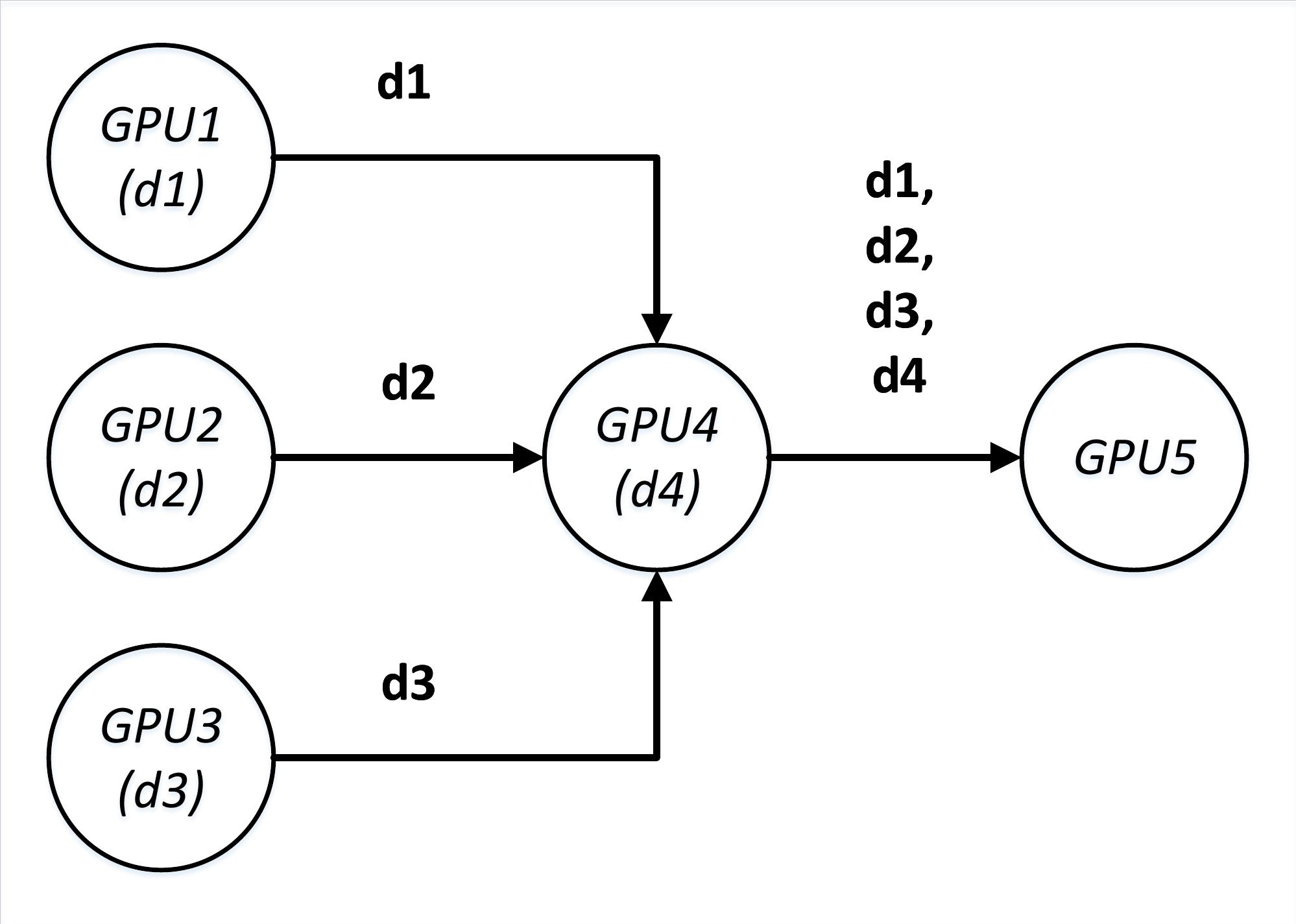}}
    \subfigure[Fan-in reduce+forward]{\label{fig:fan-in-add-forward} 
     \includegraphics[width=0.32\textwidth]{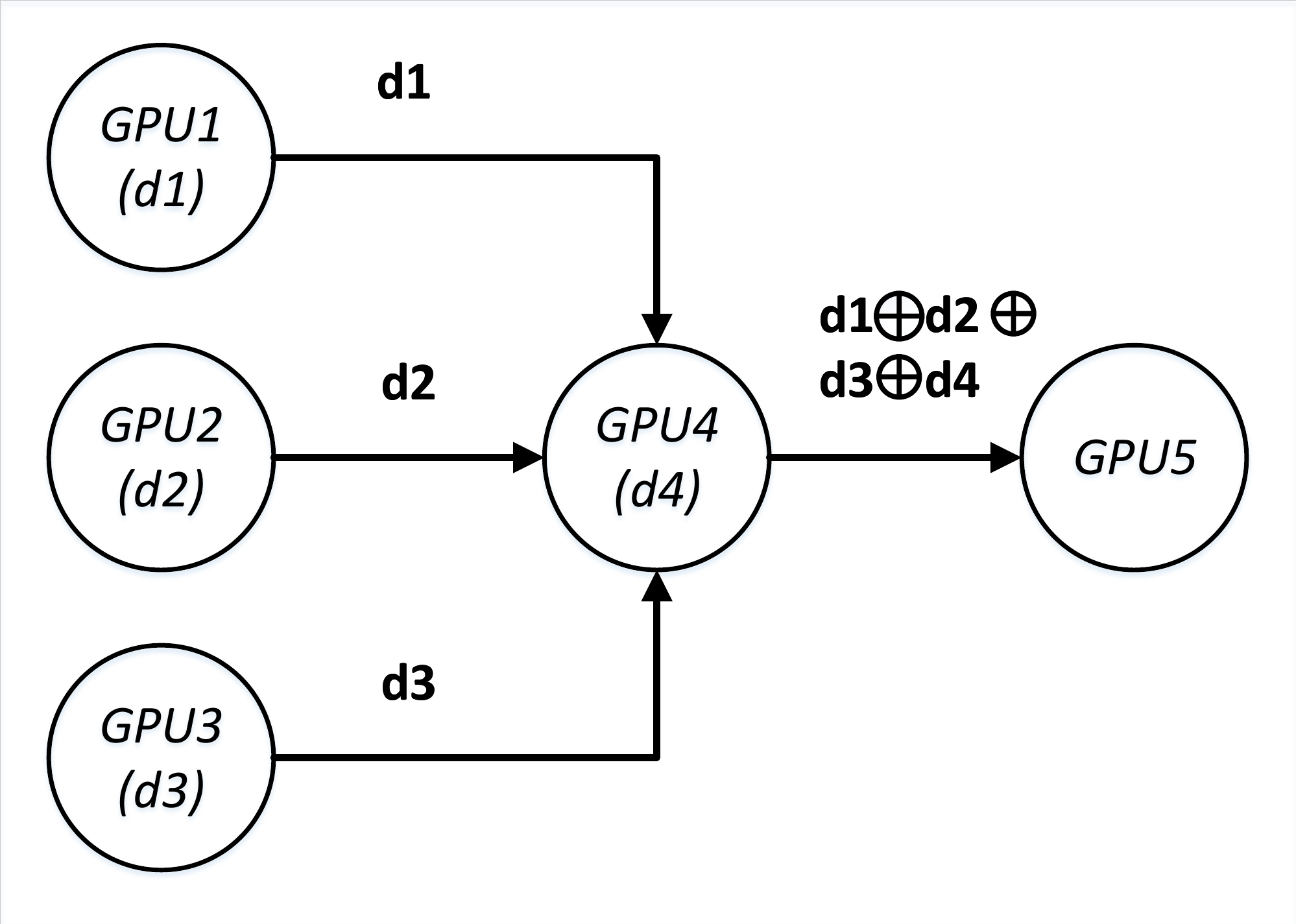}}
     \subfigure[Fan-out forward]{\label{fig:fan-out-forward} 
     \includegraphics[width=0.32\textwidth]{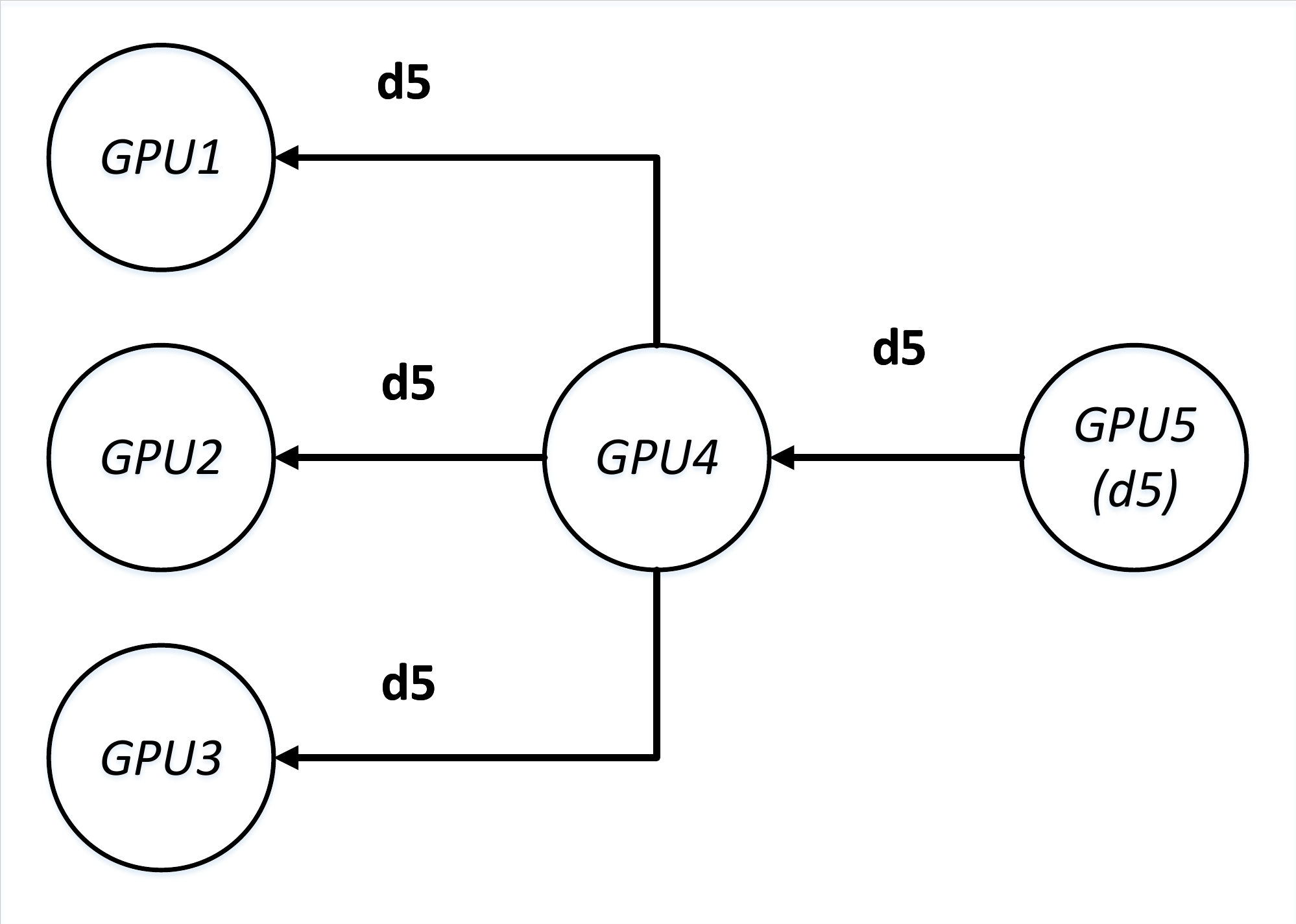}}
%   \vspace{-4mm}
   \caption{Breadth test of data forward, reduce+forward in fan-in and fan-out topologies.}
   \label{fig:breadth}
 \end{minipage}\hfill
% \vspace{-4mm}
\end{figure*}

 As illustrated in Figure~\ref{fig:fan-in-forward}, in fan-in forward, a center node (i.e. GPU4) collects data from multiple nodes and then forwards the collected data to its successor. Instead of just forwarding data, in the case of fan-in reduce+forward (Figure~\ref{fig:fan-in-add-forward}), the center node computes a reduction function over the incoming data and its own data, then forwards the result to it successor. Fan-out forward (Figure~\ref{fig:fan-out-forward}), is just the reverse of fan-in forward, in which the center node receives data from one node (i.e. GPU5), then multicasts the received data to its successors (i.e. GPU 1,2,3). 

\begin{figure*}
 \centering
 \subfigure[Fan-in forward throughput]{\label{fig:fan-in-forward-tput}
 \includegraphics[width=0.32\textwidth]{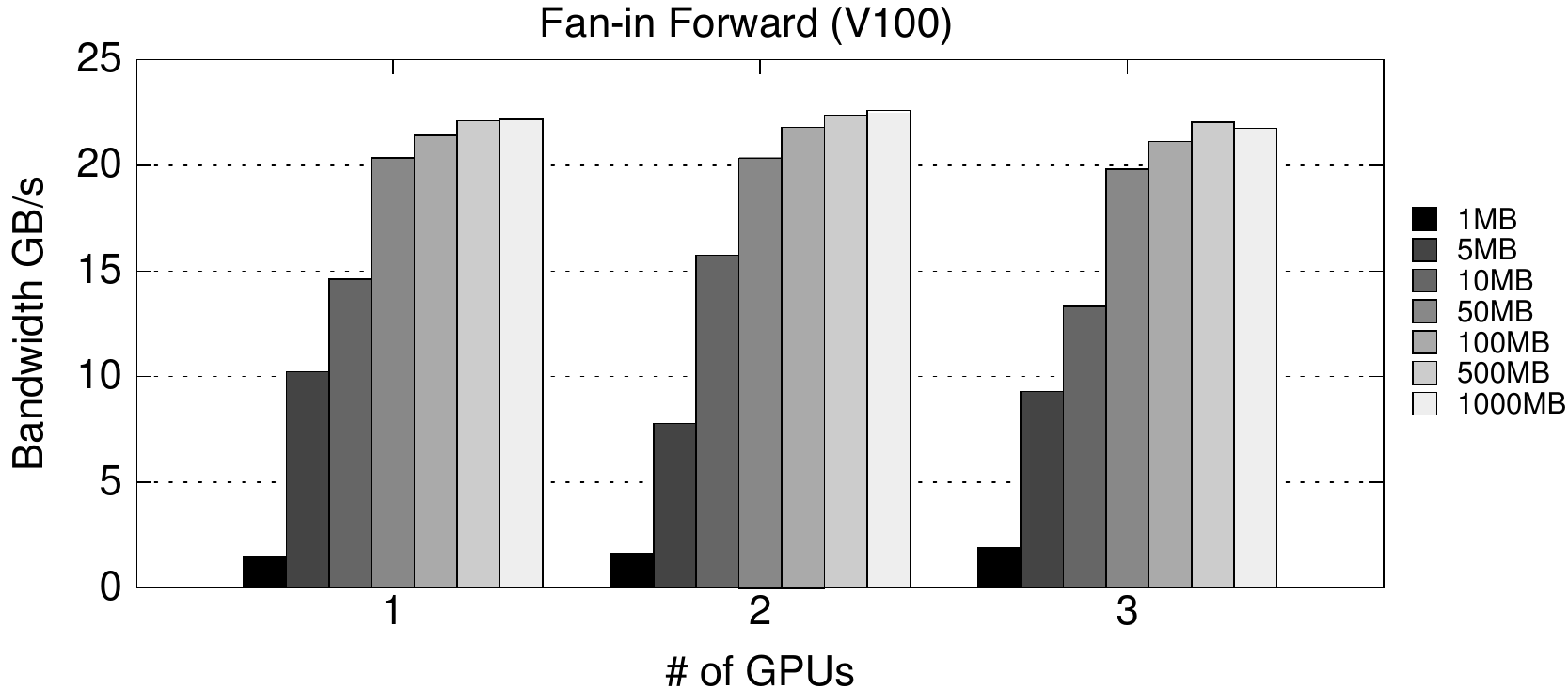}}
 \subfigure[Fan-in reduce+forward throughput]{\label{fig:fan-in-add-forward-tput} 
 \includegraphics[width=0.32\textwidth]{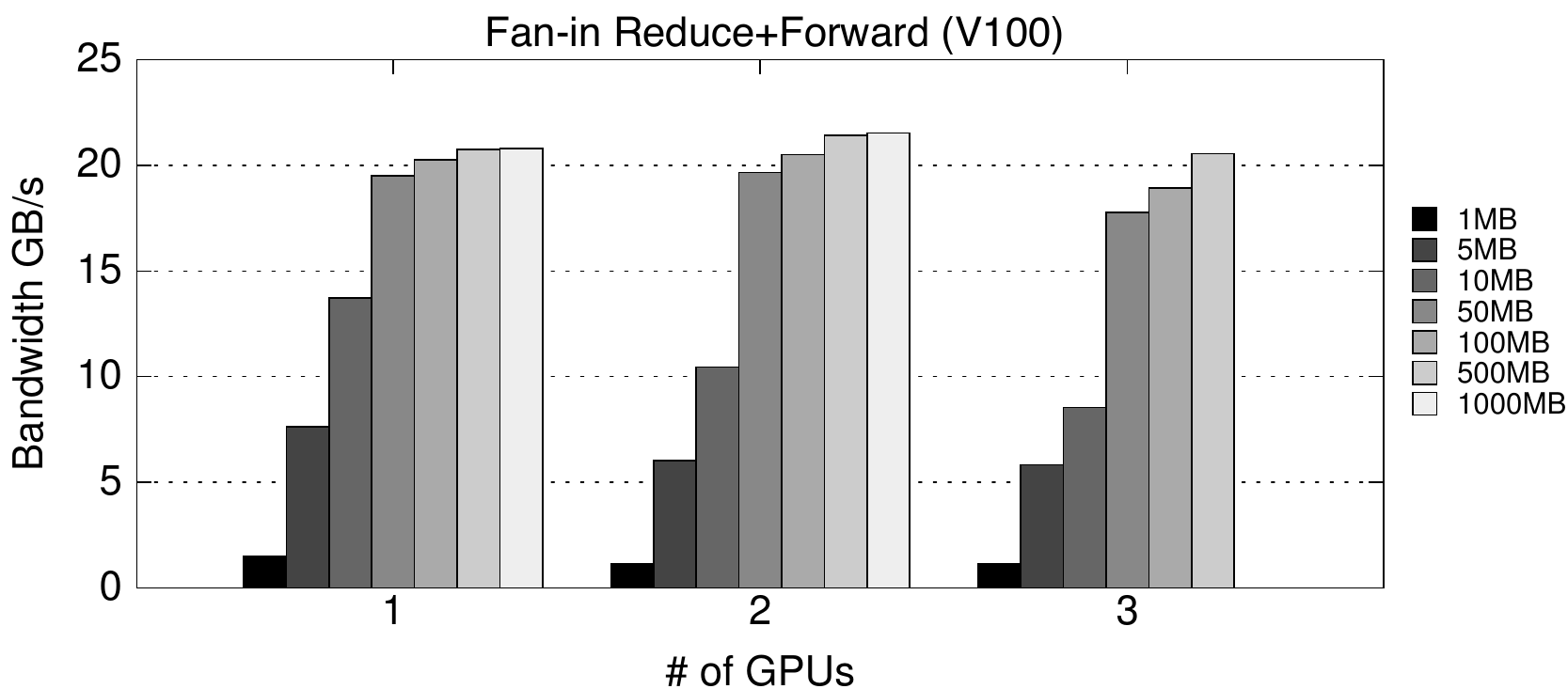}}
 \subfigure[Fan-out forward throughput]{\label{fig:fan-out-forward-tput} 
 \includegraphics[width=0.32\textwidth]{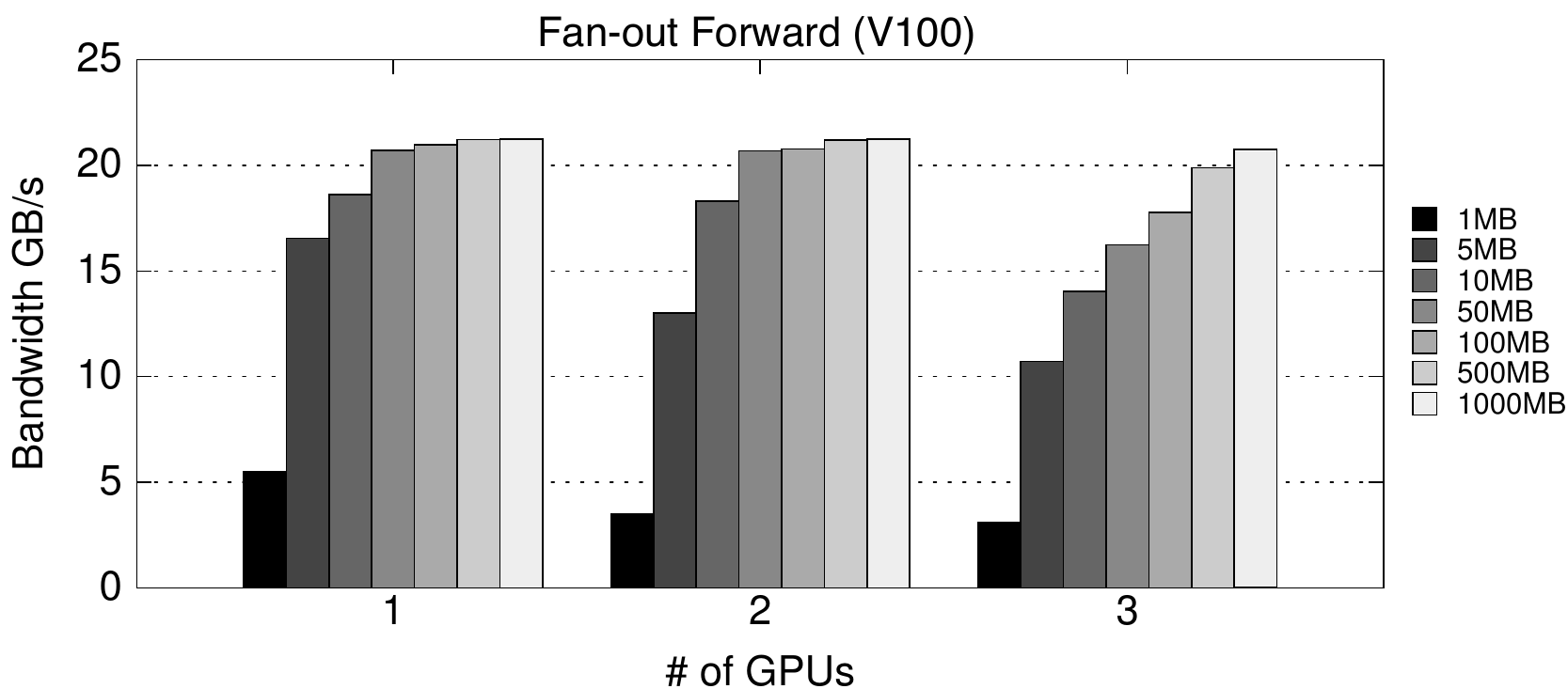}}
 %\vspace{-4mm}
 \caption{Breadth test throughput for Fan-in forward, Fan-in reduce+forward, Fan-out forward.}
 \label{fig:breadth-tput}
% \vspace{-4mm}
\end{figure*}

 We experiment with different data size as we vary the number of GPUs that serve as fan-in source nodes or fan-out destination nodes. For DGX-1s (i.e. both \dgxp and \dgxv), the maximum fan-in and fan-out degrees are limited to three.
 For brevity, we omit the graphs and highlight the key findings.  Similar to the depth tests, with data size >50MB, fan-in and fan-out forward achieves near maximum throughput. Compared with fan-in forward, the throughput of fan-in reduce+forward decreases 1-2 GB/s on average due to the latency of launching reduction function kernels on the center node (GPU4).

 Figure~\ref{fig:breadth} depicts result of breadth tests with different data size as we vary the number of GPUs that serve as fan-in source nodes or fan-out destination nodes. We'd like to note that for the given topology of \dgxv, the maximum fan-in and fan-out degrees are limited to three.
% % (e.g. 3 GPUs (GPU 1,2,3) in Figure~\ref{fig:breadth})
 In Figure~\ref{fig:fan-in-forward}, with data size >50MB, in all three cases, fan-in forward achieves near maximum throughput. Compared with fan-in forward, the throughput of fan-in reduce+forward (in Figure~\ref{fig:fan-in-add-forward}) decreases 1-2 GB/s on average due to the latency of launching reduction function kernels on the center node (GPU4). We also note that running with 1000MB and a fan-in of $3$ requires allocating memory for each incoming link and this exceeds the amount of memory available. Finally, for fan-out forward in Figure~\ref{fig:fan-out-forward}, the throughput is again close to the peak link bandwidth.

\end{document}